\documentclass[sigconf]{acmart}
\AtBeginDocument{%
  \providecommand\BibTeX{{%
    \normalfont B\kern-0.5em{\scshape i\kern-0.25em b}\kern-0.8em\TeX}}}

\copyrightyear{2024}
\acmYear{2024}
\setcopyright{rightsretained}
\acmConference[CHI '24]{Proceedings of the CHI Conference on Human Factors in Computing Systems}{May 11--16, 2024}{Honolulu, HI, USA}
\acmBooktitle{Proceedings of the CHI Conference on Human Factors in Computing Systems (CHI '24), May 11--16, 2024, Honolulu, HI, USA}
\acmDOI{10.1145/3613904.3642278}
\acmISBN{979-8-4007-0330-0/24/05}

\usepackage{multirow}
\usepackage{xspace}
\usepackage{xcolor}

\newcommand{\paragraphBold}[1]{\paragraph{\textbf{\emph{#1}}}}

\begin{document}

\title{Wikibench: Community-Driven Data Curation for AI Evaluation on Wikipedia}

\author{Tzu-Sheng Kuo}
\email{tzushenk@cs.cmu.edu}
\orcid{0000-0002-1504-7640}
\affiliation{%
  \institution{Carnegie Mellon University}
  \city{Pittsburgh}
  \state{PA}
  \country{USA}
}

\author{Aaron Halfaker}
\email{aaron.halfaker@gmail.com}
\orcid{0000-0001-8907-6367}
\affiliation{%
  \institution{Microsoft}
  \city{Redmond}
  \state{WA}
  \country{USA}
}

\author{Zirui Cheng}
\authornote{These authors, listed in alphabetical order by last name, contributed equally.}
\email{chengzr19@mails.tsinghua.edu.cn}
\orcid{0000-0001-7087-8220}
\affiliation{%
  \institution{Tsinghua University}
  \city{Beijing}
  \country{China}
}

\author{Jiwoo Kim}
\authornotemark[1]
\email{jk4671@columbia.edu}
\orcid{0009-0008-0898-8371}
\affiliation{%
 \institution{Columbia University}
 \city{New York}
 \state{NY}
 \country{USA}
}

\author{Meng-Hsin Wu}
\authornotemark[1]
\email{soniawu2302@gmail.com}
\orcid{0009-0007-4919-0801}
\affiliation{%
  \institution{Carnegie Mellon University}
  \city{Pittsburgh}
  \state{PA}
  \country{USA}
}

\author{Tongshuang Wu}
\email{sherryw@cs.cmu.edu}
\orcid{0000-0003-1630-0588}
\affiliation{%
  \institution{Carnegie Mellon University}
  \city{Pittsburgh}
  \state{PA}
  \country{USA}
}

\author{Kenneth Holstein}
\authornote{Co-senior authors contributed equally.}
\email{kjholste@cs.cmu.edu}
\orcid{0000-0001-6730-922X}
\affiliation{%
  \institution{Carnegie Mellon University}
  \city{Pittsburgh}
  \state{PA}
  \country{USA}
}

\author{Haiyi Zhu}
\authornotemark[2]
\email{haiyiz@cs.cmu.edu}
\orcid{0000-0001-7271-9100}
\affiliation{%
  \institution{Carnegie Mellon University}
  \city{Pittsburgh}
  \state{PA}
  \country{USA}
}

\renewcommand{\shortauthors}{Kuo et al.}

\begin{abstract}
AI tools are increasingly deployed in community contexts. However, datasets used to evaluate AI are typically created by developers and annotators outside a given community, which can yield misleading conclusions about AI performance. How might we empower communities to drive the intentional design and curation of evaluation datasets for AI that impacts them? We investigate this question on Wikipedia, an online community with multiple AI-based content moderation tools deployed. We introduce Wikibench, a system that enables communities to collaboratively curate AI evaluation datasets, while navigating ambiguities and differences in perspective through discussion. A field study on Wikipedia shows that datasets curated using Wikibench can effectively capture community consensus, disagreement, and uncertainty. Furthermore, study participants used Wikibench to shape the overall data curation process, including refining label definitions, determining data inclusion criteria, and authoring data statements. Based on our findings, we propose future directions for systems that support community-driven data curation.
\end{abstract}



\begin{teaserfigure}
  \includegraphics[width=\textwidth]{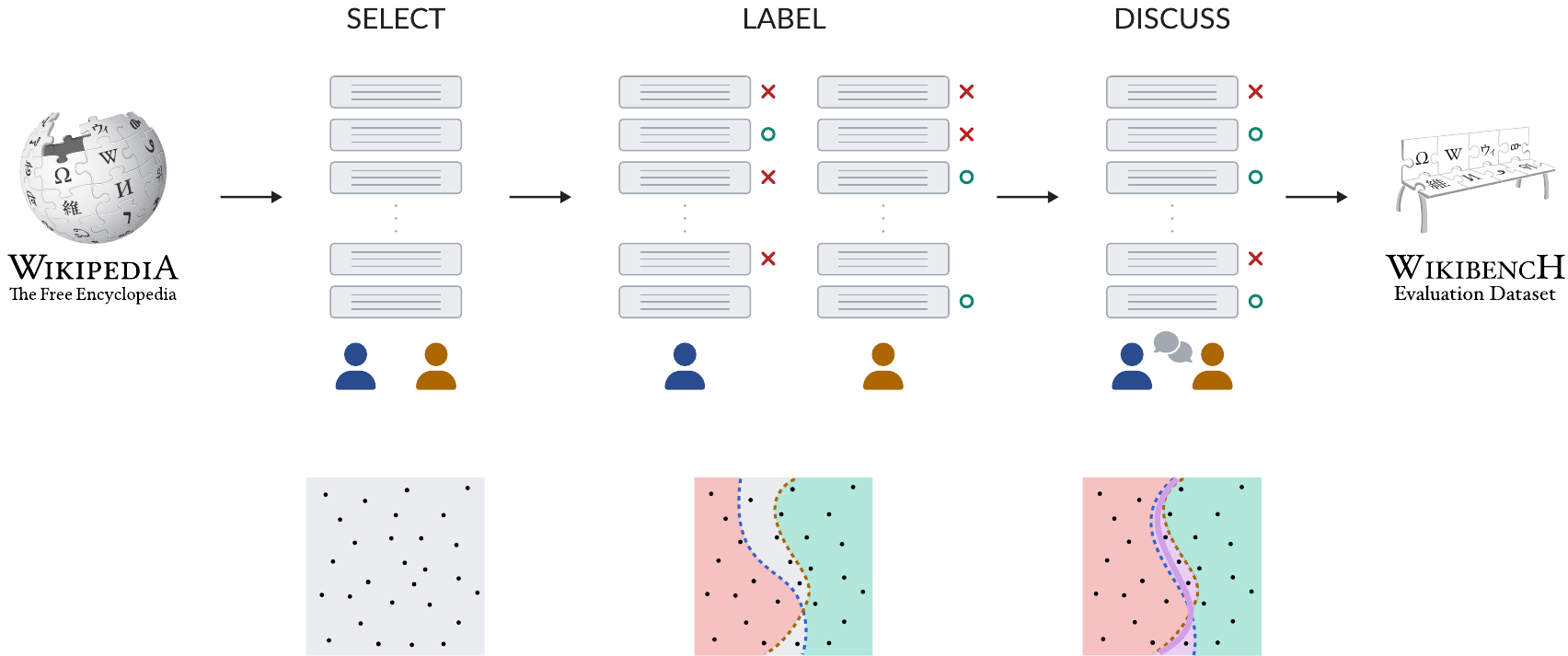}
  \caption{An overview of Wikibench's approach to supporting community-driven data curation. The top row illustrates community members' use of Wikibench to \textit{select} data points (e.g., edits on Wikipedia) for inclusion in the dataset, \textit{label} data points with ``individual'' labels based on their own initial judgments, and then \textit{discuss} their perspectives and collectively decide on a ``primary'' label for the data point. The bottom row represents data points in a conceptual 2D space. As each community member labels data points, their labels form \textit{decision boundaries} in aggregate (orange and blue dotted curves). Through discussion, participants may resolve some disagreements or clarify ambiguities in labeling, leading to changes in their individual labels. In addition, community members decide on a primary label for each data point, forming a consensus-based decision boundary (purple curve). Wikibench datasets preserve information about disagreement among community members (purple shaded region). The Wikipedia logo is licensed by Wikimedia Foundation, CC BY-SA 3.0, via Wikimedia Commons.
  }
  \Description{An illustration with two rows. The top row illustrates how community members select, label, and discuss data points using Wikibench. The bottom row illustrates how these actions reflect a conceptual 2D space.}
  \label{figure:teaser}
\end{teaserfigure}

\maketitle

\section{Introduction}

AI tools are increasingly deployed in \textit{community contexts}. For example, AI-based content moderation tools have been deployed in online communities such as Wikipedia and Reddit~\cite{halfaker2020ores, jhaver2019human}. AI-based decision-making tools have also been adopted by local governments to prioritize public services, such as allocating local housing resources~\cite{kuo2023understanding, showkat2023right}. However,
the datasets used to evaluate AI performance are typically designed, curated, and labeled by developers and data annotators outside of a given community, which can lead to misleading conclusions about AI systems' ``fit for use''~\cite{geiger2020garbage}. In turn, the deployment of poorly-fit AI tools can yield compromised user experiences or even cause harm to vulnerable populations~\cite{muller2021designing,raji2021ai,sap2021annotators}. For example, research shows that crowdsourced datasets systematically label innocuous phrases in African American English (AAE) dialects as toxic~\cite{sap2019risk}. As a consequence, if such datasets were used to prospectively evaluate content moderation tools' fit for use in a community that uses AAE, they would \textit{underestimate} the tools' false positive rates, compared with what the community would experience in deployment~\cite{sap2021annotators}.

Given that what constitutes ``good performance'' on tasks such as content moderation can be highly community-specific, recent work has argued that HCI and machine learning research should explore more \textit{community-driven} approaches to AI dataset development. For instance, in a position paper, \citet{jo2020lessons} propose that AI should draw lessons from archive and library studies, where archives are often directly contributed and curated by the communities they are meant to represent, instead of by community-outsiders. These community archives, such as the Feminist Archive and the Working Class Movement Library, are motivated by the need to represent the voices of non-elites and the marginalized~\cite{flinn2007community}. The authors argue that these traditions should inspire new approaches to AI \textit{data curation} that allow communities greater voice in specifying their collective desires for AI performance.

In the context of AI evaluation, \textit{data curation} refers to the process of designing the ``ground truth'' against which AI models' performance will be evaluated~\cite{muller2021designing}. This involves an intentional process of selecting \textit{which data points} should be included in a dataset and, in the case of labeled datasets, deciding \textit{how each data point should be labeled}~\cite{muller2019data}. For example, when developing an AI dataset for content moderation tools on Wikipedia, a ``data point'' could be an edit to an article, and its ``label'' could be a judgment of whether the edit should be considered ``damaging'' to the article or not~\cite{halfaker2020ores}. The intentional curation of AI evaluation datasets stands in stark contrast with what \citet{jo2020lessons} term ``laissez faire'' approaches to dataset development, which indiscriminately take data in masses by crawling trace data on the web~\cite{geiger2020garbage}. On their own, such datasets simply capture how people \textit{have behaved} in the past. However, they often fail to capture communities’ normative beliefs about how decisions \textit{should be made}, for evaluation purposes~\cite{jo2020lessons,raji2021ai}.

Realizing the vision of \textit{community-driven data curation} of AI datasets in practice poses numerous open challenges. For example, while a community may share broad norms and values~\cite{fiesler2018reddit,chandrasekharan2018internet}, individual community members may disagree about how specific data points should be labeled (e.g., whether a given post should be considered ``toxic'')~\cite{cabitza2023toward,gordon2021disagreement,goyal2022your}. In some cases, these disagreements may represent substantive differences in perspective, while in other cases, a brief discussion between individuals could reveal that they actually agree more than they disagree~\cite{chen2023judgment,muller2021designing}. Current approaches to account for annotator disagreement in crowdsourced datasets tend to handle disagreements post-hoc (after data have already been labeled), either by resorting to the majority vote~\cite{feffer2023moral} or by attempting to model individual subjectivity for re-weighted voting~\cite{gordon2022jury,dutta2023modeling,bakker2022fine,deng2023you,wallace2022debiased}. However, when it comes to deciding how important community decisions should be made, it is crucial that community members have opportunities to collectively build meaning and understand each other's perspectives. In contrast to prior methods, this calls for more collaborative and deliberative approaches that allow community members agency in navigating disagreements, via processes that are perceived to be fair by community members~\cite{lee2019procedural}. Furthermore, beyond selecting and labeling individual data points, it is critical to provide communities with the agency to shape higher-level decisions, such as crafting label definitions and determining data inclusion criteria. Finally, given that community members will generally have limited time and attention to contribute to the curation of AI datasets, it is important to support them in prioritizing their efforts. To the best of our knowledge, despite recent calls-to-action from the research community~\cite{muller2019data,jo2020lessons, muller2021designing,denton2021whose,raji2021ai,denton2020bringing}, there are no existing tools aimed at addressing these challenges to support the intentional, community-driven curation of AI datasets in practice.

We identify and address these challenges in the context of Wikipedia, an online community where multiple AI-based content moderation tools have been deployed, but where community members currently have limited means to prospectively assess these tools' fit for use. Through formative interviews with Wikipedia community members and AI developers, we derived a set of design requirements for systems that aim to support community-driven data curation. Based on these design requirements, we then developed Wikibench, a system that enables community members to collaboratively curate AI evaluation datasets, while navigating disagreements and ambiguities through discussion. As illustrated in Figure~\ref{figure:teaser}, community members can use Wikibench to select data points for inclusion in datasets, label data points with ``individual labels'' reflecting their personal judgments, and discuss their perspectives to decide upon a ``primary label'' for the data point. Through a field study on Wikipedia, we find that datasets curated using Wikibench can effectively capture community consensus, disagreement, and collective uncertainty. We demonstrate how Wikibench datasets can help in understanding areas of alignment and misalignment with community perspectives. Furthermore, we gain insight into the ways Wikipedia community members collaborate using Wikibench. We find that participants in our study used Wikibench to proactively shape the overall data curation process beyond labeling data, including refining label definitions, determining data inclusion criteria, and authoring data statements. 

Overall, this work demonstrates the potential of \textit{community-driven} data curation, and contributes the following:

\begin{itemize}
\item{\textbf{System}}: We introduce Wikibench, the first system that supports community-driven curation of AI datasets.
\item{\textbf{Field study}}: We present findings from a field study on Wikipedia to understand how Wikipedia community members interact with this system to collaboratively curate evaluation datasets.
\item{\textbf{Future directions}}: Based on our findings, we propose future directions for HCI systems that support community-driven data curation within and beyond the context of Wikipedia.
\end{itemize}

In the rest of the paper, we first review relevant literature and introduce our study context (Section~\ref{section:relatedwork}--\ref{section:studycontext}). Next, we present the design requirements for Wikibench and walk through the system (Section~\ref{section:designrequirements}--\ref{section:wikibench}). We then describe our evaluation of Wikibench (Section~\ref{section:evaluation}--\ref{section:finding/usage}), and conclude with future directions (Section~\ref{section:discussion}).

\section{Related Work}\label{section:relatedwork}

\subsection{Developer-Centric AI Evaluation}
AI datasets are commonly created by developers and data annotators with limited knowledge about the real-world contexts in which these AI models will be deployed, assuming a ``one-dataset-fits-all'' approach to evaluation~\cite{raji2021ai}. For example, widely used datasets for toxicity classification of online comments, such as Jigsaw's Toxic Comments~\cite{jigsaw-toxic-comment-classification-challenge} and Civil Comments~\cite{borkan2019nuanced}, are commissioned by AI developers and labeled by crowd workers. While these datasets are commonly framed as benchmarks of progress toward general abilities, such as ``toxicity detection,'' researchers argue that they are often ineffective for evaluating how an AI model will perform in real-world contexts~\cite{raji2021ai}. One reason general benchmarks can fail is that many tasks currently targeted by AI models are inherently norm- and value-laden~\cite{denton2021whose,gordon2021disagreement,kawakami2022care,kapania2023hunt,goyal2022your}. For instance, a comment that is considered ``inappropriate'' in the context of one online community may be within bounds of acceptability for those in a different community, with different norms and values~\cite{sap2019risk}. As a result, a one-dataset-fits-all approach can yield misleading conclusions when used to prospectively evaluate how well an AI model will perform in a particular community context~\cite{peng2021mitigating,sap2019risk,sap2021annotators}. These concerns have informed a recent line of research that directly involves end-users in AI evaluation and benchmark development.

\subsection{Broadening Participation in AI Evaluation}\label{section:relatedwork/userengaged}
As dataset issues cascade down to deployment~\cite{sambasivan2021everyone}, end-users have often surfaced instances of AI misbehavior through their everyday use~\cite{devos2022toward, shen2021everyday, metaxa2021auditing, noble2018algorithms, eslami2019user, eslami2017careful}. For example, Twitter users discovered that the platform's image cropping algorithm favored light-skinned over dark-skinned individuals when both are in an image~\cite{yee2021image}. Similarly, \citet{halfaker2020ores} document how various Wikipedia language communities have engaged in ad-hoc, bottom-up efforts to identify language-specific error patterns in Wikipedia's AI-based content moderation tools. As acknowledged by AI developers, end-user involvement in testing and auditing AI behavior can be extremely valuable~\cite{deng2023understanding}. Given end-users' situatedness in specific contexts where AI tools will be used, they can often surface issues that would otherwise be missed~\cite{shen2021everyday}. Recently, the HCI community has proposed several systems that support individual users in testing and auditing AI behavior~\cite{lam2022end, cabrera2021discovering,lam2023sociotechnical}. Facilitating user collaboration remains an area for further exploration~\cite{devos2022toward}.

Beyond collecting evidence of AI misbehavior, several efforts have focused on the creation of \textit{new benchmark datasets} by challenging crowdworkers and volunteers to uncover AI models' blind spots and then adding these instances to their new evaluation datasets~\cite{aroyo2021adversarial,kiela2021dynabench,attenberg2015beat, wallace2019trick, bartolo2020beat, nie2019adversarial,dinan2019build,bai2023measuring}. For example, the CATS4ML Data Challenge asked challenge participants to submit misclassified Google Open Images to create a new evaluation dataset~\cite{aroyo2021adversarial}. Similarly, the Adversarial NLI benchmark was created by challenging crowdworkers to draft 
text snippets that existing AI models could not understand~\cite{nie2019adversarial}. Dynabench and DataPerf are centralized platforms that host several of these data challenges~\cite{kiela2021dynabench,mazumder2022dataperf}. 

Existing approaches to broadening involvement in AI evaluation, such as those overviewed above, differ from our vision of community-driven AI evaluation in several ways. First, these approaches have typically focused on engaging end-users, crowdworkers, and other volunteers in identifying cases where specific AI models misbehave, rather than in proactively specifying what behavior and performance they \textit{want} to see from AI models. Second, in current approaches, individuals work independently or in competition with one another, rather than collaboratively. Finally, current approaches tend to recruit broadly, without a focus on capturing perspectives held by \textit{particular communities}. In the following subsection, we briefly overview existing scholarship relevant to the vision of \textit{community-driven} AI evaluation.

\subsection{Community-Driven AI Evaluation}\label{section:relatedwork/community}

An emerging body of research has advocated for empowering communities to shape the design of AI evaluation datasets~\cite{denton2020bringing, jo2020lessons, muller2021designing, denton2021whose, raji2021ai}. For instance, \citet{jo2020lessons} argue that AI \textcolor{black}{dataset development should learn from the rich traditions of \textit{community-driven} curation of archives in library studies.} Because community archives are motivated by the need to represent non-elites and marginalized voices~\cite{flinn2007community}, \citet{jo2020lessons} argue that they can serve as a model for how the design of AI datasets might be opened up for community input. Yet realizing the vision of \textit{community-driven} AI data curation in practice poses numerous open challenges. For example, a growing body of work in HCI and machine learning has argued that the notion of a single, objective ground truth label often does not apply when AI is deployed in complex social contexts, where different groups may have distinct perspectives~\cite{muller2021designing, denton2021whose, chen2023judgment, gordon2021disagreement, kapania2023hunt, kawakami2022care, siddarth2021ai}. In some cases, these disagreements may stem from genuine
differences in perspective, while in other cases, a brief discussion between individuals could reveal that they actually agree more than they disagree~\cite{chen2023judgment,muller2021designing}. However, current approaches to account for annotator disagreements tend to handle disagreements \textit{post-hoc} after individual labels have been gathered~\cite{gordon2022jury,dutta2023modeling,bakker2022fine,deng2023you,wallace2022debiased}, instead of facilitating discussion and deliberation among annotators. When deciding how important decisions should be made in community contexts, it is critical that community members have opportunities to discuss, understand each other's perspectives, and collectively build meaning~\cite{muller2021designing}. This calls for more collaborative, deliberative approaches that allow community members agency in navigating disagreements, through processes that they perceive to be fair and appropriate~\cite{lee2019procedural, leventhal1980should,thibaut1975procedural}.

We note that while some online platforms have existing community-driven \textit{content curation} mechanisms, their purpose is distinct from AI dataset curation. \textcolor{black}{For example, Reddit and Stack Overflow have implemented community voting systems to enable the curation of high-quality posts~\cite{gilbert2013widespread,mamykina2011design}. Similarly, Wikipedia allows community members to revert damaging edits in order to maintain article quality~\cite{geiger2010work}. These \textit{content curation} mechanisms differ from \textit{AI data curation} in two aspects. First,} in the context of AI evaluation, \textit{data curation} refers to an intentional process of designing the ``ground truth'' against which AI models' performance will be evaluated~\cite{muller2021designing}. These datasets aim to represent community members' collective beliefs about what constitutes ``good performance'' on a given task (e.g., content moderation) in the context of their community. In contrast, \textcolor{black}{votes, reverts, and other trace data generated through existing content curation processes} can carry complex meanings, which will often be misaligned with the goals of an AI evaluation~\cite{guerdan2023ground,papakyriakopoulos2023upvotes}. \textcolor{black}{For instance, on Reddit and Stack Overflow, posts may receive downvotes for reasons unrelated to the violation of the community's content moderation policy~\cite{ford2018we}. Similarly, Wikipedia edits can be reverted for reasons beyond causing damage to an article~\cite{kittur2007he,halfaker2011don}. Relying on these trace data as proxy labels for AI evaluation can introduce target variable bias, leading to misleading evaluations of AI performance~\cite{guerdan2023ground}. Second,} in the same vein, while past work has often used historical human decisions as ground truth for evaluating AI-based decision-making tools, these trace data capture only how decisions \textit{have been made} in the past---biases, errors, and all---not how a community believes decisions \textit{should be made}~\cite{geiger2020garbage}. \textcolor{black}{Therefore, recent work has argued for the necessity of intentionally curated evaluation datasets, to support meaningful and reliable AI evaluations~\cite{jo2020lessons, muller2021designing,denton2021whose,raji2021ai,denton2020bringing}. Despite these differences,} Section~\ref{section:discussion} will discuss how tools for community-driven AI data curation \textcolor{black}{may draw inspiration from} existing content curation mechanisms.

Beyond recent calls-to-action for the research community, to our knowledge no tools currently exist to address the challenges described above to support intentional, community-driven curation of AI datasets in practice. The current work is the first system in the literature aimed at supporting community-driven AI data curation.

\section{Study Context}\label{section:studycontext}
We conduct this study in the context of Wikipedia for several reasons. First, Wikipedia has a rich history of grassroots engagement to explore new modes of participation in AI development and evaluation~\cite{halfaker2020ores, smith2020keeping}, as mentioned in Section~\ref{section:relatedwork/userengaged}. However, while community members are motivated to improve the AI-based tools they use and are impacted by, there is currently no infrastructure to support them in \textit{proactively} curating datasets for AI evaluation and improvement. Thus, our research focus is well-aligned with existing interests and motivations among Wikipedia community members, and this context presents an opportunity to develop a system that is truly useful to the community. In addition, the Wikipedia context has established norms for collaborative efforts (e.g., for article editing)~\cite{forte2009decentralization, kittur2007he, beschastnikh2008wikipedian}. Our focus on Wikipedia enables us to build upon these existing community norms when exploring new mechanisms for community-driven data curation, thus bypassing the need to develop and introduce entirely new collaboration processes.

The remainder of this section briefly discuss the current AI evaluation challenges faced by the Wikipedia community and our positionality as researchers working with community members. In the remainder of this paper, we refer to Wikipedia's community members as ``Wikipedians,'' following community terminology. 

\subsection{Challenges of AI Evaluation on Wikipedia}\label{section:studycontext/challenges}
As Wikipedia scales, the community increasingly relies on AI tools for governance~\cite{muller2013work, halfaker2013rise, geiger2017operationalizing}. For example, AI-based content moderation tools are used to identify damaging edits in articles for Wikipedians to review and revert them as necessary ~\cite{halfaker2013rise, geiger2010work, geiger2013levee}. Among various content moderation tools, ORES, an AI model hosting system, is used extensively in English Wikipedia and many other languages~\cite{halfaker2020ores}. Using basic estimation, \citet{halfaker2020ores} argue that without ORES's AI model for detecting damaging edits, it would take 483 labor hours per day to review the 290k edits made to all the various language editions of Wikipedia, but with an AI model, that workload can be reduced by 90\%.

Despite the growth of AI tools, Wikipedia communities currently have limited means to evaluate particular AI tools' ``fit for use'' with respect to their collective norms and values. Currently, the curation of ORES' training and evaluation datasets relied on a system called Wikilabels\footnote{\url{https://meta.wikimedia.org/wiki/Wiki\_labels}}, which is hosted on an external website outside Wikipedia. Wikipedians can join \textit{data labeling campaigns} on Wikilabels and request a subset of data to label. However, unlike Wikipedia where each article is editable by any Wikipedian, Wikilabels assigns each data point to only a single Wikipedian for labeling in isolation. Wikilabels also doesn't enable Wikipedians to discuss labels collaboratively, unlike Wikipedia, where each article has an associated talk page\footnote{\url{https://enwp.org/WP:TALKPAGE}} for discussing its content. As such, it is less clear to what extent Wikilabels' datasets reflect the collective perspectives of the community, versus just the perspectives of some individual annotators. Besides labeling preselected data in Wikilabels, Wikipedians currently do not have a way to proactively evaluate how different AI tools perform with respect to their collective norms and values.

\subsection{Positionality and Ethical Considerations}\label{section:positionality}
Our research has dual objectives. First, we are interested in exploring new approaches to support community-driven data curation. Second, we hope our research can truly benefit the Wikipedia community, in recognition that Wikipedia is not a laboratory\footnote{\url{https://enwp.org/WP:NOTLAB}} \cite{howard2019ways}. These dual objectives guided our decision-making throughout the study, from research method to system design. In cases where these two objectives conflict, we prioritize the community's needs, preferences, and established norms over our research interests~\cite{halfaker2020ores, irani2013turkopticon}. We took several precautions to ensure that we conducted ethical research on Wikipedia\footnote{\url{https://enwp.org/WP:Ethically\_researching\_Wikipedia}}. For example, we collaborated with an experienced Wikipedian deeply involved in the development of AI tools in Wikipedia and an academic researcher with over a decade of experience studying Wikipedia. We also adhered to the norm of researching Wikipedia by creating and iteratively updating a project page on Meta-Wiki\footnote{\url{https://w.wiki/7QGb}}, where we publicly shared the study's objective, protocol, timeline, recruitment message, and our institutional affiliation for Wikipedians to access. Finally, we recruited a minimal number of Wikipedians for the study, acknowledging that the study might take their time away from their volunteer work on Wikipedia. We hope the benefit of our research has the potential to outweigh the disruptions we inevitably caused.

\section{Design Requirements}\label{section:designrequirements}

To better understand Wikipedians' desires and challenges around data curation for AI evaluation, we conducted a formative study with eight Wikipedians who had experience \textit{using} AI-based content moderation tools on the platform (e.g., edit patrollers who use AI tools), contributing to the \textit{development} of these tools (as AI data labelers or engineers), or participating in grassroots efforts to \textit{identify areas for improvement} in deployed tools (see Table~\ref{table:formative} for an overview). In this phase of our research, we aimed to recruit Wikipedians across multiple language communities, with the goal of understanding desires and challenges on Wikipedia more broadly. 
To recruit participants, we adopted a snowball sampling approach. We first recruited a Wikipedian who had been heavily involved in AI development on Wikipedia. This Wikipedian then helped us reach out to a broader set of Wikipedians by pinging them on the associated talk page of our research page on Meta-Wiki, where Wikipedians could view our study description before signing up. Seven additional Wikipedians who self-identified with at least one of the five roles we targeted (Table~\ref{table:formative}) signed up either through a form we provided or by using Wikipedia's email feature. In total, we conducted synchronous interviews with seven Wikipedians and exchanged emails with one (W8) based on their preference. The interview was semi-structured and lasted for an hour with a \$30 USD compensation. Some participants declined our compensation, viewing the study as part of their volunteer work to improve Wikipedia. Our interview questions are shown in Appendix~\ref{section:appendix/formativestudy}.

\begin{table*}[t]
  \caption{The targeted roles for recruitment in the formative study and the participants who self-identified with these roles.}
  \label{table:formative}
  \begin{tabular}{lll}
    \toprule
    Role & Role Description & Participant ID \\
    \midrule
    Community organizer & organize community efforts around AI on Wikipedia & W1, W4, W5, W6 \\
    AI evaluator & have participated in efforts to identify and report AI errors & W1, W2, W6 \\
    AI user & use AI tools for their daily work on Wikipedia & W1, W2, W5, W8 \\
    AI engineer & develop AI tools and/or organize data labeling campaigns & W1, W3, W5, W7 \\
    AI data labeler & contribute to data labeling campaigns & W1, W4, W5\\
    \bottomrule
  \end{tabular}
\end{table*}

Through a reflexive thematic analysis~\cite{braun2012thematic, braun2019reflecting} and affinity diagramming by two of the authors, we derived the following four highest-level themes as design requirements for systems that aim to support the community-driven data curation process for AI evaluation on Wikipedia. We briefly summarize each requirement below:

\paragraphBold{D1: The data curation process should be led by the community and follow their established norms.}
Participants suggested that systems for community-driven data curation should provide communities with the agency to shape the overall data curation process, beyond labeling individual data points. In addition, participants emphasized that in order to succeed, the systems need to be flexible and adaptable to the varying norms of different Wikipedia language communities: \textit{``If we're talking about Wikipedia, make sure it adapts to the local rules. English [Wikipedia] are full of categories, [whereas] in the Dutch Wikipedia it's almost a crime to have more than ten categories''} (W6).

\paragraphBold{D2: The data curation process should encourage deliberation to surface disagreements, build consensus, and promote shared understanding.}
In Wikilabels, data points were labeled by individual Wikipedians working in isolation. However, our participants argued that data curation systems should instead promote deliberation, similar to existing processes on Wikipedia for article editing. Participants highlighted the importance of deliberation due to the subjectivity of data labeling and potential disagreements among community members. Participants suggested building upon Wikipedia's existing deliberation interface and mechanisms, such as talk pages and associated norms, to build consensus for collective decision-making while ensuring individual viewpoints are fully considered. Participants also anticipated that these deliberations can have side effects that benefit the community, such as revealing otherwise hidden disagreements, gaining insights from each other, and collectively strengthening the community.

\paragraphBold{D3: The data curation process should embed within existing workflows.}
Participants believed that a key reason Wikilabels did not see sustained use was because it was hosted on an external website and required Wikipedians to leave their workflows on Wikipedia. For example, even though Wikipedians were \textit{already} reviewing edits on Wikipedia, the design of Wikilabels required them to duplicate this effort by labeling edits as damaging or not using Wikilabels only for the purpose of data labeling. Participants wished to embed the data curation process into their existing workflow on Wikipedia: \textit{``Capturing people's judgments while they're working is like sticking a waterwheel in a river. The river is already flowing, we should take advantage of that''} (W1). They also anticipated this could help curate up-to-date data instead of labeling historical data pre-selected by AI engineers when using Wikilabels: \textit{``We're going to continue to get new data, so we can update the models and continue to re-evaluate them''} (W7).

\paragraphBold{D4: The data curation process should be public and transparent to community members.}
Currently, Wikipedians have a limited and narrow view of the data curation process in Wikilabels, where each contributor only sees the data they labeled. Participants suggested that data curation systems should instead make entire datasets public and easily accessible (like most content on Wikipedia) to facilitate community-driven AI evaluation: \textit{``If I'm doing my own audit, I'm not quite sure if other people are having the same problems and benefits of this model that I am. But if we can all put it in a repository together, and have some mechanism to make sense of what's in there, then I can know how this is working for everybody. We can make decisions together about whether we want this or not in our community''} (W1).

\section{Wikibench}\label{section:wikibench}


\begin{figure*}[t!]
  \centering
  \includegraphics[width=\linewidth]{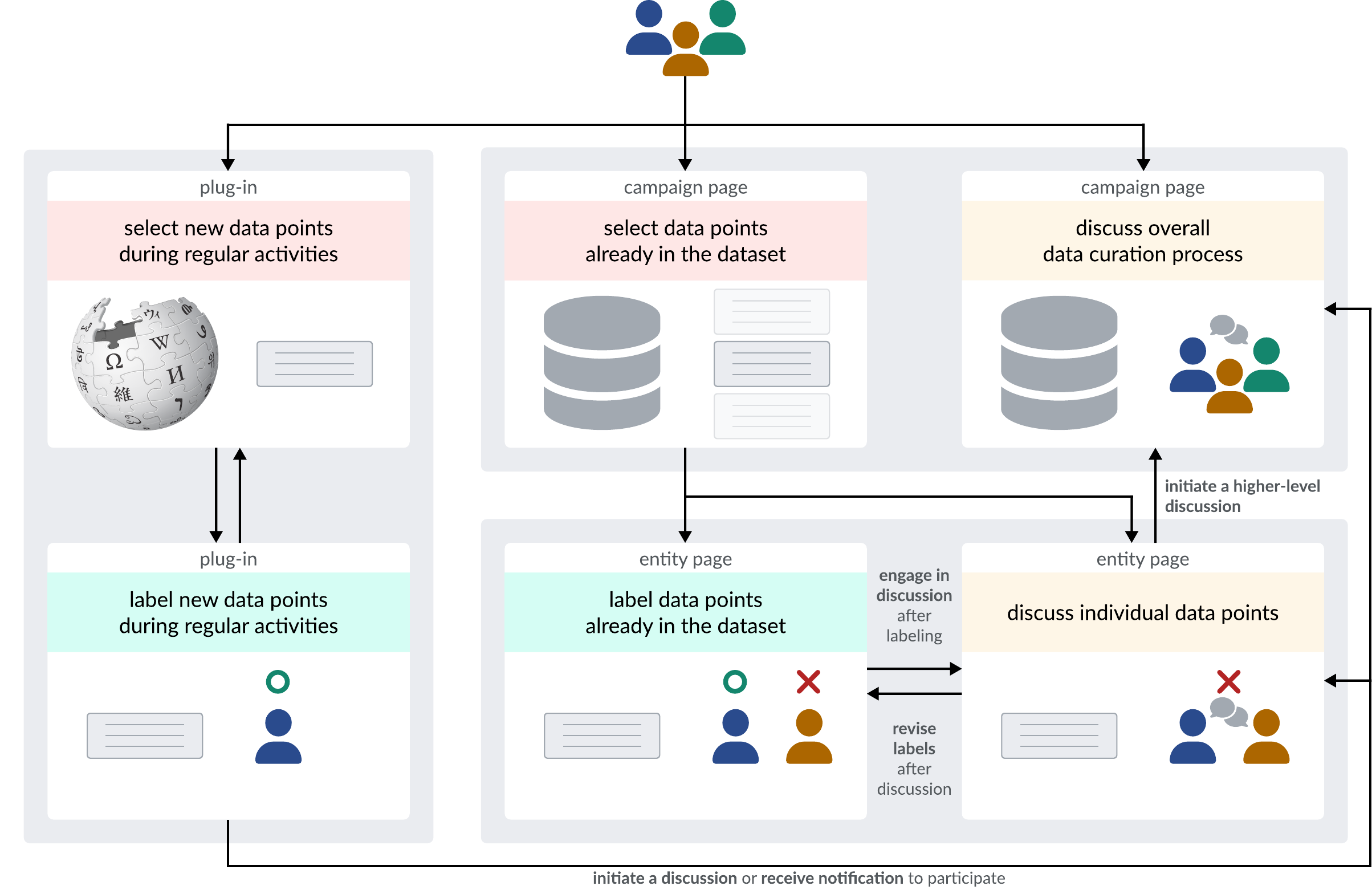}
  \caption{Wikibench's workflow. Wikibench mainly supports three actions for community-driven data curation: select, label, and discuss, each illustrated by different colors. Community members can select and label data points during their regular activities (i.e., while patrolling for damaging edits on Wikipedia) or choose from data points already collected in the dataset. They can also discuss individual data points to resolve disagreements or initiate a higher-level discussion related to the overall data curation process.}
  \Description{An illustration of Wikibench's workflow.}
  \label{figure:workflow}
\end{figure*}


Based on these design requirements, we developed Wikibench, a system that enables community members to collaboratively curate AI evaluation datasets, while navigating disagreements and ambiguities through discussion. Wikibench supports these processes through the workflow illustrated in Figure~\ref{figure:workflow}. As shown in this figure, community members can use Wikibench to \textit{select} new data points for inclusion in datasets and to \textit{label} these data points during the course of their regular, daily activities on Wikipedia (via a plug-in). Wikibench also supports community members in filtering through data points that have already been added to the dataset, to \textit{select} ones to further \textit{label} or \textit{discuss}. Through Wikibench, community members are supported in either discussing the label of individual data points, or discussing higher-level topics related to the overall data curation process.

Wikibench is designed to capture community consensus, disagreement, and uncertainty. Wikibench records two types of labels for each data point:
\begin{itemize}
\item{\textbf{Individual Label}}: Each community member can provide their unique individual label that is meant to reflect their own perspective. This label is editable only by themselves and may differ from others' labels.
\item{\textbf{Primary Label}}: Community members can collectively determine a primary label that is intended to reflect a ``consensus'' view.
\end{itemize}
Together, the individual and primary labels allows Wikibench datasets to reflect both community consensus and differing viewpoints that may underlie that consensus. Wikibench also records labelers' self-reported \textbf{confidence} associated with each individual label. In aggregate, confidence indications can provide a signal of the uncertainty associated with a data point.

In this section, we overview Wikibench's design through the specific example of dataset curation to support the evaluation of AI-based content moderation tools on Wikipedia, which are used to counter vandalism. In this context, each data point is an article edit on Wikipedia. Following the design of existing AI tools and datasets on Wikipedia, each edit has two associated labels:  \textit{edit damage}, which specifies whether the given edit is viewed as ``damaging'' to the article's quality, and \textit{user intent}, which specifies whether the edit is viewed as having been made in good or bad faith. The distinction between edit damage and user intent follows the prior data labeling campaign hosted on Wikilabels (Section~\ref{section:studycontext/challenges}), considering that damaging edits made with good intent are not considered to be vandalism\footnote{\url{https://enwp.org/WP:VAND}} on Wikipedia.

In the following subsections, we describe how Wikibench's three user interfaces on Wikipedia: plug-in, entity page, and campaign page, support data curation. Throughout this section, we link specific features of Wikibench's design to the design requirements described in the previous section, denoted as (D1)—(D4). Finally, we conclude with implementation details.

\subsection{Plug-in: Select and Label New Data Points}\label{plugin}

\begin{figure}[t]
  \centering
  \includegraphics[width=\linewidth]{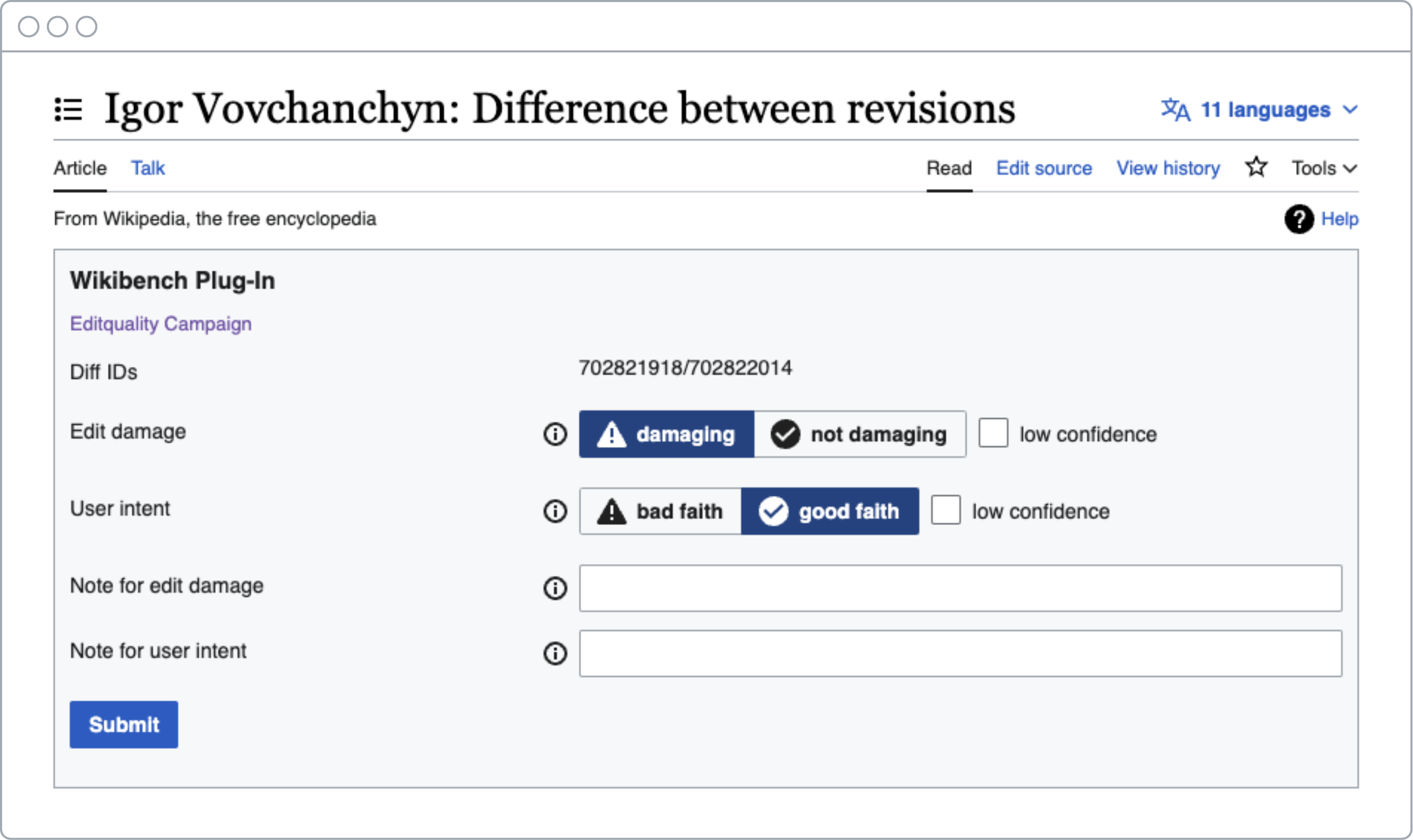}
  \caption{Wikibench's plug-in is embedded in Wikipedia's diff pages, where Wikipedians already assess edits during their regular patrolling activities. Through the plug-in, Wikipedians can label an edit's damage and user intent, specify their confidence level, and add notes if desired.}
  \Description{A screenshot of Wikibench's plug-in.}
  \label{figure:plugin}
\end{figure}

Wikipedians can use Wikibench's plug-in to select and label new edits during their regular patrolling activities on Wikipedia (\textbf{D3}). Specifically, Wikipedians who self-identify as patrollers\footnote{\url{https://enwp.org/WP:RCP}} regularly patrol edits on Wikipedia's Recent Changes page\footnote{\url{https://enwp.org/Special:RecentChanges}} and assess selected edits by opening its ``diff'' page\footnote{\url{https://enwp.org/WP:DIFF}}. Wikibench embeds a plug-in on these diff pages so that Wikipedians can label edits while they are already in the midst of assessing them, as shown in Figure~\ref{figure:plugin}. The plug-in also allows Wikipedians to specify the confidence level for their labels and include notes if desired. Overall, this design embeds the data curation process into Wikipedia's existing workflow to reduce duplication of effort and help curate up-to-date data.

\begin{figure}[t]
  \centering
  \includegraphics[width=\linewidth]{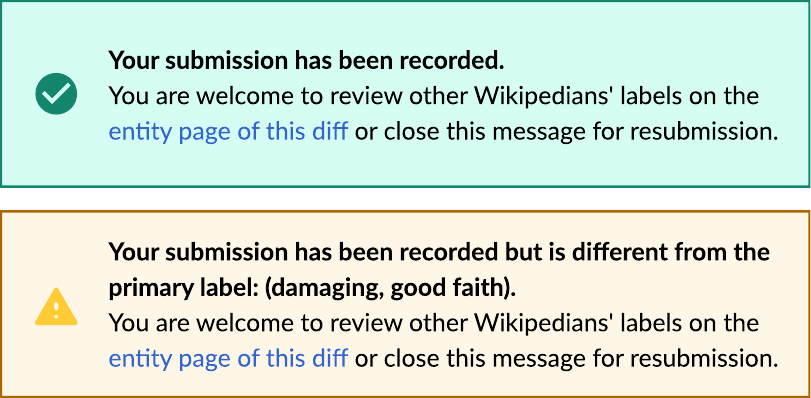}
  \caption{The message displayed after Wikipedians successfully submit their labels using the plug-in. The yellow message appears only when the submitted labels differ from the current primary label to facilitate discussion.}
  \Description{The green message shows "your submission has been recorded" at the top. The yellow message shows "your submission has been recorded but is different from the primary label" at the bottom.}
  \label{figure:nudge}
\end{figure}

After Wikipedians submit their labels, Wikibench's plug-in encourages them to engage in discussion when labeling disagreements arise (\textbf{D2}). In particular, if an individual's submitted label differs from the existing primary label of an edit, the plug-in will display the yellow message in Figure~\ref{figure:nudge} to encourage discussion. Otherwise, the green message will appear to minimize disruption to Wikipedians' regular patrolling activities. These messages encourage deliberation only when disagreements occur and minimize disruptions to existing workflows otherwise.

\subsection{Entity Page: Label and Discuss Collected Data Points}\label{entitypage}

\begin{figure*}[t]
  \centering
  \includegraphics[width=\linewidth]{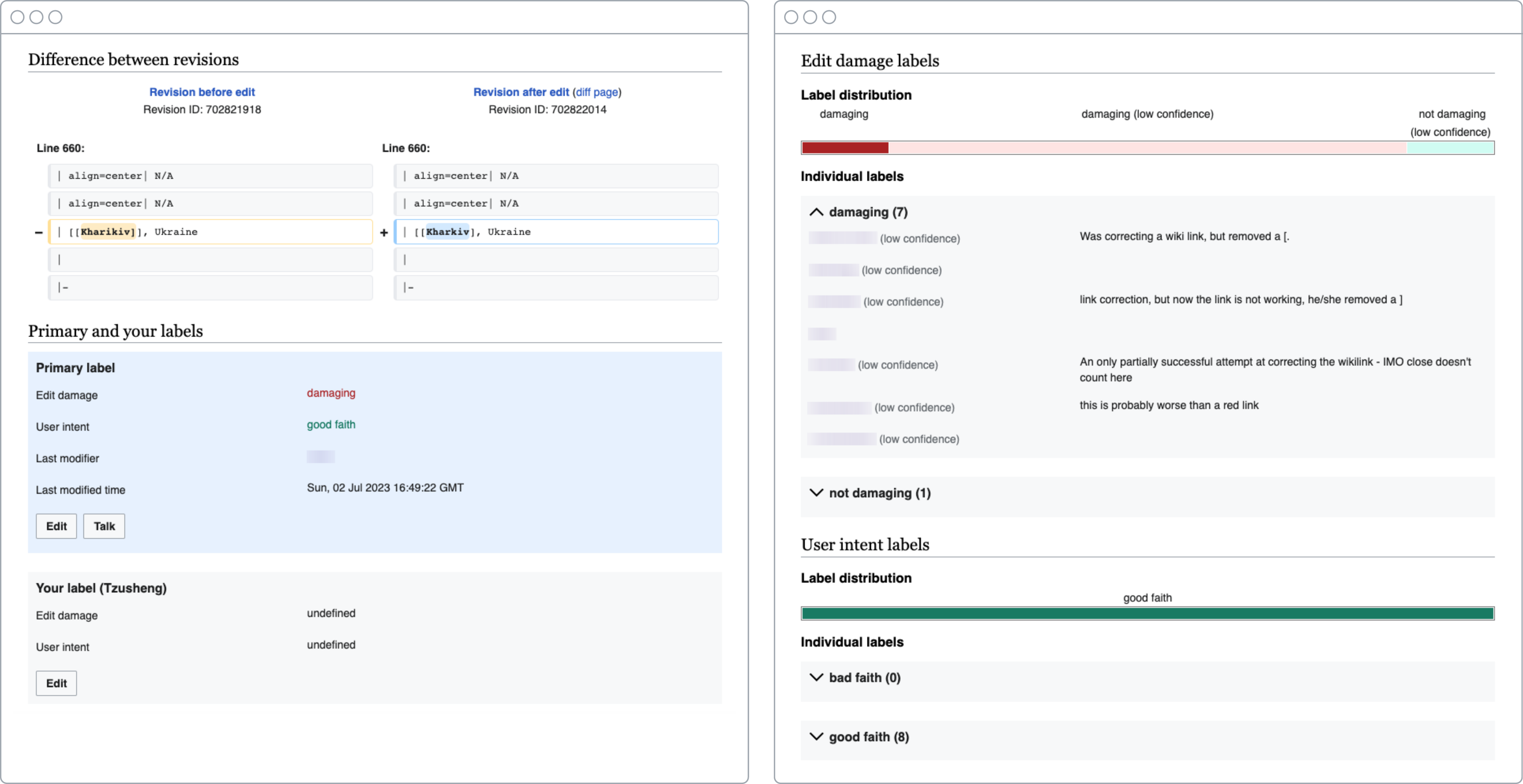}
  \caption{Wikibench's entity page for a given edit. The left side shows the top half of an entity page, featuring the edit, its primary label, and the user's individual label. The right side shows the bottom half of an entity page, containing the full set of individual labels and accompanying notes. Participants' usernames are blurred to avoid identification.}
  \Description{A screenshot of Wikibench's entity page divided into two figures side-by-side.}
  \label{figure:entity}
\end{figure*}

Wikibench's entity pages publicly show the labels of individual edits and facilitates discussions and (re-)labeling (\textbf{D2}, \textbf{D4}). As shown in Figure~\ref{figure:entity}, the top half of each entity page shows the edit, its primary label, and the user's individual label. The bottom half shows the full set of individual labels submitted by the community so far, along with any brief notes that community members may have included as rationale. This view is intended to help Wikipedians quickly understand the current level of disagreement associated with a given edit, and to examine how their own views align with or differ from others'. If Wikipedians think discussion on a given edit could be helpful, each entity page has a corresponding talk page for deliberation. This design resembles Wikipedia's article editing mechanisms, where each article has a corresponding talk page to discuss the article's content.

\begin{figure}[t!]
  \centering
  \includegraphics[width=\linewidth]{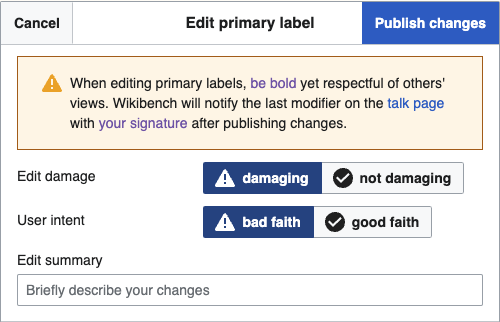}
  \caption{The message displayed when Wikipedians edit primary labels to encourage them to be bold yet respectful of others' views.}
  \Description{A screenshot of Wikibench's reminder for changing the primary label.}
  \label{figure:primarylabelchange}
\end{figure}

The mechanism by which Wikipedians choose the primary label for an edit through Wikibench is based upon the Wikipedia community's established norms for consensus-building (\textbf{D1}, \textbf{D2}). Similar to Wikipedia articles, the primary label is initially set to the value of the first submitted individual label. From that point on, it is open to modification by any Wikipedian. Wikibench does not automatically assign primary labels based the majority of individual labels, because Wikipedia follows the principle that ``polling is not a substitute for discussion'' when it comes to consensus building\footnote{\url{https://enwp.org/WP:POLL}}. When disagreements arise, Wikibench's design explicitly encourages Wikipedians to employ their well-established consensus-building processes\footnote{\url{https://enwp.org/WP:CON}} (e.g., the bold, revert, and discuss cycle\footnote{\url{https://enwp.org/WP:BOLD}}), to boldly edit primary labels and engage in discussions when others disagree with the changes (see Figure~\ref{figure:primarylabelchange}). When the primary label is changed, Wikibench also notifies previous labelers, to facilitate discussion as needed. 

\subsection{Campaign Page: Select Collected Data Points for Labeling and Discussion, or Discuss the Overall Curation Process}

\begin{figure*}[t]
  \centering
  \includegraphics[width=0.87\linewidth]{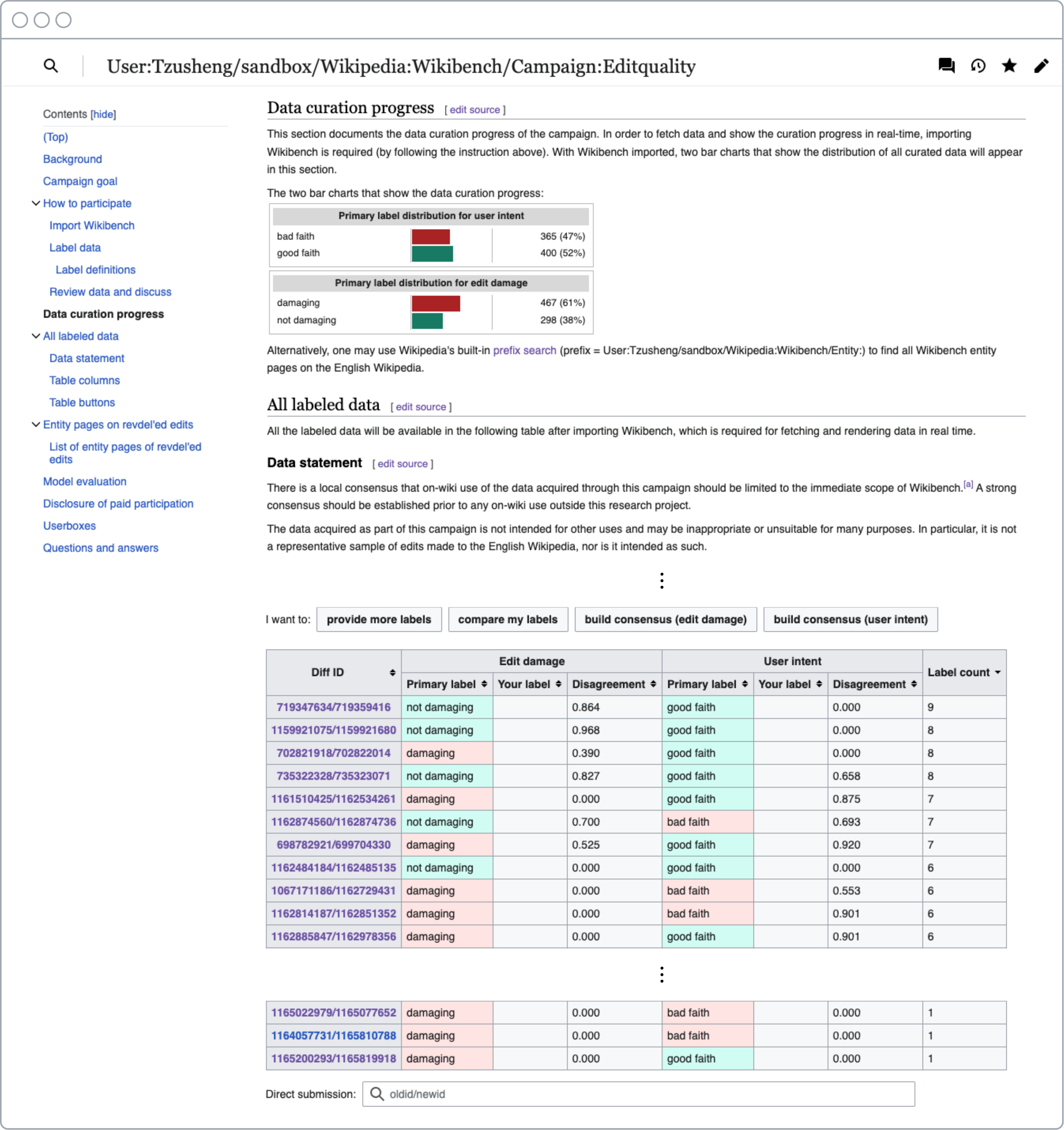}
  \caption{An excerpt of Wikibench's campaign page. The section at the top shows simple visualizations to help Wikipedians track the progress of a data curation campaign. The section at the bottom includes a table that helps Wikipedians navigate the entire dataset. The buttons above the table allow Wikipedians to sort the table and identify edits that may benefit from additional labels or discussion, including edits for which the current primary label differs from their own individual label.}
  \Description{A screenshot of Wikibench's campaign page.}
  \label{figure:campaign}
\end{figure*}

The campaign page publicly shows the entire dataset and surfaces edits that could benefit from additional attention, including edits with high disagreement and edits that could benefit from additional labelers (\textbf{D2}, \textbf{D4}). As shown in Figure~\ref{figure:campaign}, each row in the table is a link to an entity page for an edit and its label information. The four buttons above the table assist Wikipedians in sorting the table to easily find edits that may benefit from more labels or discussions. For example, the \textit{provide more labels} button helps Wikipedians find and contribute to edits with fewer individual labels. The \textit{build consensus} button surfaces edits that have high disagreement across community members' individual labels, to promote discussion among community members. The disagreement is measured as the standard deviation of encoded individual labels, with $\pm1$ for damaging/not damaging, $\pm0.5$ for damaging/not damaging submitted with low confidence; likewise for user intent.

The campaign page is also designed to enable Wikipedians to discuss and coordinate about the overall data curation process (\textbf{D1}, \textbf{D2}). 
In addition to the table, the campaign page serves as a living datasheet~\cite{gebru2021datasheets} that provides comprehensive information about the data curation campaign, such as label definitions and data statements, as outlined in Figure~\ref{figure:campaign} on the left. Similar to Wikipedia articles, the campaign page can be edited by any Wikipedian and has an associated talk page for discussion.

\subsection{Implementation}\label{section:wikibench/implementation}

The current implementation of Wikibench is built upon Wikipedia's infrastructure to ensure its user interfaces and norms are familiar to Wikipedians. In the back-end, both entity and campaign pages, used for storing labels and campaign information, are standard Wikipedia article pages with built-in talk pages. Wikibench uses Wikipedia's user script feature\footnote{\url{https://enwp.org/WP:JAVASCRIPT}} to re-render these article pages on the front-end. The plug-in is also a front-end element embedded in Wikipedia's existing diff page. To ensure that the front-end elements are familiar to Wikipedians and coherent with Wikipedia's existing interface, Wikibench uses Wikipedia's OOUI\footnote{\url{https://www.mediawiki.org/wiki/OOUI}} and design system\footnote{\url{https://design.wikimedia.org/style-guide}}. The creation and revision of Wikibench's labels are enabled through MediaWiki API\footnote{\url{https://www.mediawiki.org/wiki/API:Main\_page}}. Importantly, we adhere to our positionality statement by keeping Wikibench's campaign and entity pages within an author's user sandbox\footnote{\url{https://enwp.org/WP:SAND}}, a designated area for experimentation on Wikipedia, to minimize disruption to the site. This deep integration with Wikipedia also enables Wikipedians to easily use Wikibench by importing Wikibench into their Wikipedia account through a user script\footnote{\url{https://enwp.org/Wikipedia:User_scripts}}. Wikibench is open-sourced on Wikipedia\footnote{\url{https://en.wikipedia.org/wiki/User:Tzusheng/Wikibench-Editquality.js}} and available to all Wikipedians.
 
\section{Evaluation Study}\label{section:evaluation}

To understand how Wikipedians use Wikibench in practice, we conducted a two-part evaluation study on English Wikipedia, a highly active and extensively studied Wikipedia language community~\cite{bryant2005becoming}. Prior research has demonstrated that community needs and norms for content moderation vary across different Wikipedia language communities~\cite{halfaker2020ores,hwang2022rules}. In the current study we focus on understanding how Wikibench can support community-driven data curation on one language community\footnote{Note that we focus at the level of a \textit{language community} because this is the level at which AI-based content moderation tools are adopted on Wikipedia.}, before expanding the system to multiple communities.

In the remainder of the paper, we refer to English Wikipedia as ``Wikipedia'' for simplicity. We first conducted a one-week field study in which participants used Wikipedia to collectively curate a dataset. We then conducted a validation study with a separate set of participants, aimed at understanding whether labels generated collaboratively, through Wikibench, better reflect community consensus than those generated through Wikilabels.

\subsection{Field Study}

We conducted a one-week field study to observe how Wikipedians use Wikibench in the course of their regular activities on Wikipedia. 

\subsubsection{Study protocol}\label{fieldstudy/protocol}
The study began with a one-hour, one-on-one onboarding session that introduced the study and the data curation campaign. As part of this onboarding, participants imported Wikibench into their Wikipedia user accounts and a researcher walked them through the system's features. At the end of the onboarding, participants received instructions for the week-long field study. We set minimal participation requirements to ensure that (1) participants would have ample opportunities for interaction during the field study period, while also (2) providing participants with flexibility to decide when and how much they want to contribute (cf.~\cite{zhang2018making}). Each participant was asked to submit a minimum of 10 labels and to engage in at least 3 discussions per day using Wikibench, for 5 days out of the week. Finally, we conducted a 30-minute exit interview with each participant once they completed the field study, to learn about their experiences and gather feedback. As part of this exit interview, each participant was shown a randomly selected set of 10 edits from the dataset, and were invited to explore how Wikibench's labels compare with the predictions of ORES (described in Section~\ref{section:studycontext/challenges}) and a new AI tool\footnote{\url{https://wikitech.wikimedia.org/wiki/Machine_Learning/LiftWing}} that is currently under development by the Wikimedia Foundation (with more details in Section~\ref{section:findings/dataset/evaluation}). Our exit interview protocol is shown in Appendix~\ref{section:appendix/exitinterview}.

\begin{table*}[t]
  \caption{Field study participant demographics, including their self-identified experience and frequency of patrolling edits, registration year, edit count on English Wikipedia, and geographic location.}
  \label{table:participants}
  \begin{tabular}{lccccl}
    \toprule
    Participant ID & Patrol Experience & Patrol Frequency & Registered Since & Edit Count & Location \\
    \midrule
    P1  & Months & Daily & 2023 & 9.1k & United States \\
    P2  & Years & Daily & 2018 & 24k & Indonesia \\
    P3  & Months & Weekly & 2023 & 1.6k & United States \\
    P4  & Years & Daily & 2006 & 48k & Singapore \\
    P5  & Years & Daily & 2021 & 2.6k & Germany \\
    P6  & Months & Daily & 2018 & 7.4k & Italy \\
    P7  & Years & Daily & 2013 & 0.9k & Hungary \\
    P8  & Months & Daily & 2023 & 4.2k & United States \\
    P9  & Years & Yearly & 2013 & 9.7k & Australia \\
    P10 & Years & Daily & 2014 & 23k & United States \\
    P11 & Years & Daily & 2019 & 7.1k & Ireland \\
    P12 & Years & Daily & 2020 & 18k & United Kingdom \\
    \bottomrule
  \end{tabular}
\end{table*}

\subsubsection{Recruitment}

We adhered to our positionality statement and followed the norm of researching Wikipedia by including the recruitment message in our main project page on Meta-Wiki, where Wikipedians could review study details before signing up. We then shared this project page with Wikipedians through multiple channels, including English Wikipedia's Village Pump\footnote{\url{https://enwp.org/WP:VP}}, r/wikipedia subreddit, and the Discord servers for the Wikimedia Community and Anti-Vandalism. We also reached out to several Wikipedians who were actively patrolling edits by leaving messages on their user talk pages---an approach that aligns with existing norms for communication on Wikipedia.

In total, we recruited 12 Wikipedians with diverse experiences and backgrounds, as shown in Table~\ref{table:participants}. We conducted onboarding sessions and exit interviews via Zoom, with the exception of one participant who preferred to participate via text over Discord. We provided \$150 USD as compensation, including \$30 for onboarding, \$100 for the field study (\$20 per day for 5 days), and \$20 for the exit interview. These compensation amounts align with prior studies on the English Wikipedia \cite{ye2021wikipedia}, as well as prior HCI research that has conducted similar week-long field studies \cite{zhang2018making}. As in our formative study, some participants declined compensation at the end of the exit interview, viewing the study as part of their voluntary work to improve Wikipedia.

\subsection{Validation Study}\label{validation}

To understand whether Wikibench helped curate labels that more consistently reflect community consensus, compared with the previous approach (Wikilabels), we conducted a small-scale validation study following the conclusion of the field study. In particular, we recruited a separate group of Wikipedians to collaboratively label a subset of the edits using Wikipedia's default article and talk pages. These participants labeled edits anew, without knowledge of the labels each had previously received through either Wikibench or Wikilabels. While using the default interfaces for labeling and discussion without Wikibench's support was more involved and time-consuming for participants, this approach mirrors the standard process Wikipedians use to reach consensus on article pages. This validation study helped us understand whether Wikibench's primary labels, which are intended to be reflective of community consensus, are indeed aligned with the labels generated through Wikipedia's standard consensus-building process. Our validation study aimed to compare the consensus labels generated through this process, by an independent group of participants, with those generated through Wikibench and Wikilabels.

\subsubsection{Edit selection.}\label{validation/considerations}
We first sampled 90 edits that had previously received labels through Wikilabels. We then had field study participants label them using Wikibench during onboarding sessions without being told about the validation study to prevent them from overly focusing on these edits more than they would naturally do. These edits then underwent the standard labeling and consensus-building process in Wikibench. Following the conclusion of the field study, we identified 33 edits where Wikilabels and Wikibench's primary labels differed. The resulting edits were selected for the validation study. Additional details of our sampling procedure of the 90 edits are available in Appendix \ref{section:appendix/edit}.

\subsubsection{Study protocol.}

We created a standard article page in an author's user sandbox for the validation study. The page provided study purpose, instruction, and compensation information, along with a table where each row was an edit, and each column was available for one Wikipedian to provide their individual label. In addition, we asked Wikipedians to enter their label consensus in two extra columns, one for edit damage and another for user intent. They were also encouraged to use the talk page for discussion. The entire validation study lasted a week to ensure sufficient time for participants to label and discuss edits asynchronously.

\subsubsection{Recruitment.}

We recruited five additional Wikipedians who had signed up or expressed interest in the field study but were unavailable during the study period. We shared the link to the article page we set up for the validation study and provided \$90 USD as compensation at the end, with the estimated time required around three hours. As in the prior two studies, some participants declined compensation, viewing their participation as part of their voluntary work to improve Wikipedia. Participant demographics are shown in Table~\ref{table:validationparticipants} in Appendix~\ref{section:appendix/demographics}.

\subsection{Data Analysis}
We adopted a mixed-method approach to analyze our data. We employed a top-down approach to quantitatively analyze the resulting community-curated dataset for indicators of quality, such as primary label composition, individual label variation, and label contributor diversity. We adopted a reflexive thematic analysis approach \cite{braun2012thematic, braun2019reflecting} to qualitatively analyze both participants' interview data and their interactions with one another through Wikibench. In particular, two authors conducted open coding on 12 exit interviews and all discussions on Wikibench's campaign and entity talk pages. This process resulted in a total of 249 codes. Through iterative discussions, we synthesized higher-level themes using affinity diagramming. In total, we identified 64 first-level themes, 17 second-level themes, and 7 top-level themes through this process. Finally, we triangulated across findings from our quantitative and qualitative analyses. We report the results of this combined analysis in the next sections.

\section{Overview of Findings}\label{section:finding/overview}
We present findings from our evaluation study in the following two sections. Section \ref{section:finding/dataset} examines the quality and properties of the resulting dataset curated using Wikibench (Section \ref{section:findings/quality}) and explores its use in evaluating different AI models' alignment with community norms and values (Section \ref{section:findings/dataset/evaluation}). 
In total, Wikipedians curated 757 edits using Wikibench, with a relatively balanced primary label composition (61\% of primary labels are ``damaging'' and 48\% are ``bad faith''), which can be useful in evaluating how well an AI model's decision boundary aligns with community-specified decision boundaries~\cite{gardner2020evaluating}. Overall, we find evidence that the dataset collected through Wikibench is broadly reflective of Wikipedians' perspectives, while also capturing ambiguity and disagreement among community members. We demonstrate how the resulting dataset can be used to investigate the relative strengths and limitations of two different AI models used on Wikipedia. The dataset is publicly available on Wikibench's campaign page and on GitHub\footnote{\url{https://github.com/tskuo/Wikibench}}. 
Section 9 describes how participants used Wikibench throughout the study, and how they collectively steered the overall data curation process, in addition to labeling and discussing individual data points. Participants appreciated how Wikibench's design embedded seamlessly into their workflow (Section~\ref{section:finding/workflow}). They organically drove the overall data curation process, beyond just labeling data (Section~\ref{section:findings/agency}) and believed that the collaborative approach supported by Wikibench was beneficial both for dataset quality and for community building (Section~\ref{section:findings/collaboration}).

Taken together, our findings indicate potential for the approach embodied by Wikibench to support the curation of AI datasets that reflect community norms and values.

\section{Findings: Quality of the Community-Curated Dataset}\label{section:finding/dataset}

A dataset curated by and for a community for the purpose of AI evaluation should meet the following criteria, based on our design requirements:
\begin{itemize} 
    \item \textbf{Label Quality}: Low-quality labels diminish overall dataset quality~\cite{jain2020overview}. From a community standpoint, we consider high-quality primary labels as being reflective of community values, shared understanding, and consensus. This data criterion aligns with Design Requirements \textbf{D1} and \textbf{D2}. 
    \item \textbf{Disagreement and Uncertainty}: Meanwhile, capturing the disagreement and uncertainty behind primary labels is equally important, to ensure that the dataset reflects both (1) substantive differences in perspective across individuals and (2) inherent ambiguity of a given data point~\cite{gordon2021disagreement}. This criterion corresponds to Design Requirement \textbf{D2}.
    \item \textbf{Collaborative Labeling}: Ideally, data points should be labeled and curated by multiple community members rather than being dominated by just a few voices. This criterion corresponds to Design Requirements \textbf{D2}, \textbf{D3}, and \textbf{D4}.
\end{itemize}

In the following subsections, we describe how the dataset curated through Wikibench aligns with the criteria above (Section~\ref{section:findings/quality}) and showcase how the dataset can be used by comparing the community alignment of two AI models currently deployed on Wikipedia (Section~\ref{section:findings/dataset/evaluation}).

\subsection{Dataset Quality}\label{section:findings/quality}

\subsubsection{\textbf{Label Quality: Wikibench's labels reflect consensus among a broader set of community members.}}\label{findings/consensus}

\begin{figure}[t]
  \centering
  \includegraphics[width=\linewidth]{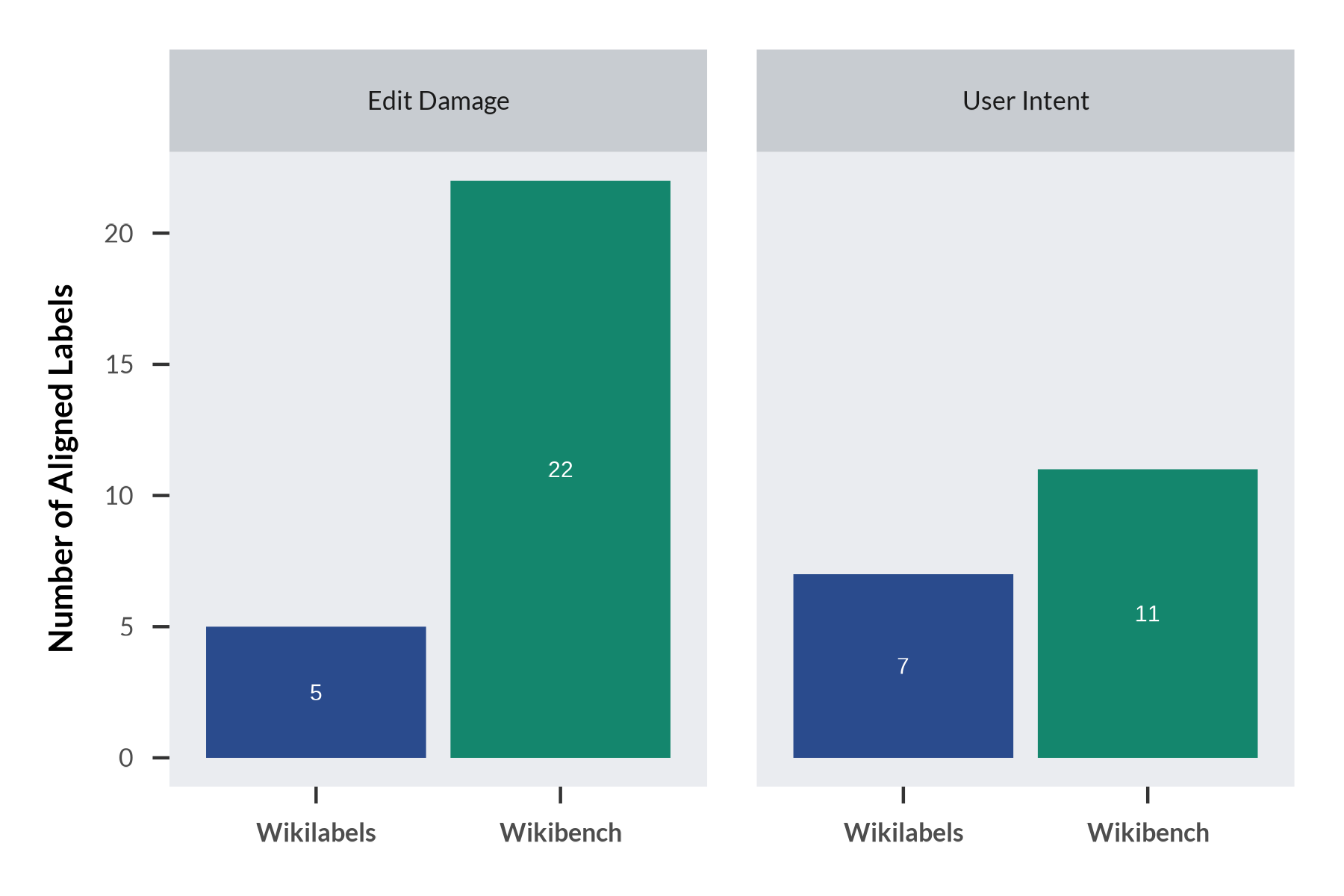}
  \caption{Counts of labels generated through Wikilabels versus Wikibench that align with validation study participants' consensus.}
  \Description{Two bar charts. The left one has 5 for Wikilabels and 22 for Wikibench. The right one has 7 for Wikilabels and 11 for Wikibench.}
  \label{figure:C27.1}
\end{figure}

To understand whether Wikibench helped curate primary labels that better reflect community consensus, compared with Wikilabels, we examined results of the validation study (Section~\ref{validation}). In this study, a set of participants who had not participated in the field study were shown a sample of 33 edits where Wikilabels and Wikibench's labels differed in edit damage (27 edits) and/or user intent (18 edits). Participants in the validation study collectively labeled edits anew through Wikipedia's standard consensus-building process, without knowledge of the labels each edit had received previously through either Wikilabels or Wikibench. Figure~\ref{figure:C27.1} shows the counts of labels generated through Wikilabels versus Wikibench that align with the labels produced in the validation study. As shown, compared with labels generated through Wikilabels, Wikibench's primary labels tend to align with the consensus labels generated in the validation study. In exit interviews, field study participants also expressed a belief that Wikibench's collaborative approach would better reflect broader community perspectives: \textit{``I believe that you're getting more of an overall viewpoint from the community itself, whereas that may not have always been the case for Wikilabels''} (P10). 

\subsubsection{\textbf{Disagreement and Uncertainty: Wikibench captures ambiguity and differences of perspective.}}\label{findings/variability}

\begin{figure}[t]
  \centering
  \includegraphics[width=\linewidth]{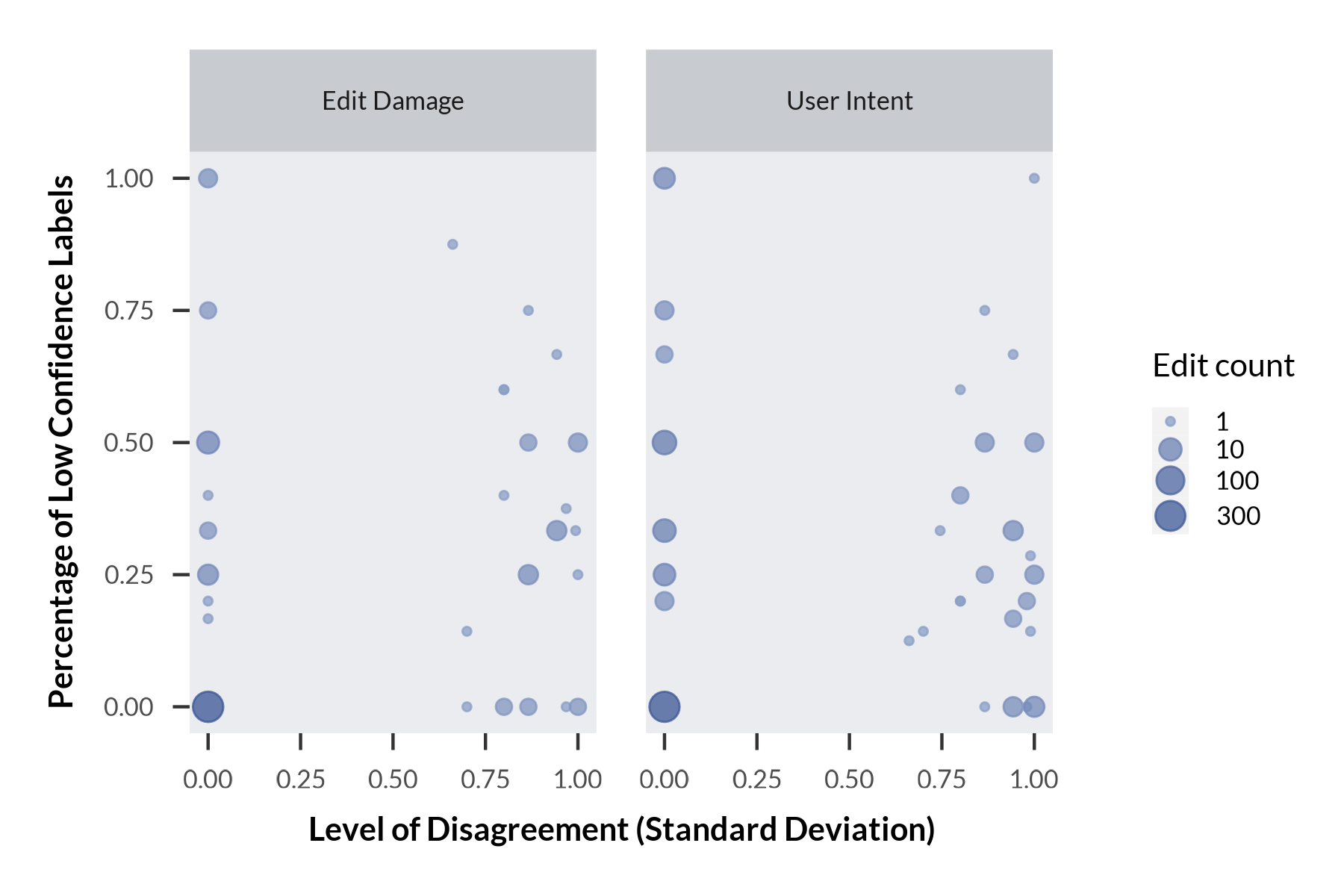}
  \caption{Edits with multiple individual labels, plotted by labeler disagreement (the standard deviation of individual labels) and confidence (the percentage of individual labels submitted with ``low confidence'').}
  \Description{Two scattered plots, one for edit damage and another for user intent.}
  \label{figure:C46.9}
\end{figure}

In addition to the primary labels, Wikibench is designed to capture disagreement and uncertainty that may underlie these collectively-determined labels by allowing each person to provide unique individual labels representing their personal perspective, along with their self-reported confidence. In line with prior research, we use these signals to distinguish between \textit{ambiguity} and \textit{genuine differences in perspective}~\cite{chen2023judgment,gordon2021disagreement}. Figure \ref{figure:C46.9} shows edits that have multiple individual labels, plotted by labeler disagreement and confidence.
For a given edit, overall confidence is measured by the proportion of individual labels that specify low confidence. Disagreement across individual labels is measured as the standard deviation of encoded individual labels, with -1 for damaging and 1 for non-damaging, and likewise for user intent\footnote{We experimented with various metrics~\cite{kader2007variability} and chose standard deviation for simplicity. All resulting charts, measured by different metrics, closely resemble Figure~\ref{figure:C46.9}.}.

As shown in Figure~\ref{figure:C46.9}, the dataset curated in our field study captures a range of qualitatively distinct cases, represented by the four corners of the plot. The majority of edits fall in the lower-left corner, with \textit{low disagreement and high confidence}. These can be interpreted as \textbf{clearer-cut cases}, where labelers tend to agree with high confidence. Edits toward the upper-right of this figure are ones with \textit{high disagreement with low confidence}. These edits may be a consequence of \textbf{inherent ambiguity} regarding what label an edit should be assigned. By contrast, edits toward the lower-right of this figure are ones with \textit{high disagreement with high confidence}. These edits are more likely to represent \textbf{genuine differences in perspective} among community members. For example, one of these edits\footnote{\url{https://enwp.org/Special:Diff/1163519177/1163595999}} is a case that divides a wiki link. While some participants argued that the edit is damaging as it violates Wikipedia's style guidelines\footnote{\url{https://enwp.org/WP:SEAOFBLUE}}, others argued in the opposite direction, noting that the edit improved readability and that the style guideline is not strictly mandatory. In exit interviews, participants expressed that Wikibench was helpful in facilitating measured discussions even in cases where community members held strong opposing viewpoints: \textit{``I was able to explain my rationale on the talk page. [...] I think that helped to make sure everyone's view was properly considered''} (P12). Finally, edits in the upper-left corner are ones with \textit{low disagreement and low confidence}. These edits may represent \textbf{agreed-upon edge cases}: cases that are more ambiguous, but where community members nonetheless tend to agree. To better communicate about and capture cases like these through Wikibench, participants expressed desires for a means to explicitly indicate their \textit{collective} confidence (or lack thereof) on a given edit: \textit{``I think in some cases, we do want to tag it as low confidence because even after discussion we're not 100\% sure''} (P9).

\subsubsection{\textbf{Collaborative Labeling: Most data points are labeled by multiple community members.}}\label{findings/diversity}

\begin{figure}[t]
  \centering
  \includegraphics[width=\linewidth]{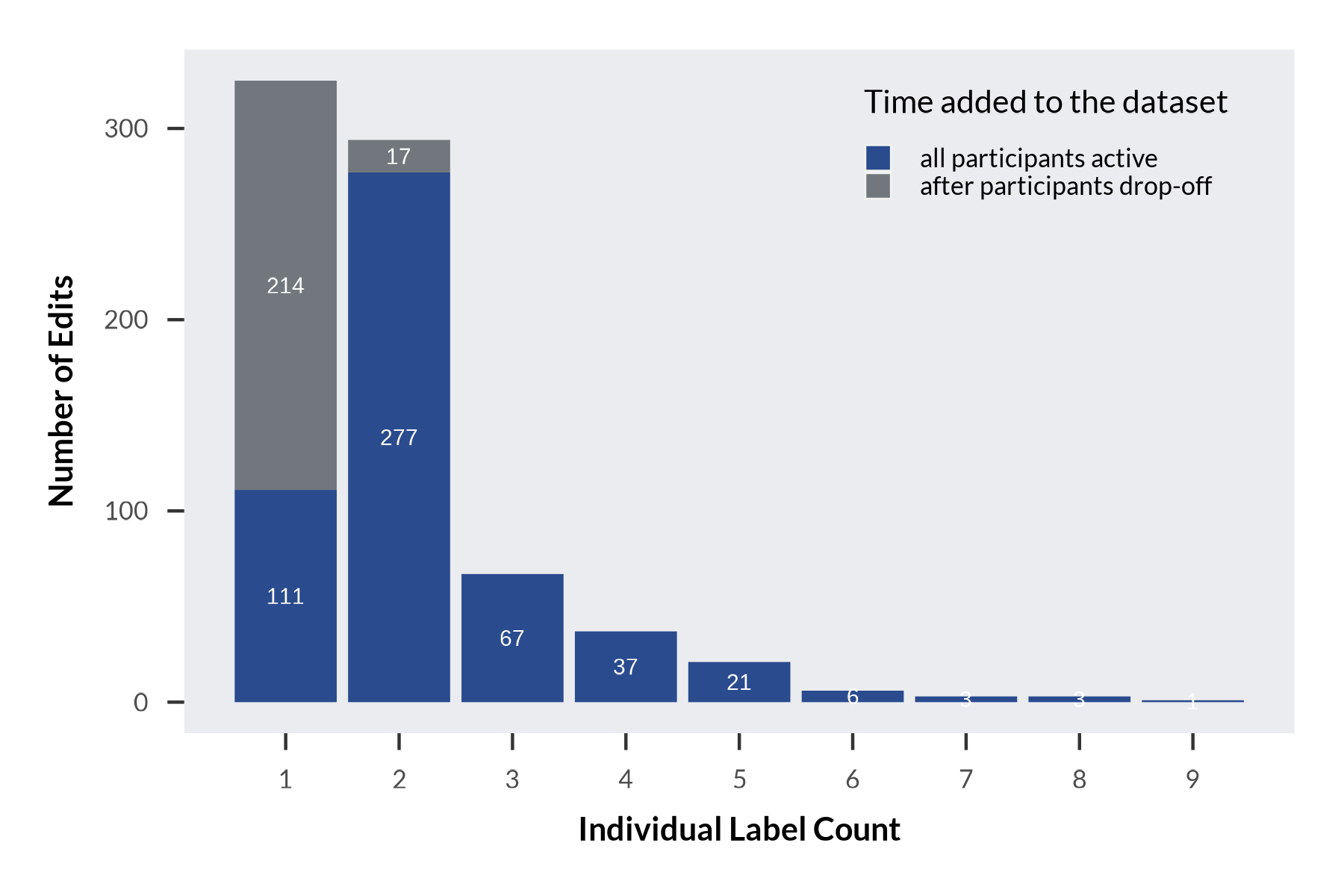}
  \caption{The count of edits with different numbers of individual labels received. Bar color denotes whether an edit was added to the dataset when all participants were active or after participant drop-off following exit interviews.}
  \Description{A bar chart for the count of edits with different numbers of individual labels received.}
  \label{figure:C26.1}
\end{figure}

\begin{figure}[t]
  \centering
  \includegraphics[width=\linewidth]{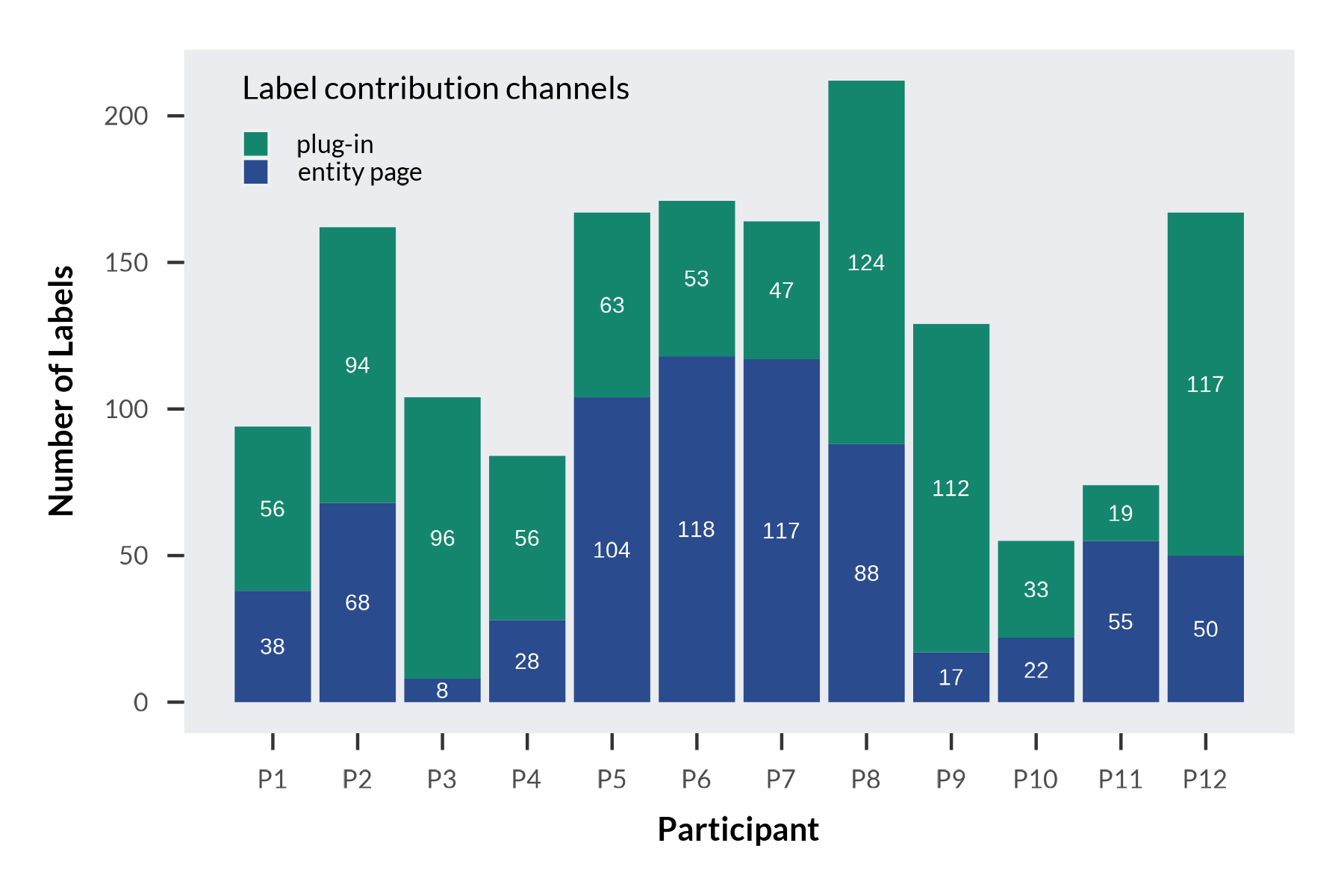}
  \caption{The number of labels contributed by each participant in the field study (bar heights), and the channels they used for contribution (colors). The upper and lower stacked bars denote plug-in and entity page, respectively.}
  \Description{A bar chart for the number of labels contributed by each participant.}
  \label{figure:C25.3}
\end{figure}

To understand how effective Wikibench is in directing \textit{multiple} labelers to edits, we examine the number of individual labels each edit received during our field study. As indicated by the blue bars in Figure \ref{figure:C26.1}, among the edits added to the dataset during the week when all participants were active, the majority (79\%) received labels from at least two individuals. Even when considering the full set of edits, including those added to the dataset after participants began dropping off following exit interviews, over half (57\%) received labels from at least two individuals.

All participants submitted labels that exceeded the minimum participation requirements, with no one strongly dominating label contributions. Participants were asked to submit at least 10 labels per day for up to 5 days. Figure~\ref{figure:C25.3} shows the number of labels each participant submitted. All participants surpassed the minimum requirement of 50 labels, with half of them contributing over triple this amount. Furthermore, as shown in Figure~\ref{figure:C25.3}, participants not only submitted labels using the plug-in during their regular patrolling activities but also contributed substantially via entity pages. Although we had asked participants to engage in at least three daily discussions on the campaign or entity talk pages, the observation that they also actively contributed many labels outside of their usual workflow suggests that they were highly engaged in the process, and may have been intrinsically motivated to contribute. This is corroborated by participants' comments in their exit interviews: \textit{``I'm quite unsure about my [patrolling] decision sometimes, so it is good to have someone more experienced talk through why they concluded differently from me [...] It's very enlightening, this [study]''} (P11). In a longer-term deployment, these kinds of intrinsic motivations could play a pivotal role in promoting broader, sustained participation.

\subsection{Potential for Use in AI Evaluation}\label{section:findings/dataset/evaluation}

While Wikibench's current interface primarily supports data curation, we are interested in understanding the potential of resulting datasets to support more informative AI evaluations downstream. In this section, we demonstrate how the dataset generated in our study can be used to compare two AI models for counter-vandalism deployed on Wikipedia (Section~\ref{section:evaluation/results}). Participants found these model comparisons informative and saw opportunities for the design of community-facing visualization and analysis tools to support such evaluations (Section~\ref{section:evaluation/perception}).

\subsubsection{\textbf{Wikibench’s dataset can help in understanding AI models’ alignment with community perspectives.}}\label{section:evaluation/results}
To showcase the potential of Wikibench datasets for use in AI evaluation, we used the dataset from the current study to evaluate the community alignment of two AI models deployed on Wikipedia for counter-vandalism. The first model is ORES, an AI system used extensively on Wikipedia for detecting damaging edits\footnote{\url{https://www.mediawiki.org/wiki/ORES\#Edit\_quality}}. The second is the Revert-Risk model\footnote{\url{https://meta.wikimedia.org/wiki/Machine\_learning\_models/Proposed/Language-agnostic\_revert\_risk}}, which was recently developed by the Wikimedia Foundation to replace ORES. In contrast to ORES, which is trained on explicit labels of edit damage and user intent from Wikipedians, the new Revert-Risk model is trained solely on historical trace data. Specifically, the Revert-Risk model uses an edit's revert history on Wikipedia as ground truth for training and predicts the probability of an edit getting reverted. Although edits may be reverted on Wikipedia for a variety of reasons, beyond being damaging to articles~\cite{kittur2007he}, the Revert-Risk model's documentation\footnote{\url{https://wikitech.wikimedia.org/wiki/ORES}} states \textit{``the idea with Revert-Risk model is to use [edit] reverts as `implicit annotations' [...] If we consider [ORES's] model as prediction for reverts, Revert-Risk is outperforming ORES in almost all scenarios.''} Here, we present an alternative perspective made possible by a community-curated evaluation dataset. 

\begin{figure}[t]
  \centering
  \includegraphics[width=\linewidth]{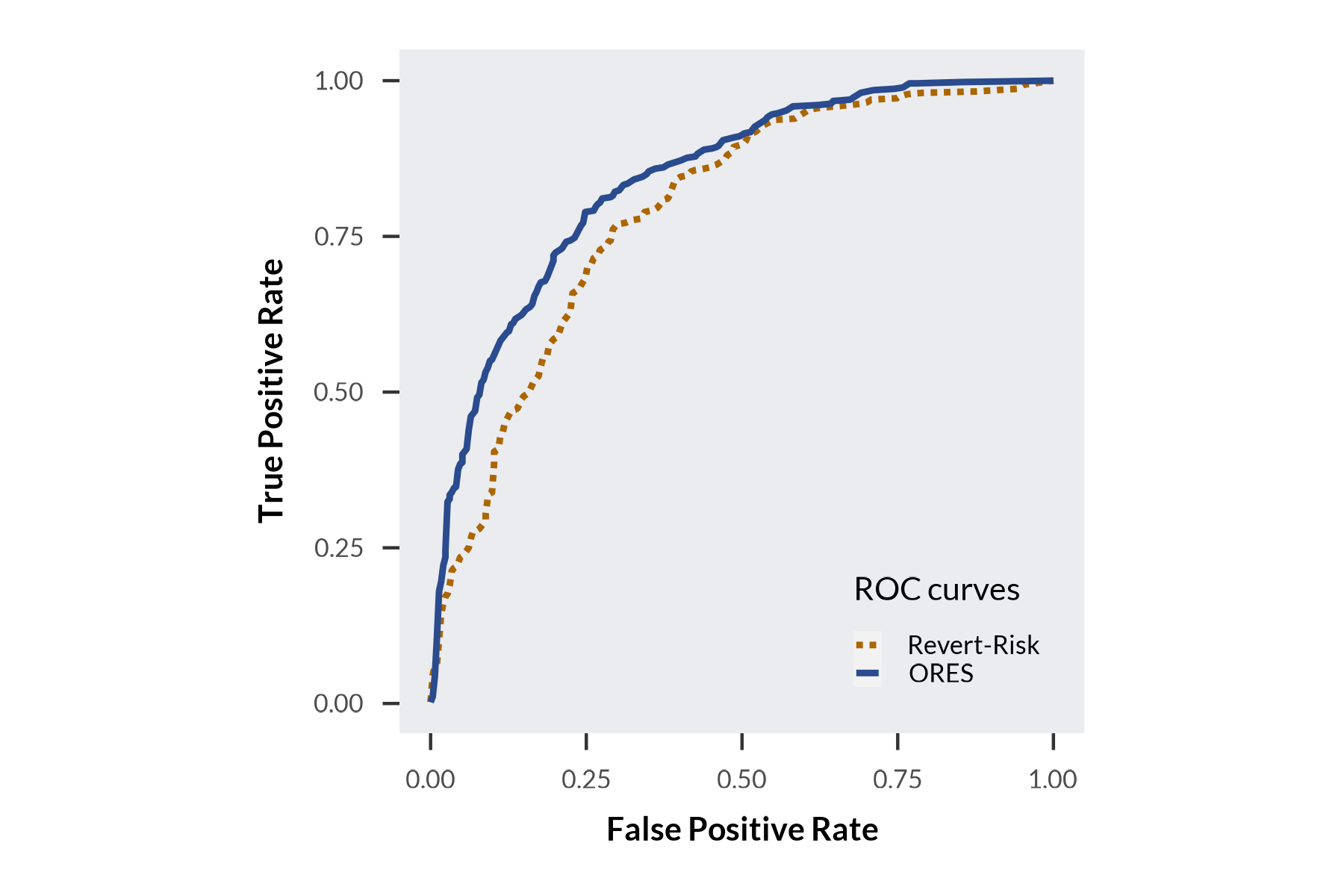}
  \caption{The ROC curves of ORES and Revert-Risk models evaluated on Wikibench's dataset. The AUC scores of ORES and Revert-Risk are 0.84 and 0.79, respectively.}
  \Description{The ROC curves of ORES and Revert-Risk.}
  \label{figure:C54.9}
\end{figure}

We evaluated each model on Wikibench's dataset to examine their relative alignment with community perspectives. Given that the Revert-Risk model is meant to replace ORES in identifying damaging edits for counter-vandalism, we used the edit damage label in Wikbench's dataset to evaluate both models. Figure~\ref{figure:C54.9} shows the ROC curves of ORES and Revert-Risk models evaluated on Wikibench's dataset. The AUC score of ORES and Revert-Risk is 0.84 and 0.79, respectively, showing that ORES performs better than Revert-Risk based on a dataset curated by community members. This result contrasts with the development team's evaluation on trace data, which had shown the opposite trend. This provides preliminary evidence that ORES's behavior aligns more closely with how our participants believe decisions \textit{should} be made, compared with the Revert-Risk model.

\begin{figure}[t]
  \centering
  \includegraphics[width=\linewidth]{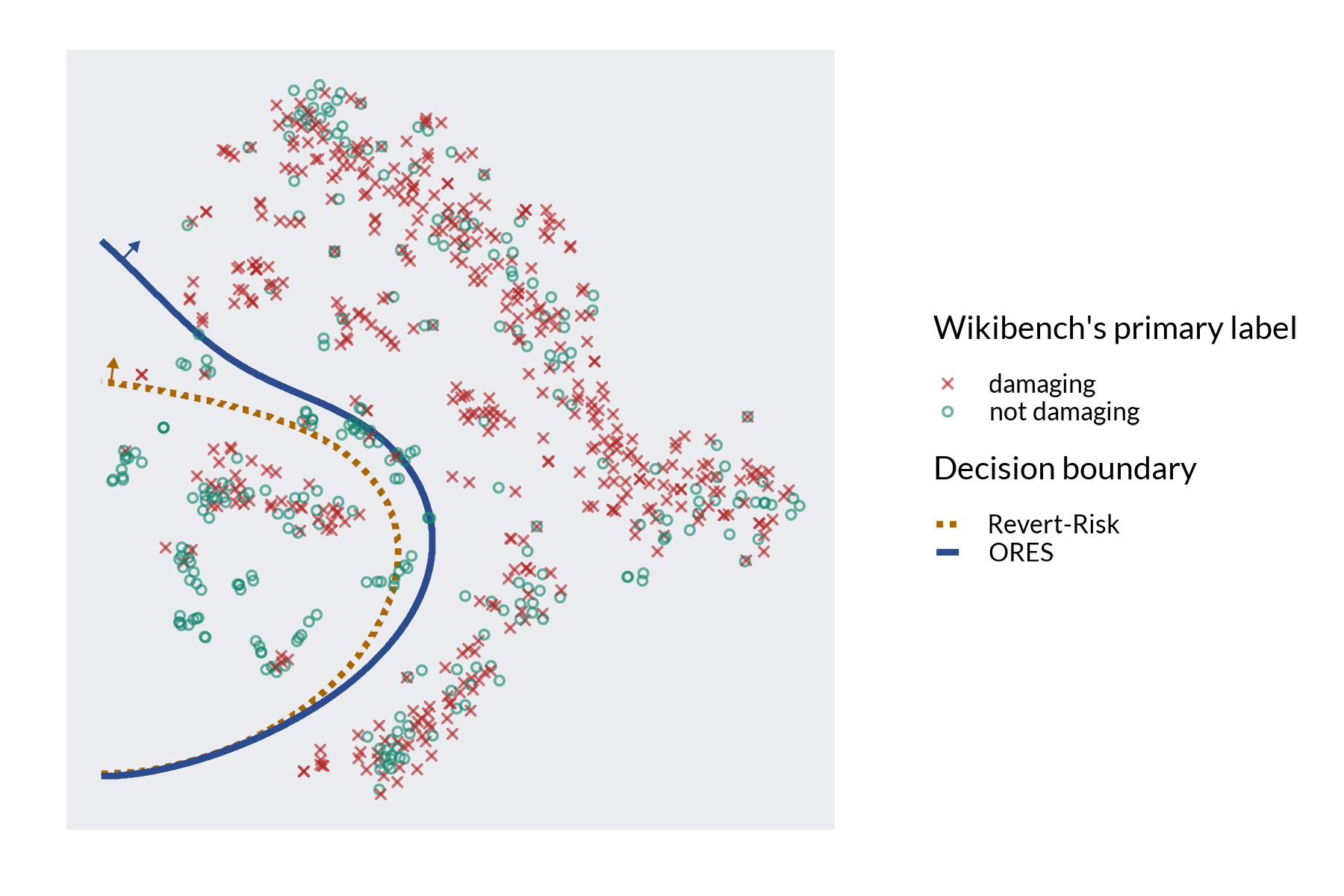}
  \caption{Edits from Wikibench's dataset projected onto a 2D space, along with decision boundaries visualized by training an SVM using the predictions from two AI models. The arrows indicate the direction in which SVMs predict damaging. As shown, the Revert-Risk model is more likely to incorrectly flag non-damaging edits.}
  \Description{A scattered plot where each dot is an edit projected onto a 2D space, along with two lines that display the decision boundaries.}
  \label{figure:C53.13}
\end{figure}

Based on the evaluation results, we can further visualize the differences between the two AI models' predictions and Wikibench's primary labels, as shown in Figure~\ref{figure:C53.13}. Each dot represents an edit in Wikibench's dataset encoded using a feature embedding\footnote{\url{https://www.mediawiki.org/wiki/ORES/Feature\_injection}}, then projected onto a 2D space through t-SNE~\cite{van2008visualizing}, and color-coded according to Wikibench's primary labels. The two lines, serving as decision boundaries for demonstrative purposes, are plotted using SVMs~\cite{chang2011libsvm} based on the respective predictions of the two AI models. The prediction thresholds for ORES and Revert-Risk are 0.3810 and 0.6513, respectively, which were selected to maximize their prediction accuracy. Edits predicted by the model with a probability above the threshold are categorized as damaging. Figure~\ref{figure:C53.13} shows that the Revert-Risk model is more likely to incorrectly flag non-damaging edits. This result is likely because Revert-Risk is trained on trace data (an edit's revert history) rather than explicit labels, increasing the likelihood of incorrectly flagging types of edits that may have historically been reverted for reasons other than being damaging. For instance, in one case, both our participants and ORES considered an edit\footnote{\url{https://enwp.org/Special:Diff/1163324435/1163324481}} non-damaging, whereas the Revert-Risk model predicted the opposite. Interestingly, the edit was eventually reverted, as predicted by Revert-Risk, but not for causing damage to the article. Instead, it resulted from an edit war, in which this edit was actually countering vandalism but was repeatedly reverted by the vandaliser. In this case, using Revert-Risk to identify vandalism might mistakenly flag edits that are actually combating vandalism, counter to its original purpose.

Overall, this analysis showcases the potential of Wikibench's dataset for use in AI evaluation. It highlights the unique value of community-curated evaluation datasets, which may help identify misalignments between community perspectives and AI behaviors that would otherwise be overlooked in the development process.

\subsubsection{\textbf{Participants believe Wikibench's dataset can help the community understand gaps between their collective values and AI models' predictions.}}\label{section:evaluation/perception}
As mentioned in Section \ref{fieldstudy/protocol}, during the exit interview we randomly selected edits from Wikibench's dataset and presented participants with both ORES's and Revert-Risk's predictions. Participants shared that \textit{``it allows us to easily compare what Wikipedians believe vs what the AI believes. Using this data, we can find patterns of mistakes of AI models''} (P1). Through these comparisons, participants also gained a better understanding of which cases may be difficult for an AI model and why: \textit{``Three people labeled it as damaging with low confidence, which might be why it wasn't picked up by the AI as well''} (P9). Participants recognized the dataset's potential to help their community evaluate among AI models: \textit{``It would serve as a good benchmark for different tools, [... and] for whatever models that people come up with afterwards''} (P8). Finally, participants saw opportunities for community-facing visualization or analysis tools to help community members conduct more systematic comparisons: \textit{``It will be really interesting if you compare these two samples, you can [visualize] nice charts [to show] how they look like''} (P7).

\section{Findings: How Wikipedians Use Wikibench}\label{section:finding/usage}

In this section we present results from our analyses of interviews with participants to better understand their usage of Wikibench,  their perceptions of how well our design requirements are fulfilled, and their visions for opportunities to improve Wikibench. We first discuss participant's perceptions of how Wikibench fits into Wikipedia's existing workflows and norms (Section~\ref{section:finding/workflow}). By cross-referencing participants' interview feedback and conversations on talk pages, we then highlight three cases where Wikibench's approach provided the community with the agency to drive the overall data curation process, beyond labeling individual edits (Section~\ref{section:findings/agency}). We conclude this section by offering further insights into the ways participants use Wikibench for collaborative data curation and their suggestions for future improvements to the system (Section~\ref{section:findings/collaboration}).

\subsection{Data Curation within Wikipedians' Existing Workflows}\label{section:finding/workflow}
\begin{quote}
\textit{``I would say it's almost like it was built-in [...] It was meant to be there''} (P10). 
\end{quote}
Compared to Wikilabels, participants find Wikibench much easier to use because it is embedded into their existing workflow: \textit{``I think that Wikibench is currently the best way to achieve this because if you still use the old labeling options you lose a lot of time to even to login and try to work with the old interface just to label some edits''} (P7). Participants also shared that they felt comfortable editing labels and engaging in discussions because Wikibench followed Wikipedia's established norms: \textit{``that creates a lot of comfort, because for me, I was navigating something that I knew very well. I know how discussions work on Wikipedia. I know what to expect. I know what's considered appropriate and what isn't. So I think it's really, really good''} (P6). For similar reasons, participants appreciated that in Wikibench's design, primary labels are established based on consensus rather than a majority vote: \textit{``That is the core of decision making on Wikipedia. [...] I think it's smart that the primary label is not automatically [set as] the majority''} (P3). 

\subsection{Community Agency over the Data Curation Process}\label{section:findings/agency}

Wikibench was designed to empower communities to drive the data curation process, beyond labeling. This section showcases three examples where participants organically shaped the overall data curation process through Wikibench, by (1) refining the label definitions, (2) determining data inclusion criteria, and (3) authoring a data statement.

\subsubsection{\textbf{Participants revised label definitions to better capture emerging nuances and community norms.}}

\begin{figure*}[t]
  \centering
  \includegraphics[width=\linewidth]{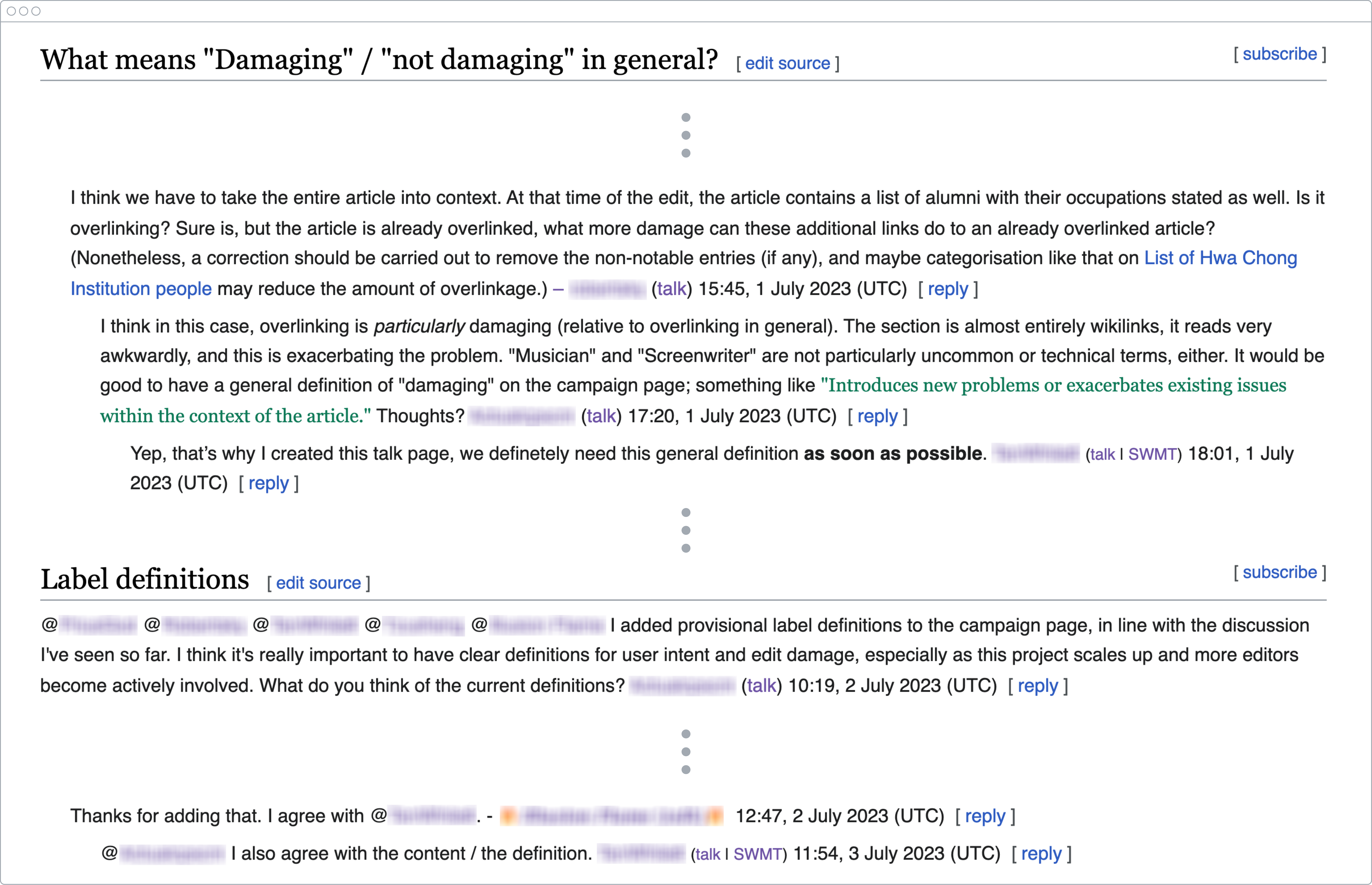}
  \caption{An abridged screenshot of the campaign-level talk page where participants organically discussed and revised label definitions.}
  \Description{An abridged screenshot of the campaign-level talk page. One discussion topic says: What means damaging / not damaging in general? Another discussion topics: Label definitions.}
  \label{figure:talkpage1}
\end{figure*}

Unexpectedly, early in our field study, participants organically identified a need for better label definitions to guide subsequent data labeling and curation. The discussion started from an edit\footnote{\url{https://enwp.org/Special:Diff/719347634/719359416}} that added more wiki links to an already overlinked article. P3 and P4 submitted opposite labels for edit damage, which attracted P2, P5, P6, and P8, from the campaign page. P6 noted: \textit{``overlinking is damaging for legibility, especially in this section that's already 80\% wikilinks.''} Meanwhile, P5 noted: \textit{``overlinking, yes, but that's not really damaging the article.''} With these differing points of view, P2 initiated a discussion on the entity talk page where five participants responded. Even though participants, such as P3, replied on the talk page that they believed \textit{``a healthy amount of differences of opinion (in the right places!) is the foundation of establishing positive consensus,''} they found it critical to have a clear label definition as a baseline to build upon. Given that general questions about label definitions went beyond the discussion about an individual edit, P5 initiated a discussion on the campaign talk page named \textit{``What means `damaging'/`not damaging' in general?''} (see Figure~\ref{figure:talkpage1}). Following a series of discussions, participants edited the campaign page to update the original label definition we provided:

\begin{quote}
\textbf{Edit damage}: The edit damage label indicates whether this edit causes damage to the article or not.\\
\textbf{User intent}: The user intent label indicates whether the edit was saved in good or bad faith.
\end{quote}
into the new label definition that better captures the nuance and community norms:
\begin{quote}
\textbf{Edit damage}: An edit is considered constructive when the post-edit revision is better than the pre-edit revision. For the purposes of evaluating edit damage, edits are not evaluated against what else could have been done to improve the article. ``Not damaging'' is a soft default; if an edit makes the article neither better nor worse at all, it is not damaging. Label data should still be provided in cases where edit damage depends on external factors. For example, an edit which introduces verifiably false information with appropriate style and formatting should be labelled as damaging.\\
\textbf{User intent}: An edit is considered ``good faith'' when it is reasonably plausible that the editor's intention was to improve the article; per WP:AGF\footnote{\url{https://enwp.org/WP:AGF}}, good faith is the default until there is a concrete reason to suspect bad faith. Not all damaging edits are made in bad faith.
\end{quote}

\subsubsection{\textbf{Participants defined inclusion criteria to ensure data was accessible by the full community.}} 
Midway through the study, participants found that some labeled edits were only visible to Wikipedia administrators. In response, they collectively decided to remove these edits from the dataset. It began when P4, P6, P9, and P10 each came across some entity pages where the edits were hidden from public view by Wikipedia administrators after being labeled. Even though these hidden edits, also known as revdel on Wikipedia\footnote{\url{https://enwp.org/WP:REVDEL}}, were still visible to some participants with administrator rights, P9 initiated a discussion on an entity talk page. P10 noticed this wasn't a one-time incident and raised the issue on the campaign talk page by creating a discussion topic \textit{``How to handle diffs that have since been revdel’d?''} While some participants wondered if \textit{``labeling them as damaging or bad faith in Wikibench''} (P8) would make sense, others argued \textit{``it might be better to have the dataset used for training and evaluating AI be entirely transparent''} (P6). Following discussions, participants collectively decided to exclude these edits from the dataset. P4, a participant with the administrator rights, volunteered to make an update so that Wikibench would no longer include these entity pages in the table on the campaign page given their updated prefix. Participants also added a new section on the campaign page named \textit{``Entity pages on revdel'ed edits,''} where they compiled a list of archived entity pages and provided instructions for people to report here when encountering other such edits.

\subsubsection{\textbf{Participants authored a data statement to specify appropriate usage of the evaluation dataset}}
During the field study, participants recognized that the resulting dataset likely would not align with the natural distribution of edits on Wikipedia. Given the observation, some participants raised the question on the campaign talk page: \textit{``I wonder how useful this dataset will be for training and evaluating AI given that it does not accurately represent the totality of edits on Wikipedia. [...] For example, damaging and bad-faith edits are significantly overrepresented''} (P6). After we explained the potential use participants added the following data statement to clarify the usage of the evaluation dataset, preventing misuse by those unfamiliar with Wikipedia's context.
\begin{quote}
\textbf{Data statement}: There is a local consensus that on-wiki use of the data acquired through this campaign should be limited to the immediate scope of Wikibench. A strong consensus should be established prior to any on-wiki use outside this research project. The data acquired as part of this campaign is not intended for other uses and may be inappropriate or unsuitable for many purposes. In particular, it is not a representative sample of edits made to the English Wikipedia, nor is it intended as such.
\end{quote}

\subsection{Collaborative Data Labeling}\label{section:findings/collaboration}

Participants appreciated Wikibench's collaborative approach to data labeling because it allowed contributors with complementary perspectives to build consensus and a stronger community. Participants found that Wikibench's campaign and entity pages facilitate collaborative data labeling by quickly pinpointing edits where more labels or discussions could be valuable. They also perceived Wikibench as effective in surfacing disagreements and facilitating consensus-building. In each area, participants also envisioned various opportunities to make Wikibench more effective. 

\subsubsection{\textbf{Participants prefer Wikibench's collaborative approach to data labeling}}
Participants found Wikibench enabled contributors with diverse and complementary expertise to discuss and reach a consensus on the final labels: \textit{``It's very useful to have editors with different areas of expertise within Wikipedia working together. [...] In the end, this supports getting better data''} (P6). This is particularly relevant for P6, while reviewing an edit\footnote{\url{https://enwp.org/Special:Diff/1128051255/1163570328}} about a coffee produced in Southeast Asia, initially found the primary label confusing: \textit{``Adding this image seems fine to me. Damaging? Why?''} After reading P4's note, a participant from Singapore, P6 agreed with the label: \textit{``Copyright? Oh, okay, this would be very difficult to investigate properly. It is a copyright violation because it has a company logo.''} Without P4's local knowledge, P6 would have missed the copyright violation and labeled the edit as not damaging. Participants find these notes and discussions facilitated by Wikibench helpful because it allows them to learn new things from others: \textit{``This allows me to discover things I might not have thought of''} (P1). They also find these discussions help the community collectively reflect on their patrolling standards: \textit{``It allows the community to establish better consensus as to what typically should be reverted and what might require more care''} (P3). In turn, they believe these interactions could help build a stronger community: \textit{``I haven't really talked to many of the other patrollers before this project. For people who are interested in a tighter community, Wikibench would be absolutely a great option''} (P11).

\subsubsection{\textbf{Participants find Wikibench helpful in quickly identifying edits where more labels or discussion could be valuable.}}
For example, some participants found the campaign page helpful in identifying edits with fewer labels to ensure that edits are collaboratively labeled: \textit{``I try to work on those which were alone because I previously complained about that the previous system allowed only one person to evaluate.''} (P7).
Some participants were particularly interested in checking edits with high disagreements and helping build consensus: \textit{``I'm very interested in seeing where people differ. [...] I wanted to build consensus, or at least to try''} (P11). Given the desire to more effectively locate edits for contribution, participants suggested: \textit{``Maybe you can add a label `discussion: no/yes’ to the big table so you can see where is a discussion and join there. The disagreement is a good way to find discussions, but isn't perfect''} (P5).

Once participants located an edit and opened its entity page, participants found the entity page provided an at-a-glance overview that helped quickly understand the current level of disagreement: \textit{``You can see the colors and see one damaging, one not damaging low confidence, [...] The preview was really good to see''} (P5). Another participant echoed: \textit{``Wikibench's feature of showing exactly where different editors align, and that one user cannot force the primary label, helped to facilitate discussions''} (P12). When deeper discussion was needed, participants found the talk page helpful: \textit{``If I still have a question, I can open the talk page''} (P5). However, a participant also expresses concern that the transparent labeler information might affect people's judgment in undesirable ways, in some cases: \textit{``If it wasn't transparent, you wouldn't know who did the labeling. [...] But on the other hand, if the quotation marks `big guys' go on that way, then maybe the small guys will follow''} (P7).

\subsubsection{\textbf{Participants find Wikibench effective in surfacing disagreements and facilitating consensus-building.}}
Participants perceived that discussions on Wikibench were typically sparked by disagreements, but concluded with a shared consensus: \textit{``Most of the time, one of me or the other one says: Okay, I think you have the better argument, and I'm switching to your position, or it's okay for me''} (P5). Even when people hold strong opposing opinions, participants found Wikibench helpful in facilitating more productive discussions and avoiding emotionally charged confrontations: \textit{``Wikibench provides an avenue for people to calmly discuss stuff because it's nothing personal''} (P4). Participants also shared that they adopted different strategies for consensus building, with some followed Wikibench's nudge to boldly edit primary labels: \textit{``I changed the primary label before I actually brought up the point, just mostly so I can actually get the person's attention''} (P11). Meanwhile, some participants prefer to initiate a discussion and wait for consensus to form first: \textit{``'I might change it if I think the other editor has made a mistake, rather than made a decision that I disagree with. If it's a disagreement, we let consensus form first'} (P8). Overall, participants appreciated the design where the primary label was not the majority vote but could be edited by anyone: \textit{``I like being able to edit the primary label to help better reflect community consensus. I find the warning helpful as it reminds users that even when being bold, your changes should reflect community consensus, not your personal opinions''} (P1).

\subsubsection{\textbf{Participants see value in the openness of Wikibench's datasets.}} Similar to Wikipedia articles, Wikibench's dataset is not owned by specific individuals but is open to the public. Participants appreciated the openness of Wikibench's dataset, and emphasized that data transparency is essential for the evaluation results to be trustworthy: \textit{``These tools do so much. They're very highly trusted. [...] If we're evaluating that kind of tool, there needs to be an additional level of trust for the dataset we're using to do that''} (P6).

\section{Discussion}\label{section:discussion}

As AI tools are increasingly developed for community contexts, it is critical to ensure that they are aligned with community needs and values. In this paper, we introduce Wikibench, a system that enables community members to collectively curate evaluation datasets for AI tools that will be used in their communities. We conducted a field study on Wikipedia to understand how a real-world community might use Wikibench in practice. Overall, we found that Wikipedians' use of Wikibench yielded labels that are reflective of consensus among a broader range of community members, while also capturing ambiguities and differences in perspective among community members (Section~\ref{section:findings/quality}). We find promising evidence of the utility of Wikibench's community-curated datasets in understanding areas of alignment and misalignment with community perspectives (Section~\ref{section:findings/dataset/evaluation}). Finally, we present examples of how community members use Wikibench to shape the overall data curation process, and discuss their experiences using the system for data labeling and curation (Section~\ref{section:finding/usage}). In this section, we highlight key takeaways and propose future directions for HCI systems that support community-driven data curation and AI evaluation.

\subsection{Supporting Community-Driven Data Curation beyond Wikipedia}
As the largest, most successful platform for collective knowledge-building and curation, we believe there is much to learn from Wikipedia for the design of effective community-driven data curation processes. Given that Wikibench was designed around Wikipedia's processes and norms, we anticipate that several aspects of Wikibench's design may be useful for community-driven data curation in other contexts. In particular, we expect that the four design requirements discussed in Section~\ref{section:designrequirements} are broadly relevant for the design of tools intended to support community-driven data curation. We also expect that the overall workflow embodied by Wikibench (Figure~\ref{figure:workflow}) will be generalizable to other community platforms, particularly those operating within the middle level of a multi-level governance structure~\cite{jhaver2023decentralizing}, such as subreddits, Facebook Groups, Mastodon Nodes, other Wikipedia language editions, and more. 

At the same time, we expect that the specific implementation of Wikibench's plug-in, entity page, and campaign page will require careful adaptation to align with the established norms of other communities. For example, imagine a subreddit community interested in adapting Wikibench to curate posts as data points and labels indicating whether a given post ought to be flagged for removal. In this context, integrating the data curation process into existing work practices (D3), would likely mean tailoring the specific implementation of a plug-in to fit the workflow of the ``Knights of New''\footnote{A group of volunteers that review new posts instead of already popular posts.} on Reddit~\cite{gilbert2013widespread,massanari2015participatory}. Similarly, to encourage deliberation (D2), the entity and campaign page would likely need to be adapted to better align with existing collaboration and consensus-building processes on Reddit~\cite{chandrasekharan2018internet}. The system may also need to incorporate the community's established norms (D1) for safeguarding against bad actors. On Wikipedia, this may mean restricting access to Wikibench to registered users, whereas on Reddit this may mean considering a Reddit user's karma scores\footnote{A score representing the positive social signals that a user's activity (e.g., posts and comments) has received.}. Future research should investigate how systems for community-driven data curation can be designed for use by other communities outside of Wikipedia, and what mechanisms in Wikibench's current design are most readily transferable across community contexts.

\subsection{Balancing Costs and Benefits in Community-Driven Data Curation}

\textcolor{black}{Prior approaches to account for annotator disagreements tend to handle disagreements post-hoc, after individual labels have been gathered~\cite{gordon2022jury,dutta2023modeling,bakker2022fine,deng2023you,wallace2022debiased}, instead of facilitating discussion and deliberation among annotators. By contrast, Wikibench provides community members the agency to navigate and resolve disagreements, leading to various benefits that would not be achievable by algorithmic methods alone. For example, discussions among community members can help resolve ambiguities in labeling, facilitate consensus building, and promote collective reflection on community standards, as discussed in Section~\ref{section:findings/collaboration}.} However, given that community members will generally have limited time and attention to contribute to the curation of AI datasets, further research is needed to find the right balance between \textit{community agency} over curation processes, on the one hand, and time and labor \textit{efficiency} on the other. The current version of Wikibench aims to make more effective use of community members' time by embedding the plug-in within their everyday workflows (D3) instead of asking them to use Wikilabels just for labeling. Wikibench also automatically surfaces edits that may benefit most from additional attention on the campaign page. Future research could explore ways to better optimize the use of community members' time, such as algorithmically prioritizing data points that are predicted as more likely to prompt disagreements for community discussions~\cite{baumler2023examples}. Relatedly, future research could investigate when the benefits of further community engagement in data curation becomes marginal, as the number of participants and contributions increase.

\subsection{Advancing Pluralistic Approaches to AI Evaluation}
Our demonstration of Wikibench's use to compare different AI models' community alignment employed the primary labels collected through Wikibench in our field study. However, a major strength of Wikibench's community-curated datasets is their ability to capture additional signals such as ambiguity in labeling and differences in perspective among community members. Recent research has suggested the benefits of evaluating AI models using datasets that reflect diverse perspectives, with multiple labelers per data point (e.g., ~\cite{gordon2021disagreement,gordon2022jury,chen2023judgment,kapania2023hunt,goyal2022your}). Where disagreements arise, these may represent ambiguity inherent to an edit or noise in labeling, or they may signal genuine differences in perspective among subgroups of a community whose voices deserve consideration. Wikibench datasets record signals to help AI evaluators tease apart these possibilities, which can support more nuanced and pluralistic analyses of AI models' community alignment. The development of evaluation methods and workflows to support more pluralistic approaches to AI evaluation is an emerging area of research~\cite{sorensen2024roadmap,cabitza2023toward}. \textcolor{black}{Future research could systematically compare Wikibench's process with alternative approaches to handling differences in perspective for pluralistic AI evaluation. Such comparisons could advance our understanding of trade-offs between different community-driven and algorithmic approaches to navigating disagreements in labeling. In turn, this may inform the development of new approaches that integrate complementary strengths of existing methods.} It is our hope that systems for community-driven data curation can help to accelerate progress in this area through the development of relevant datasets~\cite{oala2023dmlr,mazumder2022dataperf}.

\subsection{Designing Community-Facing Evaluation Interfaces}
While Wikibench's current interface primarily supports community-driven \textit{data curation}, future research should explore the design of community-facing interfaces that empower communities to effectively leverage the resulting datasets to inform decision-making about AI design and adoption. We envision that, in more complex evaluation scenarios that require caution in interpretation, community-driven AI evaluations may sometimes be facilitated through partnerships with technical experts~\cite{costanza2020design}. Beyond supporting AI evaluation, our participants found value in Wikibench's collaborative data curation process during our field study because it helped them to reflect, both individually and collectively, on their edit patrolling standards. Thus, an additional promising direction for future research, in the Wikipedia context and beyond, is to explore how community-driven data curation processes can be more explicitly designed to support such reflection. This may include the design of interfaces that help community members leverage community-curated datasets to reflect upon \textit{their own} decision-making---both individually and as a community---and identify potential areas for improvement (cf.~\cite{zhang2023deliberating}).

\subsection{Supporting Communities in Steering Overall Dataset Composition}
The current version of Wikibench was designed to deeply embed into community members' regular patrolling activities on Wikipedia. This ``lowers the floor'' of effort required to contribute to Wikibench datasets. However, a consequence of this design is that the data points included in Wikibench datasets may tend to reflect the distribution of edits that Wikipedians encounter while patrolling, which are not necessarily representative of \textit{all} edits made to English Wikipedia. Indeed, as the participant-authored data statement from our field study acknowledges, the dataset curated through this study was not intended to be representative in this sense. This makes the dataset more useful for some evaluation purposes than others. Future research should explore how to design mechanisms for community-driven curation that can assist communities in steering datasets, depending on their specific goals, toward desired distributional properties (e.g., specific notions of representativeness, or oversampling along particular dimensions of interest). It is possible that to some extent, this may require community members to spend more time contributing outside of their regular workflows. However, it may be possible to design new mechanisms that minimize the disruption required. For example, in the context of a community member's regular patrolling activities on Wikipedia, a hypothetical future browser plug-in might be designed to occasionally present edits that they \textit{would not have otherwise encountered} for labeling purposes, intended to help the community achieve their overall distributional goals for the dataset. We envision that, in many community contexts, decision-making regarding the distributional properties a campaign should aim for may benefit from accessibly-designed ``explainers'' that summarize relevant consideration (cf. \cite{kuo2023understanding}), and/or through partnerships with relevant technical experts~\cite{costanza2020design}.

\subsection{Drawing Inspiration from Existing Content Curation Mechanisms}
\textcolor{black}{
Several of Wikibench’s \textit{AI data} curation mechanisms are inspired by existing \textit{content} curation mechanisms from Wikipedia and other online platforms. For example, Stack Overflow users can vote on and discuss individual posts~\cite{mamykina2011design} or have higher-level discussions about community norms on Meta Stack Overflow~\cite{fang2023people}. Similarly, Wikipedians can use Wikibench to label and discuss individual data points on entity pages or initiate higher-level conversations about the overall data curation process on the campaign page. We see opportunities for future versions of Wikibench, or other community-driven data curation platforms, to draw further inspiration from the design of existing content curation mechanisms. For example, a feature inspired by Reddit's post flairs (short tags attached to each post)~\cite{jhaver2019did}, could be used to categorize data points and facilitate dataset navigation (e.g., via data slices~\cite{chen2019slice}). In addition, future work in this space can take inspiration from HCI research related to content curation. For example, recent social media research has proposed curating content not solely based on engagement signals such as votes but on a community's shared visions~\cite{he2023cura} and values~\cite{jia2023embedding}. In the context of data curation, these ideas may inform new approaches to the community-driven prioritization of data points for AI evaluation.
}

\section{Conclusion}
In this work, we have demonstrated the potential for new approaches to community-driven curation of AI evaluation datasets, through the introduction of the Wikibench system and a field study investigating its use. Our findings demonstrate that community-driven curation on Wikibench can produce datasets that capture community consensus, disagreement, and uncertainty, while enabling community members to shape the overall data curation process. Building on this work, future research should explore the design of tools and processes that can support community-driven data curation across a broader range of contexts, and that can expand community agency in both the curation of datasets and their use in evaluation.

\begin{acks}
The funding for this research was provided by UL Research Institutes through the Center for Advancing Safety of Machine Intelligence, CMU's Block Center for Technology and Society, and the National Science Foundation (NSF) under Award No. 1952085 and 2001851. We thank Wikipedians for their time and input that shaped this research, especially 1TWO3Writer, Actualcpscm, Alpha3031, Bencemac, Bluerasberry, Chtnnh, Ciell, Fehufanga, FenrisAureus, Illusion Flame, Loafiewa, Matthewrb, PriusGod, Robertsky, RonnieV, Schminnte, Skarmory, TenWhile6, Vermont, Zache, Zppix, and many others (including those who chose to remain anonymous). We also appreciate Jane Hsieh, Kimi Wenzel, Luke Guerdan, Ningjing Tang, Pranav Khadpe, Seyun Kim, Tiffany Chih, and Wesley Deng for their feedback on the paper draft. Finally, we thank Isadora Krsek for designing the Wikibench logo, as shown in Figure~\ref{figure:teaser}.
\end{acks}

\bibliographystyle{ACM-Reference-Format}
\bibliography{main}


\begin{thebibliography}{97}


\ifx \showCODEN    \undefined \def \showCODEN     #1{\unskip}     \fi
\ifx \showDOI      \undefined \def \showDOI       #1{#1}\fi
\ifx \showISBNx    \undefined \def \showISBNx     #1{\unskip}     \fi
\ifx \showISBNxiii \undefined \def \showISBNxiii  #1{\unskip}     \fi
\ifx \showISSN     \undefined \def \showISSN      #1{\unskip}     \fi
\ifx \showLCCN     \undefined \def \showLCCN      #1{\unskip}     \fi
\ifx \shownote     \undefined \def \shownote      #1{#1}          \fi
\ifx \showarticletitle \undefined \def \showarticletitle #1{#1}   \fi
\ifx \showURL      \undefined \def \showURL       {\relax}        \fi
\providecommand\bibfield[2]{#2}
\providecommand\bibinfo[2]{#2}
\providecommand\natexlab[1]{#1}
\providecommand\showeprint[2][]{arXiv:#2}

\bibitem[Aroyo and Paritosh(2021)]%
        {aroyo2021adversarial}
\bibfield{author}{\bibinfo{person}{Lora~Mois Aroyo} {and}
  \bibinfo{person}{Praveen~Kumar Paritosh}.} \bibinfo{year}{2021}\natexlab{}.
\newblock \showarticletitle{Adversarial Test Set for Image Classification:
  Lessons Learned from CATS4ML Data Challenge}.
\newblock  (\bibinfo{year}{2021}).
\newblock


\bibitem[Attenberg et~al\mbox{.}(2015)]%
        {attenberg2015beat}
\bibfield{author}{\bibinfo{person}{Joshua Attenberg}, \bibinfo{person}{Panos
  Ipeirotis}, {and} \bibinfo{person}{Foster Provost}.}
  \bibinfo{year}{2015}\natexlab{}.
\newblock \showarticletitle{Beat the machine: Challenging humans to find a
  predictive model's “unknown unknowns”}.
\newblock \bibinfo{journal}{\emph{Journal of Data and Information Quality
  (JDIQ)}} \bibinfo{volume}{6}, \bibinfo{number}{1} (\bibinfo{year}{2015}),
  \bibinfo{pages}{1--17}.
\newblock


\bibitem[Bai et~al\mbox{.}(2023)]%
        {bai2023measuring}
\bibfield{author}{\bibinfo{person}{Yuanchen Bai}, \bibinfo{person}{Raoyi
  Huang}, \bibinfo{person}{Vijay Viswanathan}, \bibinfo{person}{Tzu-Sheng Kuo},
  {and} \bibinfo{person}{Tongshuang Wu}.} \bibinfo{year}{2023}\natexlab{}.
\newblock \showarticletitle{Measuring Adversarial Datasets}.
\newblock \bibinfo{journal}{\emph{arXiv preprint arXiv:2311.03566}}
  (\bibinfo{year}{2023}).
\newblock


\bibitem[Bakker et~al\mbox{.}(2022)]%
        {bakker2022fine}
\bibfield{author}{\bibinfo{person}{Michiel Bakker}, \bibinfo{person}{Martin
  Chadwick}, \bibinfo{person}{Hannah Sheahan}, \bibinfo{person}{Michael
  Tessler}, \bibinfo{person}{Lucy Campbell-Gillingham}, \bibinfo{person}{Jan
  Balaguer}, \bibinfo{person}{Nat McAleese}, \bibinfo{person}{Amelia Glaese},
  \bibinfo{person}{John Aslanides}, \bibinfo{person}{Matt Botvinick},
  {et~al\mbox{.}}} \bibinfo{year}{2022}\natexlab{}.
\newblock \showarticletitle{Fine-tuning language models to find agreement among
  humans with diverse preferences}.
\newblock \bibinfo{journal}{\emph{Advances in Neural Information Processing
  Systems}}  \bibinfo{volume}{35} (\bibinfo{year}{2022}),
  \bibinfo{pages}{38176--38189}.
\newblock


\bibitem[Bartolo et~al\mbox{.}(2020)]%
        {bartolo2020beat}
\bibfield{author}{\bibinfo{person}{Max Bartolo}, \bibinfo{person}{Alastair
  Roberts}, \bibinfo{person}{Johannes Welbl}, \bibinfo{person}{Sebastian
  Riedel}, {and} \bibinfo{person}{Pontus Stenetorp}.}
  \bibinfo{year}{2020}\natexlab{}.
\newblock \showarticletitle{Beat the AI: Investigating adversarial human
  annotation for reading comprehension}.
\newblock \bibinfo{journal}{\emph{Transactions of the Association for
  Computational Linguistics}}  \bibinfo{volume}{8} (\bibinfo{year}{2020}),
  \bibinfo{pages}{662--678}.
\newblock


\bibitem[Baumler et~al\mbox{.}(2023)]%
        {baumler2023examples}
\bibfield{author}{\bibinfo{person}{Connor Baumler}, \bibinfo{person}{Anna
  Sotnikova}, {and} \bibinfo{person}{Hal Daum{\'e}~III}.}
  \bibinfo{year}{2023}\natexlab{}.
\newblock \showarticletitle{Which examples should be multiply annotated? active
  learning when annotators may disagree}. In \bibinfo{booktitle}{\emph{Findings
  of the Association for Computational Linguistics: ACL 2023}}.
  \bibinfo{pages}{10352--10371}.
\newblock


\bibitem[Beschastnikh et~al\mbox{.}(2008)]%
        {beschastnikh2008wikipedian}
\bibfield{author}{\bibinfo{person}{Ivan Beschastnikh}, \bibinfo{person}{Travis
  Kriplean}, {and} \bibinfo{person}{David McDonald}.}
  \bibinfo{year}{2008}\natexlab{}.
\newblock \showarticletitle{Wikipedian self-governance in action: Motivating
  the policy lens}. In \bibinfo{booktitle}{\emph{Proceedings of the
  International AAAI Conference on Web and Social Media}},
  Vol.~\bibinfo{volume}{2}. \bibinfo{pages}{27--35}.
\newblock


\bibitem[Borkan et~al\mbox{.}(2019)]%
        {borkan2019nuanced}
\bibfield{author}{\bibinfo{person}{Daniel Borkan}, \bibinfo{person}{Lucas
  Dixon}, \bibinfo{person}{Jeffrey Sorensen}, \bibinfo{person}{Nithum Thain},
  {and} \bibinfo{person}{Lucy Vasserman}.} \bibinfo{year}{2019}\natexlab{}.
\newblock \showarticletitle{Nuanced metrics for measuring unintended bias with
  real data for text classification}. In \bibinfo{booktitle}{\emph{Companion
  proceedings of the 2019 world wide web conference}}.
  \bibinfo{pages}{491--500}.
\newblock


\bibitem[Braun and Clarke(2012)]%
        {braun2012thematic}
\bibfield{author}{\bibinfo{person}{Virginia Braun} {and}
  \bibinfo{person}{Victoria Clarke}.} \bibinfo{year}{2012}\natexlab{}.
\newblock \bibinfo{booktitle}{\emph{Thematic analysis.}}
\newblock \bibinfo{publisher}{American Psychological Association}.
\newblock


\bibitem[Braun and Clarke(2019)]%
        {braun2019reflecting}
\bibfield{author}{\bibinfo{person}{Virginia Braun} {and}
  \bibinfo{person}{Victoria Clarke}.} \bibinfo{year}{2019}\natexlab{}.
\newblock \showarticletitle{Reflecting on reflexive thematic analysis}.
\newblock \bibinfo{journal}{\emph{Qualitative research in sport, exercise and
  health}} \bibinfo{volume}{11}, \bibinfo{number}{4} (\bibinfo{year}{2019}),
  \bibinfo{pages}{589--597}.
\newblock


\bibitem[Bryant et~al\mbox{.}(2005)]%
        {bryant2005becoming}
\bibfield{author}{\bibinfo{person}{Susan~L Bryant}, \bibinfo{person}{Andrea
  Forte}, {and} \bibinfo{person}{Amy Bruckman}.}
  \bibinfo{year}{2005}\natexlab{}.
\newblock \showarticletitle{Becoming Wikipedian: transformation of
  participation in a collaborative online encyclopedia}. In
  \bibinfo{booktitle}{\emph{Proceedings of the 2005 ACM International
  Conference on Supporting Group Work}}. \bibinfo{pages}{1--10}.
\newblock


\bibitem[Cabitza et~al\mbox{.}(2023)]%
        {cabitza2023toward}
\bibfield{author}{\bibinfo{person}{Federico Cabitza}, \bibinfo{person}{Andrea
  Campagner}, {and} \bibinfo{person}{Valerio Basile}.}
  \bibinfo{year}{2023}\natexlab{}.
\newblock \showarticletitle{Toward a perspectivist turn in ground truthing for
  predictive computing}. In \bibinfo{booktitle}{\emph{Proceedings of the AAAI
  Conference on Artificial Intelligence}}, Vol.~\bibinfo{volume}{37}.
  \bibinfo{pages}{6860--6868}.
\newblock


\bibitem[Cabrera et~al\mbox{.}(2021)]%
        {cabrera2021discovering}
\bibfield{author}{\bibinfo{person}{{\'A}ngel~Alexander Cabrera},
  \bibinfo{person}{Abraham~J Druck}, \bibinfo{person}{Jason~I Hong}, {and}
  \bibinfo{person}{Adam Perer}.} \bibinfo{year}{2021}\natexlab{}.
\newblock \showarticletitle{Discovering and validating ai errors with
  crowdsourced failure reports}.
\newblock \bibinfo{journal}{\emph{Proceedings of the ACM on Human-Computer
  Interaction}} \bibinfo{volume}{5}, \bibinfo{number}{CSCW2}
  (\bibinfo{year}{2021}), \bibinfo{pages}{1--22}.
\newblock


\bibitem[Chandrasekharan et~al\mbox{.}(2018)]%
        {chandrasekharan2018internet}
\bibfield{author}{\bibinfo{person}{Eshwar Chandrasekharan},
  \bibinfo{person}{Mattia Samory}, \bibinfo{person}{Shagun Jhaver},
  \bibinfo{person}{Hunter Charvat}, \bibinfo{person}{Amy Bruckman},
  \bibinfo{person}{Cliff Lampe}, \bibinfo{person}{Jacob Eisenstein}, {and}
  \bibinfo{person}{Eric Gilbert}.} \bibinfo{year}{2018}\natexlab{}.
\newblock \showarticletitle{The Internet's hidden rules: An empirical study of
  Reddit norm violations at micro, meso, and macro scales}.
\newblock \bibinfo{journal}{\emph{Proceedings of the ACM on Human-Computer
  Interaction}} \bibinfo{volume}{2}, \bibinfo{number}{CSCW}
  (\bibinfo{year}{2018}), \bibinfo{pages}{1--25}.
\newblock


\bibitem[Chang and Lin(2011)]%
        {chang2011libsvm}
\bibfield{author}{\bibinfo{person}{Chih-Chung Chang} {and}
  \bibinfo{person}{Chih-Jen Lin}.} \bibinfo{year}{2011}\natexlab{}.
\newblock \showarticletitle{LIBSVM: a library for support vector machines}.
\newblock \bibinfo{journal}{\emph{ACM transactions on intelligent systems and
  technology (TIST)}} \bibinfo{volume}{2}, \bibinfo{number}{3}
  (\bibinfo{year}{2011}), \bibinfo{pages}{1--27}.
\newblock


\bibitem[Chen and Zhang(2023)]%
        {chen2023judgment}
\bibfield{author}{\bibinfo{person}{Quan~Ze Chen} {and} \bibinfo{person}{Amy~X
  Zhang}.} \bibinfo{year}{2023}\natexlab{}.
\newblock \showarticletitle{Judgment Sieve: Reducing Uncertainty in Group
  Judgments through Interventions Targeting Ambiguity versus Disagreement}.
\newblock \bibinfo{journal}{\emph{arXiv preprint arXiv:2305.01615}}
  (\bibinfo{year}{2023}).
\newblock


\bibitem[Chen et~al\mbox{.}(2019)]%
        {chen2019slice}
\bibfield{author}{\bibinfo{person}{Vincent Chen}, \bibinfo{person}{Sen Wu},
  \bibinfo{person}{Alexander~J Ratner}, \bibinfo{person}{Jen Weng}, {and}
  \bibinfo{person}{Christopher R{\'e}}.} \bibinfo{year}{2019}\natexlab{}.
\newblock \showarticletitle{Slice-based learning: A programming model for
  residual learning in critical data slices}.
\newblock \bibinfo{journal}{\emph{Advances in neural information processing
  systems}}  \bibinfo{volume}{32} (\bibinfo{year}{2019}).
\newblock


\bibitem[cjadams et~al\mbox{.}(2017)]%
        {jigsaw-toxic-comment-classification-challenge}
\bibfield{author}{\bibinfo{person}{cjadams}, \bibinfo{person}{Jeffrey
  Sorensen}, \bibinfo{person}{Julia Elliott}, \bibinfo{person}{Lucas Dixon},
  \bibinfo{person}{Mark McDonald}, \bibinfo{person}{nithum}, {and}
  \bibinfo{person}{Will Cukierski}.} \bibinfo{year}{2017}\natexlab{}.
\newblock \bibinfo{title}{Toxic Comment Classification Challenge}.
\newblock
\newblock
\urldef\tempurl%
\url{https://kaggle.com/competitions/jigsaw-toxic-comment-classification-challenge}
\showURL{%
\tempurl}


\bibitem[Costanza-Chock(2020)]%
        {costanza2020design}
\bibfield{author}{\bibinfo{person}{Sasha Costanza-Chock}.}
  \bibinfo{year}{2020}\natexlab{}.
\newblock \bibinfo{booktitle}{\emph{Design justice: Community-led practices to
  build the worlds we need}}.
\newblock \bibinfo{publisher}{The MIT Press}.
\newblock


\bibitem[Deng et~al\mbox{.}(2023b)]%
        {deng2023you}
\bibfield{author}{\bibinfo{person}{Naihao Deng}, \bibinfo{person}{Siyang Liu},
  \bibinfo{person}{Xinliang~Frederick Zhang}, \bibinfo{person}{Winston Wu},
  \bibinfo{person}{Lu Wang}, {and} \bibinfo{person}{Rada Mihalcea}.}
  \bibinfo{year}{2023}\natexlab{b}.
\newblock \showarticletitle{You Are What You Annotate: Towards Better Models
  through Annotator Representations}.
\newblock \bibinfo{journal}{\emph{arXiv preprint arXiv:2305.14663}}
  (\bibinfo{year}{2023}).
\newblock


\bibitem[Deng et~al\mbox{.}(2023a)]%
        {deng2023understanding}
\bibfield{author}{\bibinfo{person}{Wesley~Hanwen Deng}, \bibinfo{person}{Boyuan
  Guo}, \bibinfo{person}{Alicia Devrio}, \bibinfo{person}{Hong Shen},
  \bibinfo{person}{Motahhare Eslami}, {and} \bibinfo{person}{Kenneth
  Holstein}.} \bibinfo{year}{2023}\natexlab{a}.
\newblock \showarticletitle{Understanding Practices, Challenges, and
  Opportunities for User-Engaged Algorithm Auditing in Industry Practice}. In
  \bibinfo{booktitle}{\emph{Proceedings of the 2023 CHI Conference on Human
  Factors in Computing Systems}}. \bibinfo{pages}{1--18}.
\newblock


\bibitem[Denton et~al\mbox{.}(2021)]%
        {denton2021whose}
\bibfield{author}{\bibinfo{person}{Emily Denton}, \bibinfo{person}{Mark
  D{\'\i}az}, \bibinfo{person}{Ian Kivlichan}, \bibinfo{person}{Vinodkumar
  Prabhakaran}, {and} \bibinfo{person}{Rachel Rosen}.}
  \bibinfo{year}{2021}\natexlab{}.
\newblock \showarticletitle{Whose ground truth? accounting for individual and
  collective identities underlying dataset annotation}.
\newblock \bibinfo{journal}{\emph{arXiv preprint arXiv:2112.04554}}
  (\bibinfo{year}{2021}).
\newblock


\bibitem[Denton et~al\mbox{.}(2020)]%
        {denton2020bringing}
\bibfield{author}{\bibinfo{person}{Emily Denton}, \bibinfo{person}{Alex Hanna},
  \bibinfo{person}{Razvan Amironesei}, \bibinfo{person}{Andrew Smart},
  \bibinfo{person}{Hilary Nicole}, {and} \bibinfo{person}{Morgan~Klaus
  Scheuerman}.} \bibinfo{year}{2020}\natexlab{}.
\newblock \showarticletitle{Bringing the people back in: Contesting benchmark
  machine learning datasets}.
\newblock \bibinfo{journal}{\emph{arXiv preprint arXiv:2007.07399}}
  (\bibinfo{year}{2020}).
\newblock


\bibitem[DeVos et~al\mbox{.}(2022)]%
        {devos2022toward}
\bibfield{author}{\bibinfo{person}{Alicia DeVos}, \bibinfo{person}{Aditi
  Dhabalia}, \bibinfo{person}{Hong Shen}, \bibinfo{person}{Kenneth Holstein},
  {and} \bibinfo{person}{Motahhare Eslami}.} \bibinfo{year}{2022}\natexlab{}.
\newblock \showarticletitle{Toward User-Driven Algorithm Auditing:
  Investigating users’ strategies for uncovering harmful algorithmic
  behavior}. In \bibinfo{booktitle}{\emph{Proceedings of the 2022 CHI
  Conference on Human Factors in Computing Systems}}. \bibinfo{pages}{1--19}.
\newblock


\bibitem[Dinan et~al\mbox{.}(2019)]%
        {dinan2019build}
\bibfield{author}{\bibinfo{person}{Emily Dinan}, \bibinfo{person}{Samuel
  Humeau}, \bibinfo{person}{Bharath Chintagunta}, {and} \bibinfo{person}{Jason
  Weston}.} \bibinfo{year}{2019}\natexlab{}.
\newblock \showarticletitle{Build it break it fix it for dialogue safety:
  Robustness from adversarial human attack}.
\newblock \bibinfo{journal}{\emph{arXiv preprint arXiv:1908.06083}}
  (\bibinfo{year}{2019}).
\newblock


\bibitem[Dutta et~al\mbox{.}(2023)]%
        {dutta2023modeling}
\bibfield{author}{\bibinfo{person}{Senjuti Dutta}, \bibinfo{person}{Sid
  Mittal}, \bibinfo{person}{Sherol Chen}, \bibinfo{person}{Deepak
  Ramachandran}, \bibinfo{person}{Ravi Rajakumar}, \bibinfo{person}{Ian
  Kivlichan}, \bibinfo{person}{Sunny Mak}, \bibinfo{person}{Alena Butryna},
  {and} \bibinfo{person}{Praveen Paritosh}.} \bibinfo{year}{2023}\natexlab{}.
\newblock \showarticletitle{Modeling subjectivity (by Mimicking Annotator
  Annotation) in toxic comment identification across diverse communities}.
\newblock \bibinfo{journal}{\emph{arXiv preprint arXiv:2311.00203}}
  (\bibinfo{year}{2023}).
\newblock


\bibitem[Eslami et~al\mbox{.}(2017)]%
        {eslami2017careful}
\bibfield{author}{\bibinfo{person}{Motahhare Eslami}, \bibinfo{person}{Kristen
  Vaccaro}, \bibinfo{person}{Karrie Karahalios}, {and} \bibinfo{person}{Kevin
  Hamilton}.} \bibinfo{year}{2017}\natexlab{}.
\newblock \showarticletitle{“Be careful; things can be worse than they
  appear”: Understanding Biased Algorithms and Users’ Behavior around Them
  in Rating Platforms}. In \bibinfo{booktitle}{\emph{Proceedings of the
  international AAAI conference on web and social media}},
  Vol.~\bibinfo{volume}{11}. \bibinfo{pages}{62--71}.
\newblock


\bibitem[Eslami et~al\mbox{.}(2019)]%
        {eslami2019user}
\bibfield{author}{\bibinfo{person}{Motahhare Eslami}, \bibinfo{person}{Kristen
  Vaccaro}, \bibinfo{person}{Min~Kyung Lee}, \bibinfo{person}{Amit Elazari
  Bar~On}, \bibinfo{person}{Eric Gilbert}, {and} \bibinfo{person}{Karrie
  Karahalios}.} \bibinfo{year}{2019}\natexlab{}.
\newblock \showarticletitle{User attitudes towards algorithmic opacity and
  transparency in online reviewing platforms}. In
  \bibinfo{booktitle}{\emph{Proceedings of the 2019 CHI Conference on Human
  Factors in Computing Systems}}. \bibinfo{pages}{1--14}.
\newblock


\bibitem[Fang et~al\mbox{.}(2023)]%
        {fang2023people}
\bibfield{author}{\bibinfo{person}{Jingchao Fang}, \bibinfo{person}{Jia-Wei
  Liang}, {and} \bibinfo{person}{Hao-Chuan Wang}.}
  \bibinfo{year}{2023}\natexlab{}.
\newblock \showarticletitle{How People Initiate and Respond to Discussions
  Around Online Community Norms: A Preliminary Analysis on Meta Stack Overflow
  Discussions}. In \bibinfo{booktitle}{\emph{Companion Publication of the 2023
  Conference on Computer Supported Cooperative Work and Social Computing}}.
  \bibinfo{pages}{221--225}.
\newblock


\bibitem[Feffer et~al\mbox{.}(2023)]%
        {feffer2023moral}
\bibfield{author}{\bibinfo{person}{Michael Feffer}, \bibinfo{person}{Hoda
  Heidari}, {and} \bibinfo{person}{Zachary~C Lipton}.}
  \bibinfo{year}{2023}\natexlab{}.
\newblock \showarticletitle{Moral Machine or Tyranny of the Majority?}
\newblock \bibinfo{journal}{\emph{arXiv preprint arXiv:2305.17319}}
  (\bibinfo{year}{2023}).
\newblock


\bibitem[Fiesler et~al\mbox{.}(2018)]%
        {fiesler2018reddit}
\bibfield{author}{\bibinfo{person}{Casey Fiesler}, \bibinfo{person}{Jialun
  Jiang}, \bibinfo{person}{Joshua McCann}, \bibinfo{person}{Kyle Frye}, {and}
  \bibinfo{person}{Jed Brubaker}.} \bibinfo{year}{2018}\natexlab{}.
\newblock \showarticletitle{Reddit rules! characterizing an ecosystem of
  governance}. In \bibinfo{booktitle}{\emph{Proceedings of the International
  AAAI Conference on Web and Social Media}}, Vol.~\bibinfo{volume}{12}.
\newblock


\bibitem[Flinn(2007)]%
        {flinn2007community}
\bibfield{author}{\bibinfo{person}{Andrew Flinn}.}
  \bibinfo{year}{2007}\natexlab{}.
\newblock \showarticletitle{Community histories, community archives: Some
  opportunities and challenges}.
\newblock \bibinfo{journal}{\emph{Journal of the Society of Archivists}}
  \bibinfo{volume}{28}, \bibinfo{number}{2} (\bibinfo{year}{2007}),
  \bibinfo{pages}{151--176}.
\newblock


\bibitem[Ford et~al\mbox{.}(2018)]%
        {ford2018we}
\bibfield{author}{\bibinfo{person}{Denae Ford}, \bibinfo{person}{Kristina
  Lustig}, \bibinfo{person}{Jeremy Banks}, {and} \bibinfo{person}{Chris
  Parnin}.} \bibinfo{year}{2018}\natexlab{}.
\newblock \showarticletitle{" We Don't Do That Here" How Collaborative Editing
  with Mentors Improves Engagement in Social Q\&A Communities}. In
  \bibinfo{booktitle}{\emph{Proceedings of the 2018 CHI conference on human
  factors in computing systems}}. \bibinfo{pages}{1--12}.
\newblock


\bibitem[Forte et~al\mbox{.}(2009)]%
        {forte2009decentralization}
\bibfield{author}{\bibinfo{person}{Andrea Forte}, \bibinfo{person}{Vanesa
  Larco}, {and} \bibinfo{person}{Amy Bruckman}.}
  \bibinfo{year}{2009}\natexlab{}.
\newblock \showarticletitle{Decentralization in Wikipedia governance}.
\newblock \bibinfo{journal}{\emph{Journal of Management Information Systems}}
  \bibinfo{volume}{26}, \bibinfo{number}{1} (\bibinfo{year}{2009}),
  \bibinfo{pages}{49--72}.
\newblock


\bibitem[Gardner et~al\mbox{.}(2020)]%
        {gardner2020evaluating}
\bibfield{author}{\bibinfo{person}{Matt Gardner}, \bibinfo{person}{Yoav Artzi},
  \bibinfo{person}{Victoria Basmova}, \bibinfo{person}{Jonathan Berant},
  \bibinfo{person}{Ben Bogin}, \bibinfo{person}{Sihao Chen},
  \bibinfo{person}{Pradeep Dasigi}, \bibinfo{person}{Dheeru Dua},
  \bibinfo{person}{Yanai Elazar}, \bibinfo{person}{Ananth Gottumukkala},
  {et~al\mbox{.}}} \bibinfo{year}{2020}\natexlab{}.
\newblock \showarticletitle{Evaluating models' local decision boundaries via
  contrast sets}.
\newblock \bibinfo{journal}{\emph{arXiv preprint arXiv:2004.02709}}
  (\bibinfo{year}{2020}).
\newblock


\bibitem[Gebru et~al\mbox{.}(2021)]%
        {gebru2021datasheets}
\bibfield{author}{\bibinfo{person}{Timnit Gebru}, \bibinfo{person}{Jamie
  Morgenstern}, \bibinfo{person}{Briana Vecchione},
  \bibinfo{person}{Jennifer~Wortman Vaughan}, \bibinfo{person}{Hanna Wallach},
  \bibinfo{person}{Hal~Daum{\'e} Iii}, {and} \bibinfo{person}{Kate Crawford}.}
  \bibinfo{year}{2021}\natexlab{}.
\newblock \showarticletitle{Datasheets for datasets}.
\newblock \bibinfo{journal}{\emph{Commun. ACM}} \bibinfo{volume}{64},
  \bibinfo{number}{12} (\bibinfo{year}{2021}), \bibinfo{pages}{86--92}.
\newblock


\bibitem[Geiger and Halfaker(2013)]%
        {geiger2013levee}
\bibfield{author}{\bibinfo{person}{R~Stuart Geiger} {and}
  \bibinfo{person}{Aaron Halfaker}.} \bibinfo{year}{2013}\natexlab{}.
\newblock \showarticletitle{When the levee breaks: without bots, what happens
  to Wikipedia's quality control processes?}. In
  \bibinfo{booktitle}{\emph{Proceedings of the 9th International Symposium on
  Open Collaboration}}. \bibinfo{pages}{1--6}.
\newblock


\bibitem[Geiger and Halfaker(2017)]%
        {geiger2017operationalizing}
\bibfield{author}{\bibinfo{person}{R~Stuart Geiger} {and}
  \bibinfo{person}{Aaron Halfaker}.} \bibinfo{year}{2017}\natexlab{}.
\newblock \showarticletitle{Operationalizing conflict and cooperation between
  automated software agents in wikipedia: A replication and expansion of'even
  good bots fight'}.
\newblock \bibinfo{journal}{\emph{Proceedings of the ACM on Human-Computer
  Interaction}} \bibinfo{volume}{1}, \bibinfo{number}{CSCW}
  (\bibinfo{year}{2017}), \bibinfo{pages}{1--33}.
\newblock


\bibitem[Geiger and Ribes(2010)]%
        {geiger2010work}
\bibfield{author}{\bibinfo{person}{R~Stuart Geiger} {and}
  \bibinfo{person}{David Ribes}.} \bibinfo{year}{2010}\natexlab{}.
\newblock \showarticletitle{The work of sustaining order in Wikipedia: The
  banning of a vandal}. In \bibinfo{booktitle}{\emph{Proceedings of the 2010
  ACM conference on Computer supported cooperative work}}.
  \bibinfo{pages}{117--126}.
\newblock


\bibitem[Geiger et~al\mbox{.}(2020)]%
        {geiger2020garbage}
\bibfield{author}{\bibinfo{person}{R~Stuart Geiger}, \bibinfo{person}{Kevin
  Yu}, \bibinfo{person}{Yanlai Yang}, \bibinfo{person}{Mindy Dai},
  \bibinfo{person}{Jie Qiu}, \bibinfo{person}{Rebekah Tang}, {and}
  \bibinfo{person}{Jenny Huang}.} \bibinfo{year}{2020}\natexlab{}.
\newblock \showarticletitle{Garbage in, garbage out? Do machine learning
  application papers in social computing report where human-labeled training
  data comes from?}. In \bibinfo{booktitle}{\emph{Proceedings of the 2020
  Conference on Fairness, Accountability, and Transparency}}.
  \bibinfo{pages}{325--336}.
\newblock


\bibitem[Gilbert(2013)]%
        {gilbert2013widespread}
\bibfield{author}{\bibinfo{person}{Eric Gilbert}.}
  \bibinfo{year}{2013}\natexlab{}.
\newblock \showarticletitle{Widespread underprovision on reddit}. In
  \bibinfo{booktitle}{\emph{Proceedings of the 2013 conference on Computer
  supported cooperative work}}. \bibinfo{pages}{803--808}.
\newblock


\bibitem[Gordon et~al\mbox{.}(2022)]%
        {gordon2022jury}
\bibfield{author}{\bibinfo{person}{Mitchell~L Gordon},
  \bibinfo{person}{Michelle~S Lam}, \bibinfo{person}{Joon~Sung Park},
  \bibinfo{person}{Kayur Patel}, \bibinfo{person}{Jeff Hancock},
  \bibinfo{person}{Tatsunori Hashimoto}, {and} \bibinfo{person}{Michael~S
  Bernstein}.} \bibinfo{year}{2022}\natexlab{}.
\newblock \showarticletitle{Jury learning: Integrating dissenting voices into
  machine learning models}. In \bibinfo{booktitle}{\emph{Proceedings of the
  2022 CHI Conference on Human Factors in Computing Systems}}.
  \bibinfo{pages}{1--19}.
\newblock


\bibitem[Gordon et~al\mbox{.}(2021)]%
        {gordon2021disagreement}
\bibfield{author}{\bibinfo{person}{Mitchell~L Gordon}, \bibinfo{person}{Kaitlyn
  Zhou}, \bibinfo{person}{Kayur Patel}, \bibinfo{person}{Tatsunori Hashimoto},
  {and} \bibinfo{person}{Michael~S Bernstein}.}
  \bibinfo{year}{2021}\natexlab{}.
\newblock \showarticletitle{The disagreement deconvolution: Bringing machine
  learning performance metrics in line with reality}. In
  \bibinfo{booktitle}{\emph{Proceedings of the 2021 CHI Conference on Human
  Factors in Computing Systems}}. \bibinfo{pages}{1--14}.
\newblock


\bibitem[Goyal et~al\mbox{.}(2022)]%
        {goyal2022your}
\bibfield{author}{\bibinfo{person}{Nitesh Goyal}, \bibinfo{person}{Ian~D
  Kivlichan}, \bibinfo{person}{Rachel Rosen}, {and} \bibinfo{person}{Lucy
  Vasserman}.} \bibinfo{year}{2022}\natexlab{}.
\newblock \showarticletitle{Is your toxicity my toxicity? Exploring the impact
  of rater identity on toxicity annotation}.
\newblock \bibinfo{journal}{\emph{Proceedings of the ACM on Human-Computer
  Interaction}} \bibinfo{volume}{6}, \bibinfo{number}{CSCW2}
  (\bibinfo{year}{2022}), \bibinfo{pages}{1--28}.
\newblock


\bibitem[Guerdan et~al\mbox{.}(2023)]%
        {guerdan2023ground}
\bibfield{author}{\bibinfo{person}{Luke Guerdan}, \bibinfo{person}{Amanda
  Coston}, \bibinfo{person}{Zhiwei~Steven Wu}, {and} \bibinfo{person}{Kenneth
  Holstein}.} \bibinfo{year}{2023}\natexlab{}.
\newblock \showarticletitle{Ground (less) Truth: A Causal Framework for Proxy
  Labels in Human-Algorithm Decision-Making}. In
  \bibinfo{booktitle}{\emph{Proceedings of the 2023 ACM Conference on Fairness,
  Accountability, and Transparency}}. \bibinfo{pages}{688--704}.
\newblock


\bibitem[Halfaker and Geiger(2020)]%
        {halfaker2020ores}
\bibfield{author}{\bibinfo{person}{Aaron Halfaker} {and}
  \bibinfo{person}{R~Stuart Geiger}.} \bibinfo{year}{2020}\natexlab{}.
\newblock \showarticletitle{Ores: Lowering barriers with participatory machine
  learning in wikipedia}.
\newblock \bibinfo{journal}{\emph{Proceedings of the ACM on Human-Computer
  Interaction}} \bibinfo{volume}{4}, \bibinfo{number}{CSCW2}
  (\bibinfo{year}{2020}), \bibinfo{pages}{1--37}.
\newblock


\bibitem[Halfaker et~al\mbox{.}(2013)]%
        {halfaker2013rise}
\bibfield{author}{\bibinfo{person}{Aaron Halfaker}, \bibinfo{person}{R~Stuart
  Geiger}, \bibinfo{person}{Jonathan~T Morgan}, {and} \bibinfo{person}{John
  Riedl}.} \bibinfo{year}{2013}\natexlab{}.
\newblock \showarticletitle{The rise and decline of an open collaboration
  system: How Wikipedia’s reaction to popularity is causing its decline}.
\newblock \bibinfo{journal}{\emph{American Behavioral Scientist}}
  \bibinfo{volume}{57}, \bibinfo{number}{5} (\bibinfo{year}{2013}),
  \bibinfo{pages}{664--688}.
\newblock


\bibitem[Halfaker et~al\mbox{.}(2011)]%
        {halfaker2011don}
\bibfield{author}{\bibinfo{person}{Aaron Halfaker}, \bibinfo{person}{Aniket
  Kittur}, {and} \bibinfo{person}{John Riedl}.}
  \bibinfo{year}{2011}\natexlab{}.
\newblock \showarticletitle{Don't bite the newbies: how reverts affect the
  quantity and quality of Wikipedia work}. In
  \bibinfo{booktitle}{\emph{Proceedings of the 7th international symposium on
  wikis and open collaboration}}. \bibinfo{pages}{163--172}.
\newblock


\bibitem[He et~al\mbox{.}(2023)]%
        {he2023cura}
\bibfield{author}{\bibinfo{person}{Wanrong He}, \bibinfo{person}{Mitchell~L
  Gordon}, \bibinfo{person}{Lindsay Popowski}, {and} \bibinfo{person}{Michael~S
  Bernstein}.} \bibinfo{year}{2023}\natexlab{}.
\newblock \showarticletitle{Cura: Curation at Social Media Scale}.
\newblock \bibinfo{journal}{\emph{Proceedings of the ACM on Human-Computer
  Interaction}} \bibinfo{volume}{7}, \bibinfo{number}{CSCW2}
  (\bibinfo{year}{2023}), \bibinfo{pages}{1--33}.
\newblock


\bibitem[Howard and Irani(2019)]%
        {howard2019ways}
\bibfield{author}{\bibinfo{person}{Dorothy Howard} {and} \bibinfo{person}{Lilly
  Irani}.} \bibinfo{year}{2019}\natexlab{}.
\newblock \showarticletitle{Ways of knowing when research subjects care}. In
  \bibinfo{booktitle}{\emph{Proceedings of the 2019 CHI Conference on Human
  Factors in Computing Systems}}. \bibinfo{pages}{1--16}.
\newblock


\bibitem[Hwang and Shaw(2022)]%
        {hwang2022rules}
\bibfield{author}{\bibinfo{person}{Sohyeon Hwang} {and} \bibinfo{person}{Aaron
  Shaw}.} \bibinfo{year}{2022}\natexlab{}.
\newblock \showarticletitle{Rules and Rule-Making in the Five Largest
  Wikipedias}. In \bibinfo{booktitle}{\emph{Proceedings of the International
  AAAI Conference on Web and Social Media}}, Vol.~\bibinfo{volume}{16}.
  \bibinfo{pages}{347--357}.
\newblock


\bibitem[Irani and Silberman(2013)]%
        {irani2013turkopticon}
\bibfield{author}{\bibinfo{person}{Lilly~C Irani} {and} \bibinfo{person}{M~Six
  Silberman}.} \bibinfo{year}{2013}\natexlab{}.
\newblock \showarticletitle{Turkopticon: Interrupting worker invisibility in
  amazon mechanical turk}. In \bibinfo{booktitle}{\emph{Proceedings of the
  SIGCHI conference on human factors in computing systems}}.
  \bibinfo{pages}{611--620}.
\newblock


\bibitem[Jain et~al\mbox{.}(2020)]%
        {jain2020overview}
\bibfield{author}{\bibinfo{person}{Abhinav Jain}, \bibinfo{person}{Hima Patel},
  \bibinfo{person}{Lokesh Nagalapatti}, \bibinfo{person}{Nitin Gupta},
  \bibinfo{person}{Sameep Mehta}, \bibinfo{person}{Shanmukha Guttula},
  \bibinfo{person}{Shashank Mujumdar}, \bibinfo{person}{Shazia Afzal},
  \bibinfo{person}{Ruhi Sharma~Mittal}, {and} \bibinfo{person}{Vitobha
  Munigala}.} \bibinfo{year}{2020}\natexlab{}.
\newblock \showarticletitle{Overview and importance of data quality for machine
  learning tasks}. In \bibinfo{booktitle}{\emph{Proceedings of the 26th ACM
  SIGKDD international conference on knowledge discovery \& data mining}}.
  \bibinfo{pages}{3561--3562}.
\newblock


\bibitem[Jhaver et~al\mbox{.}(2019a)]%
        {jhaver2019did}
\bibfield{author}{\bibinfo{person}{Shagun Jhaver},
  \bibinfo{person}{Darren~Scott Appling}, \bibinfo{person}{Eric Gilbert}, {and}
  \bibinfo{person}{Amy Bruckman}.} \bibinfo{year}{2019}\natexlab{a}.
\newblock \showarticletitle{" Did you suspect the post would be removed?"
  Understanding user reactions to content removals on Reddit}.
\newblock \bibinfo{journal}{\emph{Proceedings of the ACM on human-computer
  interaction}} \bibinfo{volume}{3}, \bibinfo{number}{CSCW}
  (\bibinfo{year}{2019}), \bibinfo{pages}{1--33}.
\newblock


\bibitem[Jhaver et~al\mbox{.}(2019b)]%
        {jhaver2019human}
\bibfield{author}{\bibinfo{person}{Shagun Jhaver}, \bibinfo{person}{Iris
  Birman}, \bibinfo{person}{Eric Gilbert}, {and} \bibinfo{person}{Amy
  Bruckman}.} \bibinfo{year}{2019}\natexlab{b}.
\newblock \showarticletitle{Human-machine collaboration for content regulation:
  The case of reddit automoderator}.
\newblock \bibinfo{journal}{\emph{ACM Transactions on Computer-Human
  Interaction (TOCHI)}} \bibinfo{volume}{26}, \bibinfo{number}{5}
  (\bibinfo{year}{2019}), \bibinfo{pages}{1--35}.
\newblock


\bibitem[Jhaver et~al\mbox{.}(2023)]%
        {jhaver2023decentralizing}
\bibfield{author}{\bibinfo{person}{Shagun Jhaver}, \bibinfo{person}{Seth Frey},
  {and} \bibinfo{person}{Amy~X Zhang}.} \bibinfo{year}{2023}\natexlab{}.
\newblock \showarticletitle{Decentralizing Platform Power: A Design Space of
  Multi-level Governance in Online Social Platforms}.
\newblock \bibinfo{journal}{\emph{Social Media+ Society}} \bibinfo{volume}{9},
  \bibinfo{number}{4} (\bibinfo{year}{2023}),
  \bibinfo{pages}{20563051231207857}.
\newblock


\bibitem[Jia et~al\mbox{.}(2023)]%
        {jia2023embedding}
\bibfield{author}{\bibinfo{person}{Chenyan Jia}, \bibinfo{person}{Michelle~S
  Lam}, \bibinfo{person}{Minh~Chau Mai}, \bibinfo{person}{Jeff Hancock}, {and}
  \bibinfo{person}{Michael~S Bernstein}.} \bibinfo{year}{2023}\natexlab{}.
\newblock \showarticletitle{Embedding democratic values into social media AIs
  via societal objective functions}.
\newblock \bibinfo{journal}{\emph{arXiv preprint arXiv:2307.13912}}
  (\bibinfo{year}{2023}).
\newblock


\bibitem[Jo and Gebru(2020)]%
        {jo2020lessons}
\bibfield{author}{\bibinfo{person}{Eun~Seo Jo} {and} \bibinfo{person}{Timnit
  Gebru}.} \bibinfo{year}{2020}\natexlab{}.
\newblock \showarticletitle{Lessons from archives: Strategies for collecting
  sociocultural data in machine learning}. In
  \bibinfo{booktitle}{\emph{Proceedings of the 2020 conference on fairness,
  accountability, and transparency}}. \bibinfo{pages}{306--316}.
\newblock


\bibitem[Kader and Perry(2007)]%
        {kader2007variability}
\bibfield{author}{\bibinfo{person}{Gary~D Kader} {and} \bibinfo{person}{Mike
  Perry}.} \bibinfo{year}{2007}\natexlab{}.
\newblock \showarticletitle{Variability for categorical variables}.
\newblock \bibinfo{journal}{\emph{Journal of statistics education}}
  \bibinfo{volume}{15}, \bibinfo{number}{2} (\bibinfo{year}{2007}).
\newblock


\bibitem[Kapania et~al\mbox{.}(2023)]%
        {kapania2023hunt}
\bibfield{author}{\bibinfo{person}{Shivani Kapania}, \bibinfo{person}{Alex~S
  Taylor}, {and} \bibinfo{person}{Ding Wang}.} \bibinfo{year}{2023}\natexlab{}.
\newblock \showarticletitle{A hunt for the Snark: Annotator Diversity in Data
  Practices}. In \bibinfo{booktitle}{\emph{Proceedings of the 2023 CHI
  Conference on Human Factors in Computing Systems}}. \bibinfo{pages}{1--15}.
\newblock


\bibitem[Kawakami et~al\mbox{.}(2022)]%
        {kawakami2022care}
\bibfield{author}{\bibinfo{person}{Anna Kawakami}, \bibinfo{person}{Venkatesh
  Sivaraman}, \bibinfo{person}{Logan Stapleton}, \bibinfo{person}{Hao-Fei
  Cheng}, \bibinfo{person}{Adam Perer}, \bibinfo{person}{Zhiwei~Steven Wu},
  \bibinfo{person}{Haiyi Zhu}, {and} \bibinfo{person}{Kenneth Holstein}.}
  \bibinfo{year}{2022}\natexlab{}.
\newblock \showarticletitle{“Why Do I Care What’s Similar?” Probing
  Challenges in AI-Assisted Child Welfare Decision-Making through Worker-AI
  Interface Design Concepts}. In \bibinfo{booktitle}{\emph{Designing
  Interactive Systems Conference}}. \bibinfo{pages}{454--470}.
\newblock


\bibitem[Kiela et~al\mbox{.}(2021)]%
        {kiela2021dynabench}
\bibfield{author}{\bibinfo{person}{Douwe Kiela}, \bibinfo{person}{Max Bartolo},
  \bibinfo{person}{Yixin Nie}, \bibinfo{person}{Divyansh Kaushik},
  \bibinfo{person}{Atticus Geiger}, \bibinfo{person}{Zhengxuan Wu},
  \bibinfo{person}{Bertie Vidgen}, \bibinfo{person}{Grusha Prasad},
  \bibinfo{person}{Amanpreet Singh}, \bibinfo{person}{Pratik Ringshia},
  \bibinfo{person}{Zhiyi Ma}, \bibinfo{person}{Tristan Thrush},
  \bibinfo{person}{Sebastian Riedel}, \bibinfo{person}{Zeerak Waseem},
  \bibinfo{person}{Pontus Stenetorp}, \bibinfo{person}{Robin Jia},
  \bibinfo{person}{Mohit Bansal}, \bibinfo{person}{Christopher Potts}, {and}
  \bibinfo{person}{Adina Williams}.} \bibinfo{year}{2021}\natexlab{}.
\newblock \showarticletitle{Dynabench: Rethinking Benchmarking in {NLP}}. In
  \bibinfo{booktitle}{\emph{Proceedings of the 2021 Conference of the North
  American Chapter of the Association for Computational Linguistics: Human
  Language Technologies}}. \bibinfo{publisher}{Association for Computational
  Linguistics}, \bibinfo{address}{Online}, \bibinfo{pages}{4110--4124}.
\newblock
\urldef\tempurl%
\url{https://doi.org/10.18653/v1/2021.naacl-main.324}
\showDOI{\tempurl}


\bibitem[Kittur et~al\mbox{.}(2007)]%
        {kittur2007he}
\bibfield{author}{\bibinfo{person}{Aniket Kittur}, \bibinfo{person}{Bongwon
  Suh}, \bibinfo{person}{Bryan~A Pendleton}, {and} \bibinfo{person}{Ed~H Chi}.}
  \bibinfo{year}{2007}\natexlab{}.
\newblock \showarticletitle{He says, she says: conflict and coordination in
  Wikipedia}. In \bibinfo{booktitle}{\emph{Proceedings of the SIGCHI conference
  on Human factors in computing systems}}. \bibinfo{pages}{453--462}.
\newblock


\bibitem[Kuo et~al\mbox{.}(2023)]%
        {kuo2023understanding}
\bibfield{author}{\bibinfo{person}{Tzu-Sheng Kuo}, \bibinfo{person}{Hong Shen},
  \bibinfo{person}{Jisoo Geum}, \bibinfo{person}{Nev Jones},
  \bibinfo{person}{Jason~I Hong}, \bibinfo{person}{Haiyi Zhu}, {and}
  \bibinfo{person}{Kenneth Holstein}.} \bibinfo{year}{2023}\natexlab{}.
\newblock \showarticletitle{Understanding Frontline Workers’ and Unhoused
  Individuals’ Perspectives on AI Used in Homeless Services}. In
  \bibinfo{booktitle}{\emph{Proceedings of the 2023 CHI Conference on Human
  Factors in Computing Systems}}. \bibinfo{pages}{1--17}.
\newblock


\bibitem[Lam et~al\mbox{.}(2022)]%
        {lam2022end}
\bibfield{author}{\bibinfo{person}{Michelle~S Lam}, \bibinfo{person}{Mitchell~L
  Gordon}, \bibinfo{person}{Dana{\"e} Metaxa}, \bibinfo{person}{Jeffrey~T
  Hancock}, \bibinfo{person}{James~A Landay}, {and} \bibinfo{person}{Michael~S
  Bernstein}.} \bibinfo{year}{2022}\natexlab{}.
\newblock \showarticletitle{End-User Audits: A System Empowering Communities to
  Lead Large-Scale Investigations of Harmful Algorithmic Behavior}.
\newblock \bibinfo{journal}{\emph{Proceedings of the ACM on Human-Computer
  Interaction}} \bibinfo{volume}{6}, \bibinfo{number}{CSCW2}
  (\bibinfo{year}{2022}), \bibinfo{pages}{1--34}.
\newblock


\bibitem[Lam et~al\mbox{.}(2023)]%
        {lam2023sociotechnical}
\bibfield{author}{\bibinfo{person}{Michelle~S Lam}, \bibinfo{person}{Ayush
  Pandit}, \bibinfo{person}{Colin~H Kalicki}, \bibinfo{person}{Rachit Gupta},
  \bibinfo{person}{Poonam Sahoo}, {and} \bibinfo{person}{Dana{\"e} Metaxa}.}
  \bibinfo{year}{2023}\natexlab{}.
\newblock \showarticletitle{Sociotechnical Audits: Broadening the Algorithm
  Auditing Lens to Investigate Targeted Advertising}.
\newblock \bibinfo{journal}{\emph{arXiv preprint arXiv:2308.15768}}
  (\bibinfo{year}{2023}).
\newblock


\bibitem[Lee et~al\mbox{.}(2019)]%
        {lee2019procedural}
\bibfield{author}{\bibinfo{person}{Min~Kyung Lee}, \bibinfo{person}{Anuraag
  Jain}, \bibinfo{person}{Hea~Jin Cha}, \bibinfo{person}{Shashank Ojha}, {and}
  \bibinfo{person}{Daniel Kusbit}.} \bibinfo{year}{2019}\natexlab{}.
\newblock \showarticletitle{Procedural justice in algorithmic fairness:
  Leveraging transparency and outcome control for fair algorithmic mediation}.
\newblock \bibinfo{journal}{\emph{Proceedings of the ACM on Human-Computer
  Interaction}} \bibinfo{volume}{3}, \bibinfo{number}{CSCW}
  (\bibinfo{year}{2019}), \bibinfo{pages}{1--26}.
\newblock


\bibitem[Leventhal(1980)]%
        {leventhal1980should}
\bibfield{author}{\bibinfo{person}{Gerald~S Leventhal}.}
  \bibinfo{year}{1980}\natexlab{}.
\newblock \showarticletitle{What should be done with equity theory? New
  approaches to the study of fairness in social relationships}.
\newblock In \bibinfo{booktitle}{\emph{Social exchange: Advances in theory and
  research}}. \bibinfo{publisher}{Springer}, \bibinfo{pages}{27--55}.
\newblock


\bibitem[Mamykina et~al\mbox{.}(2011)]%
        {mamykina2011design}
\bibfield{author}{\bibinfo{person}{Lena Mamykina}, \bibinfo{person}{Bella
  Manoim}, \bibinfo{person}{Manas Mittal}, \bibinfo{person}{George Hripcsak},
  {and} \bibinfo{person}{Bj{\"o}rn Hartmann}.} \bibinfo{year}{2011}\natexlab{}.
\newblock \showarticletitle{Design lessons from the fastest q\&a site in the
  west}. In \bibinfo{booktitle}{\emph{Proceedings of the SIGCHI conference on
  Human factors in computing systems}}. \bibinfo{pages}{2857--2866}.
\newblock


\bibitem[Massanari(2015)]%
        {massanari2015participatory}
\bibfield{author}{\bibinfo{person}{Adrienne~Lynne Massanari}.}
  \bibinfo{year}{2015}\natexlab{}.
\newblock \showarticletitle{Participatory culture, community, and play}.
\newblock  (\bibinfo{year}{2015}).
\newblock


\bibitem[Mazumder et~al\mbox{.}(2022)]%
        {mazumder2022dataperf}
\bibfield{author}{\bibinfo{person}{Mark Mazumder}, \bibinfo{person}{Colby
  Banbury}, \bibinfo{person}{Xiaozhe Yao}, \bibinfo{person}{Bojan
  Karla{\v{s}}}, \bibinfo{person}{William~Gaviria Rojas},
  \bibinfo{person}{Sudnya Diamos}, \bibinfo{person}{Greg Diamos},
  \bibinfo{person}{Lynn He}, \bibinfo{person}{Alicia Parrish},
  \bibinfo{person}{Hannah~Rose Kirk}, {et~al\mbox{.}}}
  \bibinfo{year}{2022}\natexlab{}.
\newblock \showarticletitle{Dataperf: Benchmarks for data-centric ai
  development}.
\newblock \bibinfo{journal}{\emph{arXiv preprint arXiv:2207.10062}}
  (\bibinfo{year}{2022}).
\newblock


\bibitem[Metaxa et~al\mbox{.}(2021)]%
        {metaxa2021auditing}
\bibfield{author}{\bibinfo{person}{Dana{\"e} Metaxa},
  \bibinfo{person}{Joon~Sung Park}, \bibinfo{person}{Ronald~E Robertson},
  \bibinfo{person}{Karrie Karahalios}, \bibinfo{person}{Christo Wilson},
  \bibinfo{person}{Jeff Hancock}, \bibinfo{person}{Christian Sandvig},
  {et~al\mbox{.}}} \bibinfo{year}{2021}\natexlab{}.
\newblock \showarticletitle{Auditing algorithms: Understanding algorithmic
  systems from the outside in}.
\newblock \bibinfo{journal}{\emph{Foundations and Trends{\textregistered} in
  Human--Computer Interaction}} \bibinfo{volume}{14}, \bibinfo{number}{4}
  (\bibinfo{year}{2021}), \bibinfo{pages}{272--344}.
\newblock


\bibitem[Muller et~al\mbox{.}(2019)]%
        {muller2019data}
\bibfield{author}{\bibinfo{person}{Michael Muller}, \bibinfo{person}{Ingrid
  Lange}, \bibinfo{person}{Dakuo Wang}, \bibinfo{person}{David Piorkowski},
  \bibinfo{person}{Jason Tsay}, \bibinfo{person}{Q~Vera Liao},
  \bibinfo{person}{Casey Dugan}, {and} \bibinfo{person}{Thomas Erickson}.}
  \bibinfo{year}{2019}\natexlab{}.
\newblock \showarticletitle{How data science workers work with data}. In
  \bibinfo{booktitle}{\emph{Conference on Human Factors in Computing
  Systems-Proceedings}}. \bibinfo{pages}{86--94}.
\newblock


\bibitem[Muller et~al\mbox{.}(2021)]%
        {muller2021designing}
\bibfield{author}{\bibinfo{person}{Michael Muller},
  \bibinfo{person}{Christine~T Wolf}, \bibinfo{person}{Josh Andres},
  \bibinfo{person}{Michael Desmond}, \bibinfo{person}{Narendra~Nath Joshi},
  \bibinfo{person}{Zahra Ashktorab}, \bibinfo{person}{Aabhas Sharma},
  \bibinfo{person}{Kristina Brimijoin}, \bibinfo{person}{Qian Pan},
  \bibinfo{person}{Evelyn Duesterwald}, {et~al\mbox{.}}}
  \bibinfo{year}{2021}\natexlab{}.
\newblock \showarticletitle{Designing ground truth and the social life of
  labels}. In \bibinfo{booktitle}{\emph{Proceedings of the 2021 CHI conference
  on human factors in computing systems}}. \bibinfo{pages}{1--16}.
\newblock


\bibitem[M{\"u}ller-Birn et~al\mbox{.}(2013)]%
        {muller2013work}
\bibfield{author}{\bibinfo{person}{Claudia M{\"u}ller-Birn},
  \bibinfo{person}{Leonhard Dobusch}, {and} \bibinfo{person}{James~D
  Herbsleb}.} \bibinfo{year}{2013}\natexlab{}.
\newblock \showarticletitle{Work-to-rule: the emergence of algorithmic
  governance in Wikipedia}. In \bibinfo{booktitle}{\emph{Proceedings of the 6th
  International Conference on Communities and Technologies}}.
  \bibinfo{pages}{80--89}.
\newblock


\bibitem[Nie et~al\mbox{.}(2019)]%
        {nie2019adversarial}
\bibfield{author}{\bibinfo{person}{Yixin Nie}, \bibinfo{person}{Adina
  Williams}, \bibinfo{person}{Emily Dinan}, \bibinfo{person}{Mohit Bansal},
  \bibinfo{person}{Jason Weston}, {and} \bibinfo{person}{Douwe Kiela}.}
  \bibinfo{year}{2019}\natexlab{}.
\newblock \showarticletitle{Adversarial NLI: A new benchmark for natural
  language understanding}.
\newblock \bibinfo{journal}{\emph{arXiv preprint arXiv:1910.14599}}
  (\bibinfo{year}{2019}).
\newblock


\bibitem[Noble(2018)]%
        {noble2018algorithms}
\bibfield{author}{\bibinfo{person}{Safiya~Umoja Noble}.}
  \bibinfo{year}{2018}\natexlab{}.
\newblock \showarticletitle{Algorithms of oppression}.
\newblock In \bibinfo{booktitle}{\emph{Algorithms of oppression}}.
  \bibinfo{publisher}{New York university press}.
\newblock


\bibitem[Oala et~al\mbox{.}(2023)]%
        {oala2023dmlr}
\bibfield{author}{\bibinfo{person}{Luis Oala}, \bibinfo{person}{Manil Maskey},
  \bibinfo{person}{Lilith Bat-Leah}, \bibinfo{person}{Alicia Parrish},
  \bibinfo{person}{Nezihe~Merve G{\"u}rel}, \bibinfo{person}{Tzu-Sheng Kuo},
  \bibinfo{person}{Yang Liu}, \bibinfo{person}{Rotem Dror},
  \bibinfo{person}{Danilo Brajovic}, \bibinfo{person}{Xiaozhe Yao},
  {et~al\mbox{.}}} \bibinfo{year}{2023}\natexlab{}.
\newblock \showarticletitle{DMLR: Data-centric Machine Learning Research--Past,
  Present and Future}.
\newblock \bibinfo{journal}{\emph{arXiv preprint arXiv:2311.13028}}
  (\bibinfo{year}{2023}).
\newblock


\bibitem[Papakyriakopoulos et~al\mbox{.}(2023)]%
        {papakyriakopoulos2023upvotes}
\bibfield{author}{\bibinfo{person}{Orestis Papakyriakopoulos},
  \bibinfo{person}{Severin Engelmann}, {and} \bibinfo{person}{Amy Winecoff}.}
  \bibinfo{year}{2023}\natexlab{}.
\newblock \showarticletitle{Upvotes? Downvotes? No Votes? Understanding the
  relationship between reaction mechanisms and political discourse on Reddit}.
  In \bibinfo{booktitle}{\emph{Proceedings of the 2023 CHI Conference on Human
  Factors in Computing Systems}}. \bibinfo{pages}{1--28}.
\newblock


\bibitem[Peng et~al\mbox{.}(2021)]%
        {peng2021mitigating}
\bibfield{author}{\bibinfo{person}{Kenny Peng}, \bibinfo{person}{Arunesh
  Mathur}, {and} \bibinfo{person}{Arvind Narayanan}.}
  \bibinfo{year}{2021}\natexlab{}.
\newblock \showarticletitle{Mitigating dataset harms requires stewardship:
  Lessons from 1000 papers}.
\newblock \bibinfo{journal}{\emph{arXiv preprint arXiv:2108.02922}}
  (\bibinfo{year}{2021}).
\newblock


\bibitem[Raji et~al\mbox{.}(2021)]%
        {raji2021ai}
\bibfield{author}{\bibinfo{person}{Inioluwa~Deborah Raji},
  \bibinfo{person}{Emily~M Bender}, \bibinfo{person}{Amandalynne Paullada},
  \bibinfo{person}{Emily Denton}, {and} \bibinfo{person}{Alex Hanna}.}
  \bibinfo{year}{2021}\natexlab{}.
\newblock \showarticletitle{AI and the everything in the whole wide world
  benchmark}.
\newblock \bibinfo{journal}{\emph{arXiv preprint arXiv:2111.15366}}
  (\bibinfo{year}{2021}).
\newblock


\bibitem[Sambasivan et~al\mbox{.}(2021)]%
        {sambasivan2021everyone}
\bibfield{author}{\bibinfo{person}{Nithya Sambasivan}, \bibinfo{person}{Shivani
  Kapania}, \bibinfo{person}{Hannah Highfill}, \bibinfo{person}{Diana Akrong},
  \bibinfo{person}{Praveen Paritosh}, {and} \bibinfo{person}{Lora~M Aroyo}.}
  \bibinfo{year}{2021}\natexlab{}.
\newblock \showarticletitle{“Everyone wants to do the model work, not the
  data work”: Data Cascades in High-Stakes AI}. In
  \bibinfo{booktitle}{\emph{proceedings of the 2021 CHI Conference on Human
  Factors in Computing Systems}}. \bibinfo{pages}{1--15}.
\newblock


\bibitem[Sap et~al\mbox{.}(2019)]%
        {sap2019risk}
\bibfield{author}{\bibinfo{person}{Maarten Sap}, \bibinfo{person}{Dallas Card},
  \bibinfo{person}{Saadia Gabriel}, \bibinfo{person}{Yejin Choi}, {and}
  \bibinfo{person}{Noah~A Smith}.} \bibinfo{year}{2019}\natexlab{}.
\newblock \showarticletitle{The risk of racial bias in hate speech detection}.
  In \bibinfo{booktitle}{\emph{Proceedings of the 57th annual meeting of the
  association for computational linguistics}}. \bibinfo{pages}{1668--1678}.
\newblock


\bibitem[Sap et~al\mbox{.}(2021)]%
        {sap2021annotators}
\bibfield{author}{\bibinfo{person}{Maarten Sap}, \bibinfo{person}{Swabha
  Swayamdipta}, \bibinfo{person}{Laura Vianna}, \bibinfo{person}{Xuhui Zhou},
  \bibinfo{person}{Yejin Choi}, {and} \bibinfo{person}{Noah~A Smith}.}
  \bibinfo{year}{2021}\natexlab{}.
\newblock \showarticletitle{Annotators with attitudes: How annotator beliefs
  and identities bias toxic language detection}.
\newblock \bibinfo{journal}{\emph{arXiv preprint arXiv:2111.07997}}
  (\bibinfo{year}{2021}).
\newblock


\bibitem[Shen et~al\mbox{.}(2021)]%
        {shen2021everyday}
\bibfield{author}{\bibinfo{person}{Hong Shen}, \bibinfo{person}{Alicia DeVos},
  \bibinfo{person}{Motahhare Eslami}, {and} \bibinfo{person}{Kenneth
  Holstein}.} \bibinfo{year}{2021}\natexlab{}.
\newblock \showarticletitle{Everyday algorithm auditing: Understanding the
  power of everyday users in surfacing harmful algorithmic behaviors}.
\newblock \bibinfo{journal}{\emph{Proceedings of the ACM on Human-Computer
  Interaction}} \bibinfo{volume}{5}, \bibinfo{number}{CSCW2}
  (\bibinfo{year}{2021}), \bibinfo{pages}{1--29}.
\newblock


\bibitem[Showkat et~al\mbox{.}(2023)]%
        {showkat2023right}
\bibfield{author}{\bibinfo{person}{Dilruba Showkat}, \bibinfo{person}{Angela~DR
  Smith}, \bibinfo{person}{Wang Lingqing}, {and} \bibinfo{person}{Alexandra
  To}.} \bibinfo{year}{2023}\natexlab{}.
\newblock \showarticletitle{“Who is the right homeless client?”: Values in
  Algorithmic Homelessness Service Provision and Machine Learning Research}. In
  \bibinfo{booktitle}{\emph{Proceedings of the 2023 CHI Conference on Human
  Factors in Computing Systems}}. \bibinfo{pages}{1--21}.
\newblock


\bibitem[Siddarth et~al\mbox{.}(2021)]%
        {siddarth2021ai}
\bibfield{author}{\bibinfo{person}{Divya Siddarth}, \bibinfo{person}{Daron
  Acemoglu}, \bibinfo{person}{Danielle Allen}, \bibinfo{person}{Kate Crawford},
  \bibinfo{person}{James Evans}, \bibinfo{person}{Michael Jordan}, {and}
  \bibinfo{person}{E Weyl}.} \bibinfo{year}{2021}\natexlab{}.
\newblock \showarticletitle{How AI fails us}.
\newblock \bibinfo{journal}{\emph{arXiv preprint arXiv:2201.04200}}
  (\bibinfo{year}{2021}).
\newblock


\bibitem[Smith et~al\mbox{.}(2020)]%
        {smith2020keeping}
\bibfield{author}{\bibinfo{person}{C~Estelle Smith}, \bibinfo{person}{Bowen
  Yu}, \bibinfo{person}{Anjali Srivastava}, \bibinfo{person}{Aaron Halfaker},
  \bibinfo{person}{Loren Terveen}, {and} \bibinfo{person}{Haiyi Zhu}.}
  \bibinfo{year}{2020}\natexlab{}.
\newblock \showarticletitle{Keeping community in the loop: Understanding
  wikipedia stakeholder values for machine learning-based systems}. In
  \bibinfo{booktitle}{\emph{Proceedings of the 2020 CHI Conference on Human
  Factors in Computing Systems}}. \bibinfo{pages}{1--14}.
\newblock


\bibitem[Sorensen et~al\mbox{.}(2024)]%
        {sorensen2024roadmap}
\bibfield{author}{\bibinfo{person}{Taylor Sorensen}, \bibinfo{person}{Jared
  Moore}, \bibinfo{person}{Jillian Fisher}, \bibinfo{person}{Mitchell Gordon},
  \bibinfo{person}{Niloofar Mireshghallah},
  \bibinfo{person}{Christopher~Michael Rytting}, \bibinfo{person}{Andre Ye},
  \bibinfo{person}{Liwei Jiang}, \bibinfo{person}{Ximing Lu},
  \bibinfo{person}{Nouha Dziri}, {et~al\mbox{.}}}
  \bibinfo{year}{2024}\natexlab{}.
\newblock \showarticletitle{A Roadmap to Pluralistic Alignment}.
\newblock \bibinfo{journal}{\emph{arXiv preprint arXiv:2402.05070}}
  (\bibinfo{year}{2024}).
\newblock


\bibitem[Thibaut and Walker(1975)]%
        {thibaut1975procedural}
\bibfield{author}{\bibinfo{person}{John~W Thibaut} {and}
  \bibinfo{person}{Laurens Walker}.} \bibinfo{year}{1975}\natexlab{}.
\newblock \showarticletitle{Procedural justice: A psychological analysis}.
\newblock \bibinfo{journal}{\emph{(No Title)}} (\bibinfo{year}{1975}).
\newblock


\bibitem[Van~der Maaten and Hinton(2008)]%
        {van2008visualizing}
\bibfield{author}{\bibinfo{person}{Laurens Van~der Maaten} {and}
  \bibinfo{person}{Geoffrey Hinton}.} \bibinfo{year}{2008}\natexlab{}.
\newblock \showarticletitle{Visualizing data using t-SNE.}
\newblock \bibinfo{journal}{\emph{Journal of machine learning research}}
  \bibinfo{volume}{9}, \bibinfo{number}{11} (\bibinfo{year}{2008}).
\newblock


\bibitem[Wallace et~al\mbox{.}(2019)]%
        {wallace2019trick}
\bibfield{author}{\bibinfo{person}{Eric Wallace}, \bibinfo{person}{Pedro
  Rodriguez}, \bibinfo{person}{Shi Feng}, \bibinfo{person}{Ikuya Yamada}, {and}
  \bibinfo{person}{Jordan Boyd-Graber}.} \bibinfo{year}{2019}\natexlab{}.
\newblock \showarticletitle{Trick me if you can: Human-in-the-loop generation
  of adversarial examples for question answering}.
\newblock \bibinfo{journal}{\emph{Transactions of the Association for
  Computational Linguistics}}  \bibinfo{volume}{7} (\bibinfo{year}{2019}),
  \bibinfo{pages}{387--401}.
\newblock


\bibitem[Wallace et~al\mbox{.}(2022)]%
        {wallace2022debiased}
\bibfield{author}{\bibinfo{person}{Shaun Wallace}, \bibinfo{person}{Tianyuan
  Cai}, \bibinfo{person}{Brendan Le}, {and} \bibinfo{person}{Luis~A Leiva}.}
  \bibinfo{year}{2022}\natexlab{}.
\newblock \showarticletitle{Debiased label aggregation for subjective
  crowdsourcing tasks}. In \bibinfo{booktitle}{\emph{CHI Conference on Human
  Factors in Computing Systems Extended Abstracts}}. \bibinfo{pages}{1--8}.
\newblock


\bibitem[Ye et~al\mbox{.}(2021)]%
        {ye2021wikipedia}
\bibfield{author}{\bibinfo{person}{Zining Ye}, \bibinfo{person}{Xinran Yuan},
  \bibinfo{person}{Shaurya Gaur}, \bibinfo{person}{Aaron Halfaker},
  \bibinfo{person}{Jodi Forlizzi}, {and} \bibinfo{person}{Haiyi Zhu}.}
  \bibinfo{year}{2021}\natexlab{}.
\newblock \showarticletitle{Wikipedia ORES explorer: Visualizing trade-offs for
  designing applications with machine learning API}. In
  \bibinfo{booktitle}{\emph{Designing Interactive Systems Conference 2021}}.
  \bibinfo{pages}{1554--1565}.
\newblock


\bibitem[Yee et~al\mbox{.}(2021)]%
        {yee2021image}
\bibfield{author}{\bibinfo{person}{Kyra Yee}, \bibinfo{person}{Uthaipon
  Tantipongpipat}, {and} \bibinfo{person}{Shubhanshu Mishra}.}
  \bibinfo{year}{2021}\natexlab{}.
\newblock \showarticletitle{Image cropping on twitter: Fairness metrics, their
  limitations, and the importance of representation, design, and agency}.
\newblock \bibinfo{journal}{\emph{Proceedings of the ACM on Human-Computer
  Interaction}} \bibinfo{volume}{5}, \bibinfo{number}{CSCW2}
  (\bibinfo{year}{2021}), \bibinfo{pages}{1--24}.
\newblock


\bibitem[Zhang et~al\mbox{.}(2023)]%
        {zhang2023deliberating}
\bibfield{author}{\bibinfo{person}{Angie Zhang}, \bibinfo{person}{Olympia
  Walker}, \bibinfo{person}{Kaci Nguyen}, \bibinfo{person}{Jiajun Dai},
  \bibinfo{person}{Anqing Chen}, {and} \bibinfo{person}{Min~Kyung Lee}.}
  \bibinfo{year}{2023}\natexlab{}.
\newblock \showarticletitle{Deliberating with AI: Improving Decision-Making for
  the Future through Participatory AI Design and Stakeholder Deliberation}.
\newblock \bibinfo{journal}{\emph{Proceedings of the ACM on Human-Computer
  Interaction}} \bibinfo{volume}{7}, \bibinfo{number}{CSCW1}
  (\bibinfo{year}{2023}), \bibinfo{pages}{1--32}.
\newblock


\bibitem[Zhang and Cranshaw(2018)]%
        {zhang2018making}
\bibfield{author}{\bibinfo{person}{Amy~X Zhang} {and} \bibinfo{person}{Justin
  Cranshaw}.} \bibinfo{year}{2018}\natexlab{}.
\newblock \showarticletitle{Making sense of group chat through collaborative
  tagging and summarization}.
\newblock \bibinfo{journal}{\emph{Proceedings of the ACM on Human-Computer
  Interaction}} \bibinfo{volume}{2}, \bibinfo{number}{CSCW}
  (\bibinfo{year}{2018}), \bibinfo{pages}{1--27}.
\newblock


\end{thebibliography}

\appendix

\section{Study Details}

\subsection{Formative Study}\label{section:appendix/formativestudy}
As described in Section~\ref{section:designrequirements}, we conducted semi-structured interviews with Wikipedians who self-identified themselves with one or more of the roles listed in Table~\ref{table:formative}. Here, we provide our interview questions for reference:
\begin{itemize}
    \item Please describe your experience with AI tools, such as ORES, for counter-vandalism, new page review, or other tasks on Wikipedia.
    \item Please share your experience as a [participant’s role] within your community on Wikipedia.
    \item How did your community make decisions about the design and use of AI tools?
    \item How were you involved in this decision-making process?
    \item How did your community evaluate whether these AI tools fit the community’s needs and values before or after the deployment?
    \item How were you involved in the evaluation process?
    \item How did community members collaborate and resolve disagreements during the evaluation process?
    \item Was the evaluation process effective or not? Why?
    \item Are there forms of support that would be particularly helpful to have from your perspective as a [participant’s role]?
    \item Are there forms of support that would be particularly helpful to have for the entire community?
    \item Did the evaluation change the community's perception and acceptance of AI tools? How?
    \item Have you ever participated in data labeling campaigns on Wikipedia? If so, would you please describe your experience?
    \item Can you envision ways to better support communities in evaluating AI tools before they are deployed to make more informed decisions about whether or not the community should adopt them?
\end{itemize}

\subsection{Field Study}
\subsubsection{Exit interview questions}\label{section:appendix/exitinterview}
As described in Section~\ref{fieldstudy/protocol}, we conducted an exit interview with each participant once they completed the field study to learn about their experiences and gather feedback. Here, we provide our interview questions for reference:
\begin{itemize}
    \item What is your overall experience using Wikibench?
    \item What is your best and worst experience?
    \item Does this process provide the community with agency over the curation of evaluation datasets?
    \item How well does this process align with Wikipedian’s norms for editing and discussion?
    \item How well does Wikibench support Wikipedians in discussing and resolving disagreements?
    \item Do you feel the primary labels are the result of community consensus?
    \item How do you think Wikibench could be better designed to support consensus building?
    \item How well does Wikibench fit into Wikipedia’s interface and workflow?
    \item How do you think Wikibench could fit better?
    \item Were you able to get a good overview of the data curation progress using Wikibench?
    \item Do you feel Wikibench shows all the data and edits with transparency? Is it good or bad?
    \item Is there anything you would like to be able to keep track of?
    \item How do you think we can improve Wikibench?
\end{itemize}

In addition to collecting feedback, we presented participants with a randomly selected set of edits from Wikibench's datasets, including their labels and the predictions from the ORES and Revert-Risk models (mentioned in Section~\ref{section:studycontext/challenges} and~\ref{section:findings/dataset/evaluation}). Specifically, we presented these edits in a table, with each row featuring an edit's ID, a link to its entity page, its primary label in Wikibench's dataset, and the models' predicted probabilities of the edit being damaging or reverted. The table displayed a random set of ten edits at a time and resampled each time as participants reloaded their browsers.

\subsubsection{Exit interview thematic analysis}\label{section:appendix/exitthemes}
In Table~\ref{table:exitinterviewthemes}, we provide a summary of the seven highest-level themes we identified through a reflexive thematic analysis of the exit interviews. We also list the sections in which each is discussed in the paper. Due to the limitation of word counts, we do not include the 17 second-level themes, 64 first-level themes, and 249 codes in the table.

\begin{table*}[t]
  \caption{The seven highest-level themes we identified through data analysis and sections where each is discussed in the paper.}
  \label{table:exitinterviewthemes}
  \begin{tabular}{p{0.65\textwidth}p{0.17\textwidth}}
    \toprule
    Highest-Level Themes & Relevant Paper Sections \\
    \midrule
    Participants perceive Wikibench as effective in surfacing disagreements and facilitating the development of a shared consensus. & 8.1.1, 8.1.2, 9.3.3 \\
    \midrule
    Participants appreciate Wikibench's collaborative approach to data labeling because it allows contributors to build consensus and a stronger community. & 8.1.3, 9.3.1 \\
    \midrule
    Participants find data produced by Wikibench helpful in understanding gaps between different AI models' predictions and community consensus. & 8.2 \\
    \midrule
    Participants perceive that Wikibench's user interface and process fit naturally into Wikipedia's existing interface, workflow, and community norms. & 9.1 \\
    \midrule
    Participants perceive that Wikibench provides them with the agency to collectively shape and reflect on the data curation process. & 9.2 \\
    \midrule
    Participants find Wikibench's campaign and entity pages helpful for quickly pinpointing edits where more labels or discussions may be valuable. & 9.3.2 \\
    \midrule
    Participants believe the transparency Wikibench provides into the data and the process by which it is curated is essential for evaluation results to be trustworthy to the community. & 9.3.4 \\
    \bottomrule
  \end{tabular}
\end{table*}

\begin{table*}[t]
  \caption{Validation study participant demographics, including their self-identified experience and frequency of patrolling edits, registration year, edit count on English Wikipedia, and geographic location. A dash indicates that they chose not to provide that information.}
  \label{table:validationparticipants}
  \begin{tabular}{lccccl}
    \toprule
    Participant ID & Patrol Experience & Patrol Frequency & Registered Since & Edit Count & Location \\
    \midrule
    V1 & Years & Weekly & 2021 & 6.6k & -- \\
    V2 & Years & Monthly & 2018 & 10k & United States \\
    V3 & Years & Daily & 2008 & 55k & United States \\
    V4 & Years & Weekly & 2010 & 12k & United States \\
    V5 & Months & Daily & 2022 & 20k & United Kingdom \\
    \bottomrule
  \end{tabular}
\end{table*}

\subsection{Validation Study}
\subsubsection{Edit selection procedure}\label{section:appendix/edit}
We selected the edits for the validation study based on a few considerations. First, these edits should have labels from Wikilabels and Wikibench for comparison. We achieved this by sampling existing, labeled edits from Wikilabel's dataset and having the field study participants provide initial labels for these edits using Wikibench during onboarding sessions, so that they would be added to Wikibench's dataset. To ensure participants did not focus on these edits more than they would otherwise, they were not told that these edits would be used in a validation study. Secondly, due to the limited onboarding time, we asked each participant to label only a small number of edits, which in turn limited the total number of edits we could sample from Wikilabels' dataset in the first place. Finally, considering the total limit and our goal of assessing Wikibench's label quality resulting from participants' navigation of consensus, disagreement, and uncertainty, we oversampled edits that were likely to spark discussion while also including straightforward ones that were more likely to receive unanimous labels, as described below. Given these three considerations, we sampled 90 edits from Wikilabels' datasets and had each of the 12 field study participants label a random subset of 15 edits. This design ensured that each edit was labeled by at least two participants, the minimum number needed to kickstart discussion.

In order to sample 90 edits from Wikilabels' dataset\footnote{\url{https://labels.wmflabs.org/stats/enwiki/41}} that were likely to spark discussion, we first identified 4,407 edits with both edit damage and user intent labels available in the dataset when we conducted the study. We excluded 127 edits that were hidden from public view by Wikipedia administrators and 18 edits that were labeled by more than one person due to Wikilabels' system race conditions. Among the remaining 4,262 edits, we categorized edits into three categories: (1) potentially ambiguous edits, (2) contested edits, and (3) other edits (cf.~\cite{chen2023judgment,gordon2021disagreement}). We considered \textit{potentially ambiguous edits} as those with the "unsure" mark specified in Wikilabels' dataset. We identified \textit{contested edits} as edits that had received higher-confidence labels (i.e., without the ``unsure'' mark), which were different from their actual reversion outcomes on Wikipedia (e.g., edits labeled as damaging in Wikilabels but didn't get reverted on Wikipedia). Given that these were cases where two Wikipedians had historically disagreed, we expected that these edits were ones for which Wikipedians are more likely differ in their perspectives. This categorization led to 365 potentially ambiguous edits, 660 contested edits, and 3,237 other edits. We sampled 30 edits from each category, resulting in 90 edits in total.

\subsubsection{Participant demographics}\label{section:appendix/demographics}
The demographic information of the five additional Wikipedians we recruited for the validation study is shown in Table~\ref{table:validationparticipants}.

\end{document}